%% file: conference_v4.tex
\documentclass[conference]{IEEEtran}
\IEEEoverridecommandlockouts

\usepackage{array,multirow}
\usepackage{booktabs} 
\usepackage{graphicx}
\usepackage{caption}
\usepackage{subcaption}
\usepackage{amssymb}
\usepackage{threeparttable}
\usepackage{booktabs}
\usepackage{cite}
\usepackage{bm}
\usepackage{amsmath,amssymb,amsfonts}
\usepackage{algorithmic}
\usepackage{graphicx}
\usepackage{mathtools}
\usepackage{textcomp}
\usepackage{xcolor}
\usepackage[hyphens]{url}

\usepackage[capitalize]{cleveref}
\crefname{section}{Sec.}{Secs.}

\def\BibTeX{{\rm B\kern-.05em{\sc i\kern-.025em b}\kern-.08em
T\kern-.1667em\lower.7ex\hbox{E}\kern-.125emX}}

\usepackage[acronyms,nonumberlist,nopostdot,nomain,nogroupskip]{glossaries}
\input{./acronyms.tex}

\usepackage{tikz}
\usepackage{pgfplots}
\usepackage{tikzscale}
\usepackage{tikz-qtree}
\input{./myTikz.tex}

\makeatletter
\def\endthebibliography{%
\def\@noitemerr{\@latex@warning{Empty `thebibliography' environment}}%
\endlist
}
\makeatother

\def \sfwidth{0.9\linewidth}
\def \sfheight {0.6\linewidth}
\def \bpheight {0.95\linewidth}

\definecolor{color0}{HTML}{FFD700}
\definecolor{color1}{HTML}{FFB14E}
\definecolor{color2}{HTML}{FA8775}
\definecolor{color3}{HTML}{EA5F94}
\definecolor{color4}{HTML}{CD34B5}
\definecolor{color5}{HTML}{9D02D7}
\definecolor{color6}{HTML}{0000FF}

\begin{document}

\title{Temporal Characterization of XR Traffic with Application to Predictive Network Slicing}


\author{\IEEEauthorblockN{Mattia Lecci$^*$, Federico Chiariotti$^\dagger$, Matteo Drago$^*$, Andrea Zanella$^*$, and Michele Zorzi$^*$
\thanks{This work was partially supported by the National Institute of Standards and Technology (NIST) under award no. 60NANB21D127 and by the IntellIoT project under the H2020 framework grant no. 957218.
The work of M. Lecci was supported by Fondazione CaRiPaRo under grant ``Dottorati di Ricerca 2018.''}}

\IEEEauthorblockA{$^*$Department of Information Engineering, University of Padova, 35131 Padua, Italy\\
$^\dagger$Department of Electronic Systems, Aalborg University, 9220 Aalborg \O{}st, Denmark\\
Emails: $^*$\texttt{\{leccimat, dragomat, zorzi\}@dei.unipd.it}, $^\dagger$\texttt{fchi@es.aau.dk}
}
}

\maketitle

\begin{abstract}
    Over the past few years, \gls{xr} has attracted increasing interest thanks to its extensive industrial and commercial applications, and its popularity is expected to rise exponentially over the next decade. However, the stringent \gls{qos} constraints imposed by \gls{xr}'s interactive nature require \gls{ns} solutions to support its use over wireless connections: in this context, quasi-\gls{cbr} encoding is a promising solution, as it can increase the predictability of the stream, making the network resource allocation easier.
    However, traffic characterization of \gls{xr} streams is still a largely unexplored subject, particularly with this encoding. In this work, we characterize \gls{xr} streams from more than 4 hours of traces captured in a real setup, analyzing their temporal correlation and proposing two prediction models for future frame size. Our results show that even the state-of-the-art H.264 \gls{cbr} mode can have significant frame size fluctuations, which can impact the \gls{ns} optimization. 
    Our proposed prediction models can be applied to different traces, and even to different contents, achieving very similar performance. 
    We also show the trade-off between network resource efficiency and \gls{xr} \gls{qos} in a simple \gls{ns} use case.
\end{abstract}


\IEEEpeerreviewmaketitle

\begin{tikzpicture}[remember picture,overlay]
\node[anchor=north,yshift=-15pt] at (current page.north) {\parbox{\dimexpr\textwidth-\fboxsep-\fboxrule\relax}{
\centering\footnotesize This paper has been submitted to IEEE WoWMoM 2022. Copyright may change without notice.}};
\end{tikzpicture}
\vspace{-10pt}
\glsresetall

\section{Introduction}
\label{sec:introduction}
\glsresetall


Over the past few years, the rapid technological development of \glspl{hmd} and the strong push towards the virtual world caused by the COVID-19 pandemic have caused an explosion of the~\gls{xr} market, which includes technologies such as \gls{vr}, \gls{ar}, and \gls{mr}. Recent studies estimate hundreds of millions of users of these technologies in a time span of just 3 years~\cite{huaweiVrArWhitePaper}, requiring millions of new devices to be developed, produced, and shipped around the world for a business in the order of billions of dollars~\cite{huaweiArInsight}.

While the latest news on the \emph{metaverse} seem to indicate that the fastest growth will be in the entertainment and social media industries, \gls{xr} is expected to make an impact in a huge variety of scenarios~\cite{oculusVr,qualcommXr}. Interactive design, marketing, healthcare, and employee training are just a few of the proposed scenarios, but industrial remote control in manufacturing and agriculture might have the largest impact, allowing human operators to remotely control machines in risky, hard to reach or unsafe environments, through a fully interactive virtual framework. 

One of the peculiarities shared by all these new applications is their interactive nature: users do not passively receive the information or stream a video, but need to manipulate the environment and affect it in meaningful ways, while maintaining an illusion of presence which requires the application to operate under very strict end-to-end delay constraints~\cite{3gpp.26.928,itu-t-f.743.10}. In particular, safety-critical and industrial applications will have stricter constraints, as the consequences of network impairments can be significantly more serious. \emph{Cybersickness} is another important issue, as a delay over 20~ms between movements and visual and auditory feedback can cause disorientation and dizziness~\cite{huaweiVrArWhitePaper,kim2017vrSickness}.

In order to fulfill these stringent latency requirements over a wireless connection, the application and the network need to cooperate. The \emph{\gls{ns}} paradigm~\cite{barakabitze20205g} allows 5G and Beyond networks to reserve resources to a given stream, defining \gls{qos} targets, but most efforts in this sense focus on relatively predictable applications. In this setting, the need for predictability in the \gls{xr} traffic becomes extremely important, leading to a resurgence of quasi-\gls{cbr} encoders, which are not used in passive streaming due to their lower picture quality stability. While some efforts have been devoted by prominent standard bodies on this topic~\cite{3gpp.26.928, itu-t-f.743.10}, the current availability of traffic models for \gls{xr} is scarce. Furthermore, to the best of our knowledge, no detailed analysis of the temporal statistics of quasi-\gls{cbr} video streams can be found in the literature, making existing scheduling schemes rely on uncertain foundations.

However, even \gls{cbr} encoders are not perfect, and the interplay between the video content and the movements and actions of the users may cause significant fluctuations. In this work, we analyze the traffic from a real \gls{vr} application using the Periodic-Intra Refresh mode of the H.264 codec, which results in relatively small differences in the frame sizes.
Modeling these imperfections, and consequently predicting the size of future frames in advance, can be extremely significant in the allocation of network resources, particularly if some critical \gls{qos} metrics have to be reached.
For example, this is the case for \textit{Cloud XR}, a new trend pursued by some major players in the telecommunication industry that moves the processing and rendering steps of the \gls{xr} content from the user to the Cloud, making the QoS requirements even more critical~\cite{nokiaCloudGaming,huawei2017cloudVr}. 

In this paper, we hence address the problem of providing a realistic stochastic characterization of an \gls{xr} traffic source.
Building upon our previous works~\cite{lecci21bursty,lecci2021open} where we collected more than 4 hours of live sessions with a real \gls{hmd} and performed basic traffic characterization, in this paper we take the analysis one step further by modeling the size of \gls{xr} frames in the stream as a correlated time series.  
We propose two parametric regression models to predict the size of future frames, and show that the behavior of the encoder can be generalized to other traces and even different applications with limited regression performance loss.
Finally, we present a simple network slicing use case, in which we show the trade-off between resource efficiency and latency for different types of resource scheduling.
All our traces as well as the analysis and simulation code is publicly available.\footnote{Code repository: {https://github.com/signetlabdei/vr-trace-analysis}}

The rest of the paper is structured as follows.
\cref{sec:soa} will discuss the current state of the art on \gls{xr} modeling, and our experimental setup and \gls{vr} application are briefly presented in~\cref{sec:xr_streaming_architecture}.
Our analysis is reported in~\cref{sec:video_trace_analysis}, while~\cref{sec:network_slicing_use_case} illustrates how our analysis can be leveraged for a simple \gls{ns} use case.
Finally, \cref{sec:conclusions} draws conclusions and presents some avenues for future work.

\section{State of the Art}
\label{sec:soa}

Despite a steady scientific interest in \gls{vr} since the 1990s~\cite{latta1994conceptual}, relatively little work has been done to characterize the details of this type of traffic.
With respect to our work, we can distinguish two main areas of research: the modeling and characterization of \gls{xr} traffic, and the scheduling and resource management of \gls{xr} data streams.

The former is closely related to 2D video content, and, even more so, to live, interactive applications such as video conferencing and gaming.
However, most of the work on the subject has focused on \gls{vbr} encoding, based either on the H.264 or the H.265 standards~\cite{tanwir2013vbrSurvey}, i.e., the customary encoding for streaming pre-generated video content.
\gls{vbr} can provide a stable visual quality, improving the user \gls{qoe}, but is also subject to significant jitter due to the large frame size fluctuations.
Transmitting \gls{vbr} videos with low latency can then be a significant challenge even over channels with constant capacity~\cite{liew98vbrOverCbr}.
On the other hand, \gls{cbr} encoding sacrifices some visual quality stability to obtain an encoded video stream with a stable transmission rate~\cite{mohsenian99cbr}.
Although the higher predictability of the encoded output makes \gls{cbr} encoding attractive for interactive video and \gls{xr} content, it is still relatively unexplored in the relevant literature.

A topic related to \gls{xr} traffic is video game streaming, also called \textit{Cloud gaming}.
These Cloud frameworks run games over a remote server, streaming the screen directly to the users without the need for client-side computation.
The stringent requirements of gaming applications, especially in terms of latency, and the need to address them with optimized protocols and new transmission strategies, have led to an increased interest in their characterization.
 
The authors of~\cite{carrascosa2020stadia} carried out an extensive measurement campaign in Google Stadia, a famous cloud gaming platform, giving an overview of its inner workings.
They studied the distributions of downlink traffic, packet size, inter-packet time under multiple settings, including different resolutions, video codecs, and network conditions.
On the other hand, in~\cite{didomenico2021analysis,graff2021gaminganalysis} direct comparisons were made between different cloud gaming platforms, mostly focusing only on the bitrate of the video stream, without including latencies or user experience.

A more comprehensive Cloud gaming testbed, including a full implementation with different network alternatives and automated trace acquisition over Ethernet, WiFi, and LTE, was presented in~\cite{pulla2021measuring}. 
This is surely an advantage in terms of reproducibility and speed of the experiments, but the unpredictability of the users' actions in gaming scenarios (and, more importantly, in \gls{xr}) is the real challenge that the network has to face, limiting the usefulness of the results.

These works represent a good starting point for the collection and modeling of \gls{xr} traffic, as it is reasonable to assume that most of these Cloud gaming companies will start providing \gls{xr} services soon.
However, most works focused specifically on \gls{xr} still consider simple applications, such as interactive data visualization~\cite{hentschel2009vrSimulation}, and do not provide much insight on the more complex scenarios.
There is an extensive literature on immersive video streaming~\cite{chiariotti2021vrSurvey}, but it has been mostly focused on passive applications in which the user is only a viewer, with different \gls{qoe} and encoding considerations.

Regarding \gls{xr} traffic scheduling and resource management, some works have already tried to propose schemes for efficient systems.
For example, in~\cite{chen18vrOverWireless,chen19correlation} game-theoretic approaches are proposed to tackle the optimization of multi-user \gls{vr} streaming over a small cell, with the help of machine learning.
The authors of~\cite{yang18mecVr} analyze the scheduling problem from the perspective of \gls{mec}, proposing scheduling strategies and analyzing communication, computing, and caching trade-offs.
While the models proposed for the network architectures considered in these works are extremely complex, there is no comparison with real-world \gls{vr} streaming.

To the best of our knowledge, our previous works, which proposed a simple architecture for collecting traffic traces from \gls{vr} games~\cite{lecci21bursty} and a simple generative model for the frame size~\cite{lecci2021open}, were the first to use real \gls{vr} traffic traces.
This work extends our previous ones by characterizing the temporal behavior of the \gls{xr} traces and drawing novel conclusions for \gls{ns} optimization.

\begin{figure*}[t!]
    \setlength\fheight{0.8\columnwidth}
    \setlength\fwidth{2\columnwidth}
    \centering
    \input{img/stem_legend.tex}

    \input{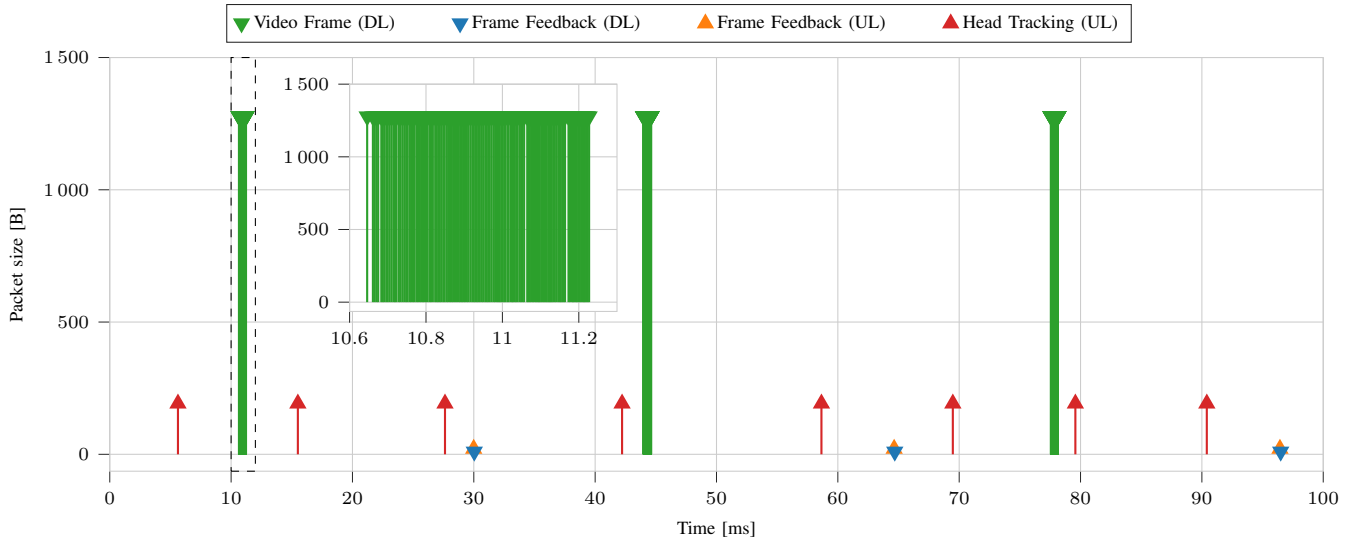}

    \caption{Portion of traffic trace from \emph{Virus Popper} (50~Mb/s, 30~FPS).
        For this trace, about 130--140 individual fragments make up each video frame burst.}
    \label{fig:stem}
\end{figure*}

\section{XR Streaming Architecture}
\label{sec:xr_streaming_architecture}

In this section, we describe the architecture of our \gls{xr} streaming acquisition and give some perspective on the full end-to-end setup.
To further understand what are the steps that most influence the \gls{xr} performance, it is useful to describe a common end-to-end \gls{xr} architecture.
First, we can start from the collection and processing of tracking information, delegated to the \gls{hmd}.
Then, this information is sent to a remote server to compose the viewport, i.e., what is actually shown to the user.
This process includes the rendering of the scene, the video encoding providing a more robust transmission towards the mobile device, and possibly some additional information e.g., the direction in which the rendered frame is supposed to be displayed.
After receiving and decoding the video stream together with all the additional meta-information, the \gls{hmd} generates the images to display at the occurring screen refresh.
These steps need to be accomplished with minimal delay to guarantee adequate \gls{qoe}.

Our experimental setup consisted of a desktop computer equipped with an NVIDIA GeForce RTX~2080~Ti graphics card acting as the rendering server, and an iPhone~XS enclosed in a \gls{vr} cardboard acting as the \gls{hmd}.
VR applications were thus run on the rendering server and streamed to the headset using the \emph{RiftCat~2.0} application (on the server), and \emph{VRidge} 2.7.7 (on the phone).\footnote{\url{https://riftcat.com/vridge}}

The application uses hardware-accelerated H.264 encoding via \gls{nvenc} as long as a compatible graphics card is present on the system.
\emph{RiftCat}'s developers disclosed that Periodic Intra-Refresh is used, a setting provided by the encoder that allows each frame to be roughly the same size, making the stream almost \gls{cbr} and thus easier to handle from a network perspective.
It does so by replacing key-frames by \emph{waves} of refreshed intra-coded blocks, i.e., blocks without any dependence on other frames, effectively spreading a key frame over multiple frames.
Image quality is balanced with resilience to packet loss by setting the \texttt{intraRefreshPeriod} parameter, which determines the period after which an intra refresh happens again, and the \texttt{intraRefreshCnt} parameter, which sets the number of frames over which the intra refresh happens~\cite{nvencApi}.
If we consider a 30~\gls{fps} video, a value of 30 for the \texttt{intraRefreshPeriod} would ensure that the frame is fully recovered every second.
On the other hand, choosing the value of \texttt{intraRefreshCnt} determines the number of frames over which the intra refresh will happen within an intra refresh period, with smaller values leading to a quicker refresh but lower quality.

Detailed information about the video encoder is of the utmost importance for our work, since different encoders typically behave differently, especially when analyzing the temporal behavior of the encoded source.
Still, we believe that our work offers network researchers a peek into the intricacies of this topic, showing some key results on how an \gls{xr} traffic flow can be analyzed for resource provisioning.

Different freely available games and applications were used to acquire our dataset, including \emph{Minecraft}, \emph{Virus Popper}, and \emph{Google Earth VR}.
Further details on the acquisition setup and our traces can be found in~\cite{lecci2021open}. In the following, we will mostly concentrate on one trace acquired using the \emph{Virus Popper} application, but the methodology holds throughout the dataset, and can be easily replicated for any of the other traces.

\begin{figure*}[t!]
    \setlength\fheight{0.55\columnwidth}
    \setlength\fwidth{0.9\columnwidth}
    \hspace*{\fill}%
    \begin{subfigure}[t]{0.45\textwidth}
        \centering
        \input{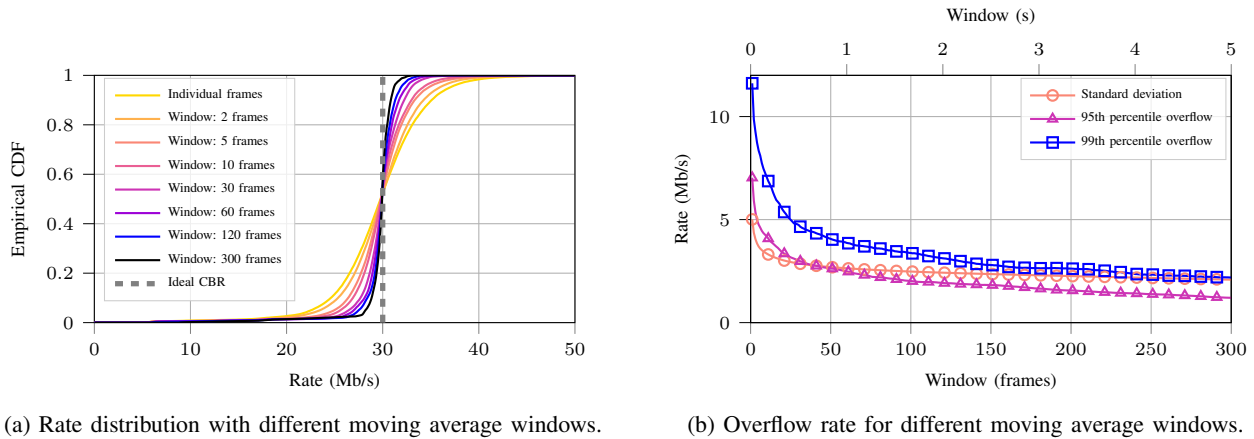}
        \caption{Rate distribution with different moving average windows.}
        \label{fig:rate_dist}
    \end{subfigure}
    \hspace*{\fill}%
    \begin{subfigure}[t]{0.45\textwidth}
        \centering
        \input{img/window_deviation_vp_30_60.tex}
        \caption{Overflow rate for different moving average windows.}
        \label{fig:rate_deviation}
    \end{subfigure}
    \hspace*{\fill}%
    \caption{Video frame size statistics for \emph{Virus Popper} (30~Mb/s, 60~FPS).}
    \label{fig:rate_window}
\end{figure*}

\section{Video Trace Analysis}
\label{sec:video_trace_analysis}

Analyzing the acquired traces, we determined that the application used \gls{udp} over IPv4. It also used an additional application-layer protocol header of variable size, which we decoded to determine the types of packets being exchanged. More specifically, synchronization and acknowledgment packets were exchanged in both directions, while the \gls{ul} stream from the \gls{hmd} to the rendering server also contained frequent and relatively small head-tracking information packets. Naturally, the \gls{dl} stream also had regular video frame packet bursts.

\cref{fig:stem} is a visual representation of a slice of bidirectional \gls{vr} streaming, showing the main data streams in both \gls{dl} and \gls{ul}.
As the figure clearly shows, most of the traffic is concentrated in \gls{dl} and is made up of packet bursts encoding video frames.
Video frame fragments were consistently found to be 1320~B long in all acquired traces, with a data size (the UDP payload) of 1278~B.

The low impact of non-video packets on the total streaming data rate, along with their strong dependence on the application setup, led us to focus exclusively on the video frame data, discarding all other packets from our analysis.
Our results can then be applied to any \gls{vr} application using the same encoder. 
By decoding the application protocol, we managed to identify frame boundaries and extract the video frame data, removing metadata and control information.
We can then consider the size of individual frames in a video trace.

The encoder makes use of the H.264 Periodic Intra-Refresh compression scheme to reduce the variation between frame sizes, so we do not expect a multimodal distribution, as would be the case for a classical keyframe-based encoding.
As we mentioned above, encoding \gls{vr} traffic as \gls{cbr} can be significantly better for network optimization, although it leads to a less stable picture quality: if all frames have the same size, it is possible for network slicing schemes to provide a guaranteed latency without wasting resources.

However, \gls{cbr} encoding is not perfect, and frames may still have variable size, although the average rate almost perfectly matches the required one.
We can use a simple \gls{ma} filter to examine the behavior of the \gls{vr} traffic on longer timescales, which is useful if resource allocation is performed at a slower pace.
Naturally, allocating resources every $N$ frames leads to a larger jitter between frames, but it can also improve the resource allocation efficiency, as size fluctuations will tend to average out over multiple frames.

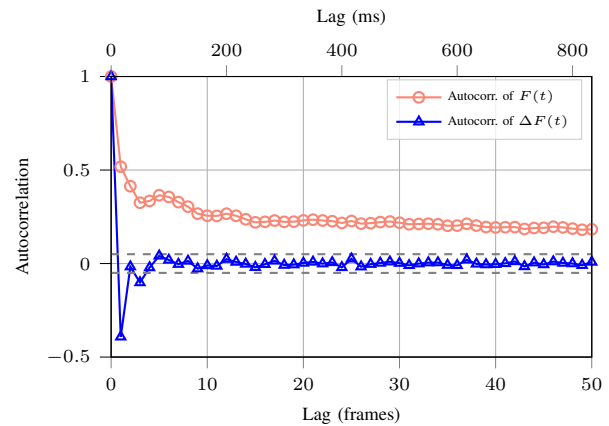
\begin{figure}[t!]
    \setlength\fheight{0.55\columnwidth}
    \setlength\fwidth{0.9\columnwidth}
        \centering
        \input{img/size_autocorr_vp_30_60.tex}
    \caption{Video frame size autocorrelation for \emph{Virus Popper} (30~Mb/s, 60~\gls{fps}).}
    \label{fig:autocorr}
\end{figure}

In order to measure this effect, we consider the \emph{Virus Popper} trace, with a required rate $R=30$~Mb/s and a $\varphi=60$~\gls{fps} refresh rate.
We only measure the video traffic, without packet headers and redundancy added by the application: this results in an average rate of 29.76~Mb/s.
Fig.~\ref{fig:rate_dist} shows the empirical \gls{cdf} of the rate, considering different \gls{ma} window sizes.
If we consider each frame individually, there is a significant variation, which gradually reduces as we increase the period over which the rate is measured.
It is also possible to notice that the frame size distribution is skewed towards smaller frame sizes, even for longer windows, as can be seen from the asymmetric tails of the distributions.

However, providing reliable service will require a significant overhead even if we relax the scheduling time: Fig.~\ref{fig:rate_deviation} shows the overflow rate (i.e., the difference between the actual rate and the expected 30~Mb/s \gls{cbr} rate) as a function of the \gls{ma} window.
The plot shows the standard deviation, as well as the 95\textsuperscript{th} and 99\textsuperscript{th} percentile overflow rates.
If our aim is to provide 99\% reliability, we need to overprovision by more than 8~Mb/s (i.e., almost 30\% of the \gls{cbr} rate) even if we consider a timescale of 100~ms for resource allocation, i.e., 6~frames.
Even averaging over periods of multiple seconds leads to worst-case rates almost 4~Mb/s higher than the average, probably corresponding to highly dynamic content in the video or to how the \gls{cbr} encoder works.
Interestingly, the standard deviation does not decay as fast as the 95\textsuperscript{th} and 99\textsuperscript{th} percentile overflows for longer averaging windows, due to the fact that the frame size distribution is skewed towards smaller sizes, as previously highlighted.

\begin{figure}[t!]
    \setlength\fheight{0.6\columnwidth}
    \setlength\fwidth{0.75\columnwidth}
        \centering
        \input{img/size_rolling_autocorr_vp_30_60.tex}
    \caption{Rolling windowed $\Delta F$ autocorrelation for \emph{Virus Popper} (30~Mb/s, 60~\gls{fps}). The windows were 600 frames (10 s) long, with a time shift of 60 frames (1 s).}
    \label{fig:autocorr_rolling}
\end{figure}
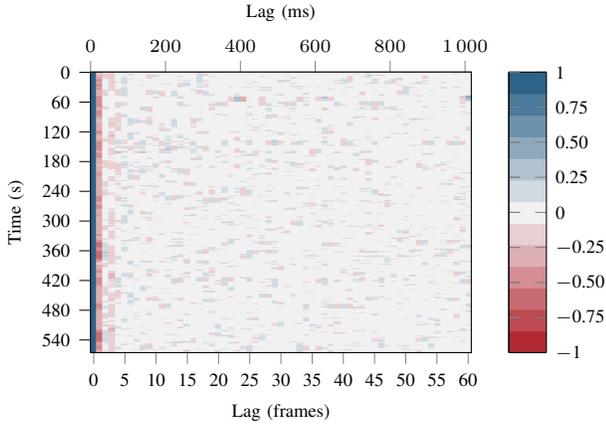

We can also analyze the autocorrelation of the frame size signal $F(t)$, to identify patterns in how the signal changes.
Fig.~\ref{fig:autocorr} shows the autocorrelation of $F(t)$ and $\Delta F(t)=F(t)-F(t-1)$.
While $F(t)$ has a strong long-term autocorrelation, due to the constant component, the $\Delta F(t)$ signal has a strong negative autocorrelation between one frame and the next, while almost all longer time differences fall within the $\pm$0.05 range.

\begin{figure*}[t!]
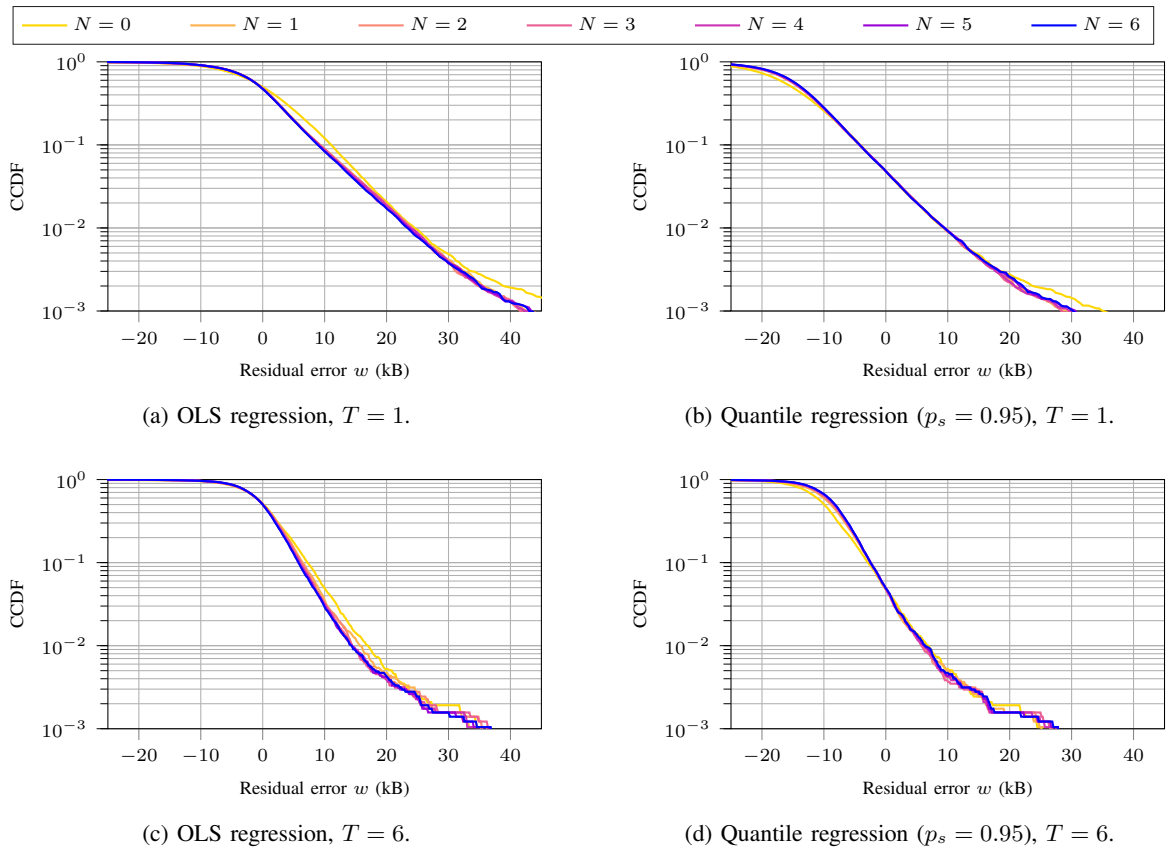

    \centering
    \begin{subfigure}[b]{\linewidth}
        \centering
	    \input{./img/legend_n.tex}
    \end{subfigure}	
    \\
       \begin{subfigure}[b]{.45\linewidth}
	    \centering
        \input{./img/ccdf_Linear_1_frames.tex}
        \caption{OLS regression, $T=1$.}
        \label{fig:o1_ccdf}
    \end{subfigure}	
	\begin{subfigure}[b]{.45\linewidth}
	    \centering
        \input{./img/ccdf_Quantile_1_frames.tex}
        \caption{Quantile regression ($p_s=0.95$), $T=1$.}
        \label{fig:q1_ccdf}
    \end{subfigure}
    \\\bigskip
    \begin{subfigure}[b]{.45\linewidth}
	    \centering
        \input{./img/ccdf_Linear_6_frames.tex}
        \caption{OLS regression, $T=6$.}
        \label{fig:o6_ccdf}
    \end{subfigure}	
	\begin{subfigure}[b]{.45\linewidth}
	    \centering
        \input{./img/ccdf_Quantile_6_frames.tex}
        \caption{Quantile regression ($p_s=0.95$), $T=6$.}
        \label{fig:q6_ccdf}
    \end{subfigure}		
    \caption{Complementary CDF of the residual error $w$ with $\tau=1$, for different values of $N$ and $T$.}
    \label{fig:ccdf}
\end{figure*}

This means that the encoder tends to balance out fluctuations between one frame and the next, such that a frame that is bigger than the previous one tends to be followed by a smaller one again.
We can check that this holds throughout the whole video by computing a rolling window autocorrelation, showed in Fig.~\ref{fig:autocorr_rolling} for $\Delta F(t)$. 
In this case, the plot clearly shows that there are no strong long-term correlations in any part of the video. 
The frame difference signal has a noticeable autocorrelation only with lag 1 and 3, confirming the result from Fig.~\ref{fig:autocorr}.

\subsection{Frame Size Prediction}

Let us consider the average size of future frames in the time interval $[t,t+T)$, given by 
\begin{equation}\label{eq:model}
    \overline{F_T}(t) = \frac{1}{T} \sum_{i=0}^{T-1} F(t+i).
\end{equation}
We denote by $\hat{F_T}(t,\tau)$ an estimate of $\overline{F_T}(t+\tau)$, $\tau>0$, i.e., considering a look-ahead of $\tau$ frames.
We focus on linear predictors based on the last $N\geq 0$ samples, so that 
\begin{equation}
    \hat{F_T}(t,\tau) =\theta_0 + \sum_{j=1}^{N} \theta_j F(t-j+1),
\end{equation}
where $\boldsymbol{\theta}=[\theta_0,\ldots,\theta_N]$ is a weight vector, which determines the accuracy of the estimate.
The difference between actual and estimated value is captured by the error term
$
w(t,\tau,T) = \overline{F_T}(t+\tau) - \hat{F_T}(t,\tau),
$
which will be denoted just as $w$ in the following, for ease of writing.

We can then consider two different regression methods to determine the value of the parameter vector $\bm{\theta}$:
\begin{itemize}
  \item \emph{\gls{ols} linear regression}: least squares regression was independently developed by Gauss and Legendre in the 19\textsuperscript{th} century~\cite{stigler1981gauss}, and is the most classic form of regression. In this case, the objective is to minimize the $\ell^2$ norm of the signal $w$. \gls{ols} regression can be useful in determining the average behavior of the underlying stochastic process, giving easily interpretable results on the quality of the prediction and the dynamics of the frame size over time;
  \item \emph{Quantile regression}~\cite{koenker1978regression}: this technique estimates $\hat F_T(t,\tau) $ so that the probability that it is higher than the real value, is not larger than $p_s$. This has obvious implications for the main objective of this paper, which is VR traffic modeling for network resource provisioning: as we are interested in providing enough resources to send a frame within the required latency with probability $p_s$, estimating the corresponding quantile might be the best way to get the required quality.
\end{itemize}
We also used \emph{Robust linear regression}~\cite{yu2017robust} to verify that the \gls{ols} prediction was not too sensitive to outliers. We considered a robust method using Huber's T norm instead of the $\ell^2$ norm: the two norms have the same quadratic behavior if the error is smaller than a threshold $\delta$, but Huber's T increases linearly for larger values. Setting the threshold to $\delta=\frac{\mathbb{E}\left[|F|\right]}{4}$, we found that the results matched exactly those of the \gls{ols} model, suggesting that outliers are not playing a relevant role in this case and thus letting us discard this model.

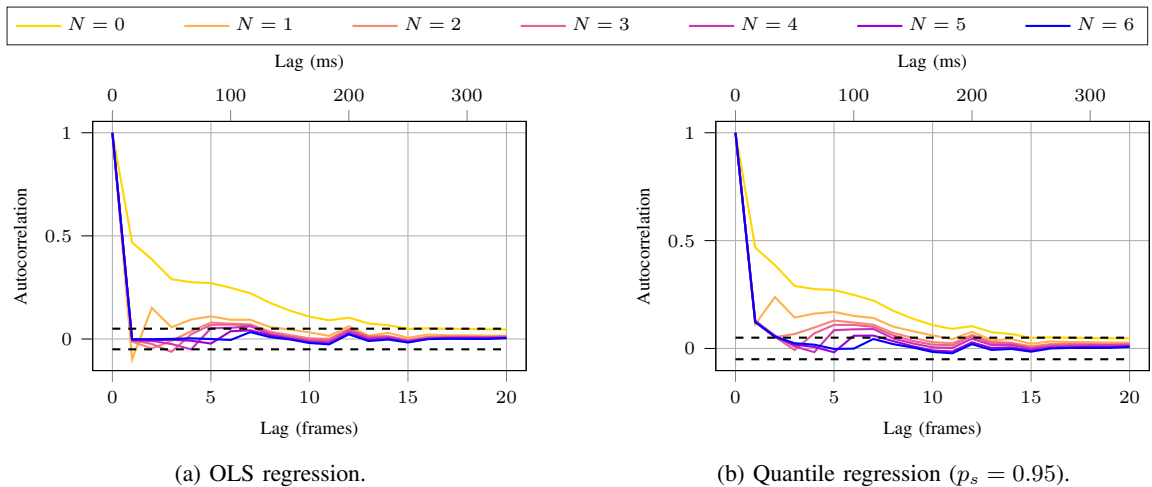
\begin{figure*}[t!]
    \centering
    \begin{subfigure}[b]{\linewidth}
        \centering
	    \input{./img/legend_n.tex}
    \end{subfigure}	
    \\
    \begin{subfigure}[b]{.45\linewidth}
	    \centering
        \input{./img/autocorr_Linear_1_frames.tex}
        \caption{OLS regression.}
        \label{fig:ols_aut}
    \end{subfigure}
	\begin{subfigure}[b]{.45\linewidth}
	    \centering
        \input{./img/autocorr_Quantile_1_frames.tex}
        \caption{Quantile regression ($p_s=0.95$).}
        \label{fig:quant_aut}
    \end{subfigure}		
    \caption{Autocorrelation of the residual error $w$ for next-frame prediction ($T=1$, $\tau=1$) for different values of $N$.}
    \label{fig:autocorr_pred}
\end{figure*}

In this section, we will show results for both the \gls{ols} and the quantile regression models. As we stated above, while the results from \gls{ols} are more immediate, quantile regression is useful when focusing on scheduling network resources for a \gls{vr} stream, which requires a model of the tail of the frame size distribution to provide latency guarantees.

We can now examine the results of the regression analysis for the \emph{Virus Popper} trace, considering a rate of 30~Mb/s and 60~\gls{fps}.
We focus on this video trace as the standard example in the paper, but other traces, even at different bitrates and frame rates, exhibit a similar behavior.
Fig.~\ref{fig:ccdf} shows the complementary \gls{cdf} of the residual error $w$, considering $\tau=1$ and two different values of $T$.
The first thing we can notice is that the error distribution has a slightly different shape for the \gls{ols} and quantile regression models, indicating that the difference in the two models is not simply a shift in the value of the intercept $\theta_0$, but instead the two predictions are meaningfully different.
We can also notice that there is some benefit from having a longer memory, although increasing $N$ yields diminishing returns.
Finally, we can confirm that the reliable transmission of this \gls{vr} content will require significant overprovisioning, even when using prediction: for $T=1$ the 95th percentile error of the \gls{ols} prediction is approximately 15~kB higher than the mean with any of the models, i.e., about 25\% of the average frame size (which is 62.5~kB for this trace).
In fact, this is close to the difference between the average predictions of the \gls{ols} and quantile models.

This difference is about halved for $T=6$, due to the fact that computing the average over multiple frames allows errors to compensate and cancel each other out.
However, provisioning over multiple frames means that only the average amount of resources will be scheduled for the stream, which will cause larger frames to have a higher latency, thus taking longer than $\frac{1}{\varphi}$ seconds to be delivered and causing additional queuing delay to subsequent frames.
Since the frame cannot be properly shown on screen until it is fully received, this translates to a higher jitter and reduces the \gls{qoe} perceived by the user, making a lower value of $T$ preferable.

Another fundamental component in evaluating the quality of a predictor is the autocorrelation of the residual error $w$: if the autocorrelation between subsequent samples of the residual error is high, the model did not capture some effect, usually due to an insufficient memory, i.e., too low a value of $N$.
Fig.~\ref{fig:autocorr_pred} shows the autocorrelation of $w$ for different values of $N$: it is easy to see that models with $N<4$, and particularly with $N=0$ and $N=1$, do not have enough memory to capture the frame size dynamics.
This is more evident in quantile regression, which shows a higher autocorrelation for these models.

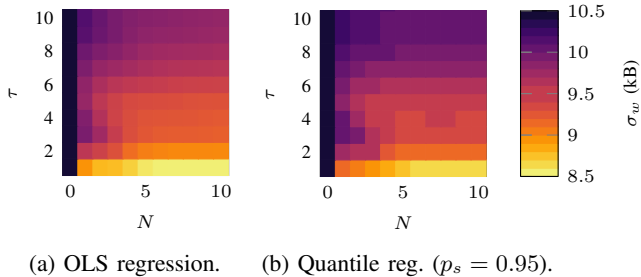
\begin{figure}[t!]
       \begin{subfigure}[b]{.45\linewidth}
	    \centering
        \input{./img/heatmap_Linear_std.tex}
        \caption{OLS regression.}
        \label{fig:oheat}
    \end{subfigure}
    \hspace{-3ex}
	\begin{subfigure}[b]{.45\linewidth}
	    \centering
        \input{./img/heatmap_Quantile_std.tex}
        \vspace{-2.75ex} 
        \caption{Quantile reg. ($p_s=0.95$).}
        \label{fig:qheat}
    \end{subfigure}	
    \caption{Heatmap of the residual error standard deviation (measured in kB) as a function of $N$ and $\tau$, with $T=1$.}
    \label{fig:heat}
\end{figure}

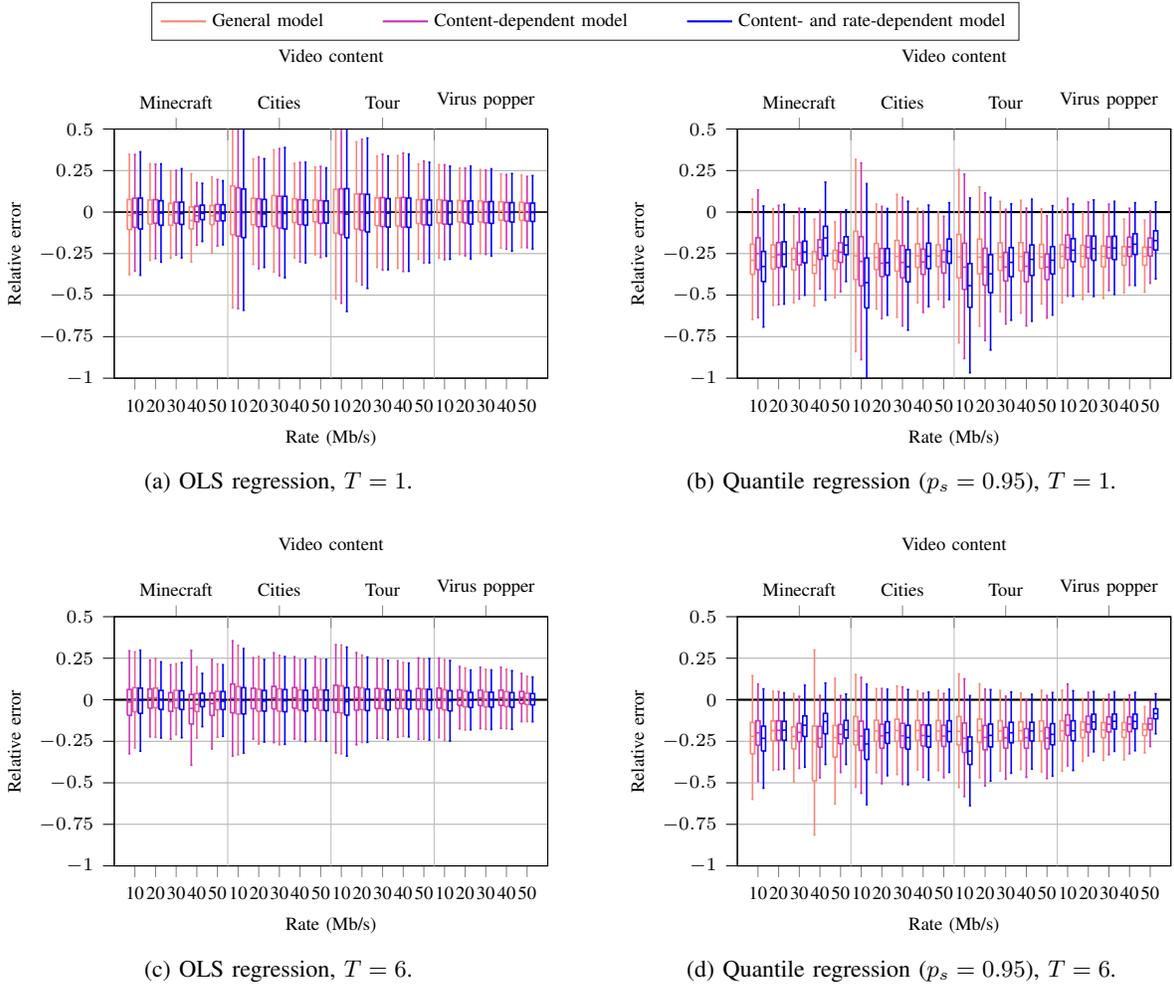
\begin{figure*}[t!]
    \centering
    \begin{subfigure}[b]{\linewidth}
        \centering
	    \input{./img/legend_gen.tex}
    \end{subfigure}	
    \\
       \begin{subfigure}[b]{.45\linewidth}
	    \centering
        \input{./img/gen_boxplot_Linear_1.tex}
        \caption{OLS regression, $T=1$.}
        \label{fig:ols_box1}
    \end{subfigure}	
    \begin{subfigure}[b]{.45\linewidth}
	    \centering
        \input{./img/gen_boxplot_Quantile_1.tex}
        \caption{Quantile regression ($p_s=0.95$), $T=1$.}
        \label{fig:quant_box1}
    \end{subfigure}
    \\\bigskip
    \begin{subfigure}[b]{.45\linewidth}
	    \centering
        \input{./img/gen_boxplot_Linear_6.tex}
        \caption{OLS regression, $T=6$.}
        \label{fig:ols_box6}
    \end{subfigure}	
	\begin{subfigure}[b]{.45\linewidth}
	    \centering
        \input{./img/gen_boxplot_Quantile_6.tex}
        \caption{Quantile regression ($p_s=0.95$), $T=6$.}
        \label{fig:quant_box6}
    \end{subfigure}		
    \caption{Boxplot of the relative residual error $\frac{\varphi w}{R}$ for different levels of generalization with $N=6$ and $\tau=1$.}
    \label{fig:box_gen}
\end{figure*}

Finally, we can examine the effect of $N$ and $\tau$ on the quality of the prediction by looking at Fig.~\ref{fig:heat}, which shows the standard deviation of the residual error $w$ as a function of these two parameters with $T=1$.
The figure clearly shows that increasing the memory of the model improves the prediction, but gives diminishing returns, as the difference between $N=6$ and $N=10$ is minimal.
Furthermore, we see an expected increase in the error if $\tau$ increases, but this is not monotonic for $N<3$: this might be due to the autocorrelation we observed in the $w$ signal, as $N<3$ is not sufficient to fully represent the state of the stochastic process, resulting in suboptimal predictions.

\subsection{Model Generalization}\label{ssec:general}

In the above, we studied how well regression models can predict future frame sizes $\hat{F_T}(t,\tau)$, but we always found the parameter vector $\bm{\theta}$ based on the same video trace. In the following, we study how prediction models perform when the regression is performed over multiple traces, with different bitrates and types of content. This has significant advantages, as finding a predictor for each specific video content requires acquiring traces for each content and quality level, while generalizing the predictor would allow for simpler deployment.

We consider $N=6$ and $\tau=1$, as we determined that $N=6$ is sufficient to capture the dynamics of the model.
In order to directly compare traces with different bitrates $R$ and frame rates $\varphi$, we normalize the video traces by the expected frame size $\varphi^{-1}R$, obtaining a normalized parameter vector $\tilde{\bm{\theta}}$, which, given the linearity of our models, can be converted back to the original parameter vector as $\bm{\theta}=\frac{R\tilde{\bm{\theta}}}{\varphi}$ of the regression model in~\eqref{eq:model}. By normalizing our frame sizes, we can train and use our models on multiple traces with different values of $R$ and $\varphi$.
We then consider three generalized models:
\begin{enumerate}
    \item A \gls{gm}, which computes $\bm{\theta}$ using the whole dataset, with different frame rates, bitrates, and video content types;
    \item A \gls{cm}, which computes $\bm{\theta}$ using a single type of content (e.g., the \emph{Virus Popper} game), but with different bitrates and frame rates;
    \item A \gls{crm}, which derives the parameter vector on a per-content, frame rate, and bitrate basis, i.e., a single trace.
\end{enumerate}

Given that different values of $R$ and $\varphi$ can have different scales of errors which can be difficult to compare directly, in \cref{fig:box_gen} we show the error normalized to the expected frame size $R/\varphi$.
As the figure shows, the model can generalize quite well: the performance of \gls{cm} is almost always similar to that obtained by \gls{crm}, making generalization across different bitrates and frame rates possible for the same video content.
On the other hand, \gls{gm} performs slightly worse, and has a large error in the Minecraft trace with $R=40$~Mb/s: it is possible that this trace involves different dynamics in the content or head movements, leading to sharp differences even with other traces with the same type of content.
On the other hand, \gls{gm} has similar performance to \gls{cm} and \gls{crm} with the \gls{ols} predictor, but shows a less consistent behavior for the quantile regressor.
For example, the \textit{Minecraft} trace with $R=40$~Mb/s shows very different performance between the three models and different values of $T$. Furthermore, the \emph{Virus Popper} trace seems to have a smaller tail, as \gls{gm} is more conservative than the models based only on that video content.

As we can see, using the quantile model leads to the prediction being between 25\% and 40\% higher than the average, skewing the error distribution.
We can also note that the relative error decreases with the bitrate: lower bitrate traces have a higher prediction error relative to the frame size, although the raw error $w$ is still larger for increasing bitrates.
As we noted above, averaging over multiple frames can also significantly reduce the error across almost all traces.

\section{Network Slicing Use Case}
\label{sec:network_slicing_use_case}

In this section, we consider an \gls{ns} use case for the models we developed in Sec.~\ref{sec:video_trace_analysis}. The \gls{vr} stream is assigned to a high-priority slice, with the objective of allowing each frame to be delivered before the generation of the next one, i.e., maintaining a latency below $1/60^{\rm th}$ of a second.
Provisioning the time and frequency resources for \gls{vr} is a critical component of Beyond 5G networks, and guaranteeing limited latency while reducing the impact on other users is an important application of our model.

We can then assume that the network slicing orchestrator is equipped with the \gls{cm} quantile regression from Sec.~\ref{ssec:general}, and can predict the frame sizes for arbitrary values of $T$ and $\tau$.
We consider an orchestrator that can make decisions on the resource allocation only at times $t=kS$, $k\in \mathbb{Z}$, i.e., every $S$~frames or, conversely, every $\Delta t = \frac{S}{\varphi}$~ms.
In the following, we consider queued bits from earlier frames in the scheduling as well.
At time $t=kS$, we consider that the previous slice might have been unable to send all the data in time, leaving in the queue $q_t$ bits that have to be sent in the following slices with an excess capacity of $q_t/\varphi$.

Recalling that our predictors are able to estimate values for $\hat{F_T}(t, \tau)$, as expressed in~\eqref{eq:model}, we thus propose two different models:
\begin{enumerate}
    \item \gls{cs}, which only allows the scheduler to set a constant slice capacity $C(t)$ for the next $S$ frames, i.e., only one prediction is performed, with $T=S$ and $\tau=1$:
    \begin{equation}
        C_{\text{CS}}(kS + \ell)=
        \hat{F_S}(kS,1) + \frac{q_t}{S\varphi}, \quad \ell=1, \dots, S,
    \end{equation}
    where the excess capacity from the queued bits is spread among the following $S$ frames;

    \item \gls{fs}, in which a different slice capacity $C(t)$ can be set for every inter-frame period in the next $S$ frames, i.e., there are $S$ independent predictions, with $T=1$ and $\tau\in\{1,\ldots,S\}$:
    \begin{equation}
        C_{\text{FS}}(kS + \ell)=\begin{cases}
            \hat{F_1}(kS, \ell) + \frac{q_t}{\varphi}, \, &\ell=1;\\
            \hat{F_1}(kS, \ell), \, &\ell=2, \dots, S,
            \end{cases}
    \end{equation}
    where the excess capacity from the queued bits is added entirely to the next frame to minimize latency.
\end{enumerate}

As we remarked in the previous section, the \gls{fs} scheme can reduce the jitter by having a more fine-grained prediction, as each frame will be allocated enough resources to be transmitted with probability $p_s$.
On the other hand, the \gls{cs} scheme has a rougher prediction, with consequently higher jitter, but will waste fewer network resources, as it can allow larger frames to be compensated by smaller ones before and after them.
Both models are realistic, as they work under different assumptions: in the first case, the resources that are allocated for each frame need to be over both time and frequency, while the second case gives the slice a constant bandwidth over the scheduling interval, the most common slicing model in the literature.

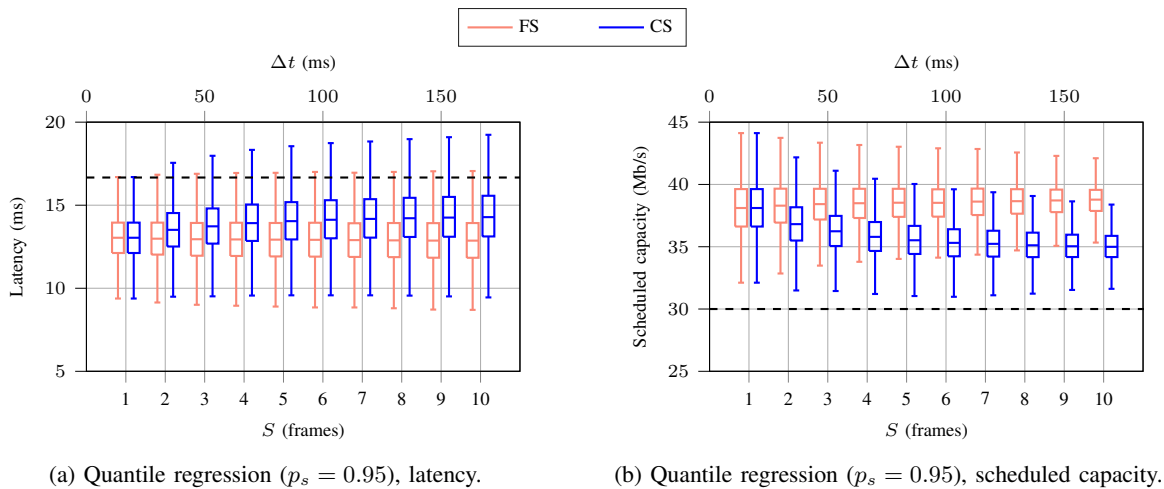
\begin{figure*}[t!]
    \centering
    \begin{subfigure}[b]{\linewidth}
        \centering
	    \input{./img/legend_sched.tex}
    \end{subfigure}	
    \\
    \begin{subfigure}[b]{.45\linewidth}
	    \centering
        \input{./img/schedule_Quantile_latency_granularity.tex}
        \caption{Quantile regression ($p_s=0.95$), latency.}
        \label{fig:quant_latbox}
    \end{subfigure}
    \begin{subfigure}[b]{.45\linewidth}
	    \centering
        \input{./img/schedule_Quantile_rate_granularity.tex}
        \caption{Quantile regression ($p_s=0.95$), scheduled capacity.}
        \label{fig:quant_schbox}
    \end{subfigure}	   
    \caption{Boxplot of scheduling performance for \gls{fs} and \gls{cs} with $\varphi=60$~\gls{fps}, $R=30$~Mb/s, $N=6$, and $\tau=1$.}
    \label{fig:latsch_gran}
\end{figure*}

\begin{figure*}[t!]
    \centering
    \begin{subfigure}[b]{.45\linewidth}
	    \centering
        \input{./img/perc_latency_6.tex}
        \caption{Latency.}
        \label{fig:perc_lat_ps}
    \end{subfigure}
    \begin{subfigure}[b]{.45\linewidth}
	    \centering
        \input{./img/perc_schedule_6.tex}
        \caption{Scheduled capacity.}
        \label{fig:perc_sch_ps}
    \end{subfigure}	   
    \caption{Average and worst-case percentiles of the latency and scheduled capacity with $S=6$ (100~ms), $N=6$, and $\tau=1$, as a function of the quantile regression parameter $p_s$.}
    \label{fig:ps_perc}
\end{figure*}
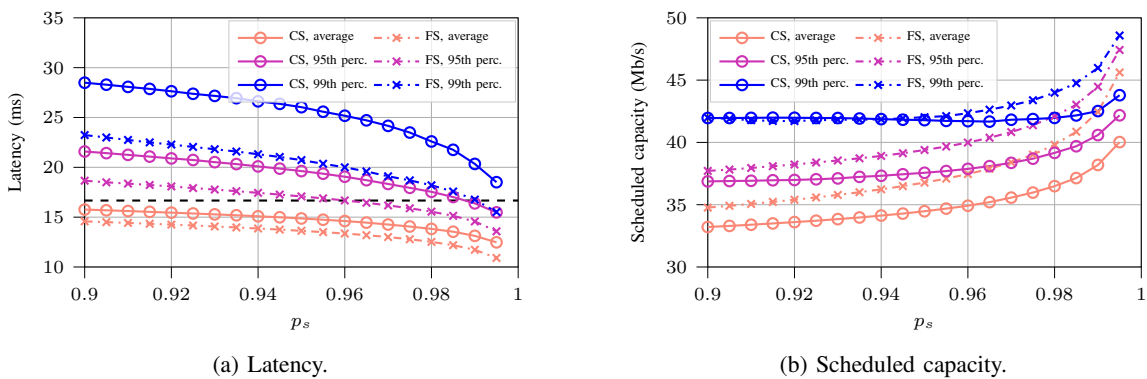

We can then look at the schedulers' performance as a function of $S$, setting $p_s=0.95$ and $N=6$:
Fig.~\ref{fig:latsch_gran} shows boxplots of the latency and scheduled capacity for \gls{fs} and \gls{cs}.
Fig.~\ref{fig:quant_latbox} clearly shows that, while the scheduler granularity has a limited effect on \gls{fs}, the
lower precision of \gls{cs} means that the longer the scheduling interval, the higher the average latency, and
the worst-case latency, represented by the upper whisker of the boxplots, increases even more.
On the other hand, as
Fig.~\ref{fig:quant_schbox} shows, the capacity required by \gls{cs} decreases as $S$ grows, while
the average capacity required by the \gls{fs} algorithm remains roughly constant irrespective of the value of $S$, but always higher than the capacity used by \gls{cs}.

This behavior is to be expected, as the errors in frame prediction can compensate over a longer window, but comes at the cost of a higher latency. Naturally, the choice between the two models depends not only on the desired point in the trade-off between \gls{qos} and resource efficiency, but also on the capabilities of the underlying system: state-of-the-art slicing frameworks often consider scheduling with a period $\Delta t=100$~ms, which would correspond to $S=6$ frames, and the granularity of the scheduling over time and frequency will dictate whether \gls{fs} is even an option.

It is also possible to simply increase the value of $C(t)$, e.g., by increasing $p_s$, in the \gls{cs} scheme to match the \gls{fs} performance in terms of latency, but \gls{cs} will always be less efficient for the same latency target. Fig.~\ref{fig:ps_perc} shows the scheduling performance as a function of the value of $p_s$. Naturally, a higher $p_s$ means a more conservative prediction of getting larger frames, which reduces the latency but increases the capacity requirements.
The closer we get to 1, the more increasing $p_s$ affects the latency, with a correspondingly larger increase in the capacity that is reserved to the \gls{xr} flow.
We can also notice that \gls{cs} requires a much higher value of $p_s$ to get the same performance as \gls{fs} in terms of latency.
A sensible example is to target a latency of one inter-frame interval, i.e., $\varphi^{-1}=16.67$~ms (the dashed line in \cref{fig:perc_lat_ps}), with a probability of 0.95 (the pink lines in \cref{fig:ps_perc}).
We notice that to meet this requirement, a value of $p_s\geq0.96$ has to be chosen for the \gls{fs} scheme, but the same requirement can only be fulfilled if $p_s\geq0.99$ using \gls{cs}.
This corresponds to an average scheduled capacity of at least 37.45~Mb/s for \gls{fs}, but 38.18~Mb/s for \gls{cs}. While the difference is not very significant, and the \gls{cs} scheme can be used without a big performance loss, choosing the correct value of $p_s$ to compensate for the scheduler's optimism is not simple, particularly in more complex network scenarios, while it is relatively straightforward for \gls{fs}.

\section{Conclusions}
\label{sec:conclusions}

This work aims at closing a gap in the literature on traffic source modeling: there are several analyses for passive streaming, both 2D and in immersive setups with \glsfirstplural{hmd}, and some for live gaming traffic in 2D, but none for interactive \gls{xr} with strict latency requirements and quasi-\gls{cbr} encoding. We analyzed live captures from a setup we devised, publishing both the dataset and the code for the analysis, and presented the performance of two regression models. The models are simple and flexible, can be generalized over different traces with limited performance loss, and can be used for provisioning. We also showed a simple \glsfirst{ns} scenario, which highlights the importance of the trade-off between resource efficiency and \gls{qoe}.

This is a first step towards fully designing an \gls{ns} system able to satisfy the stringent \gls{qos} requirements of \gls{xr} applications in industrial settings, in which the consequences of network failures are not only discomfort and nausea for the user, but also significant delays in production and even safety hazards. There are several additional analyses and opportunities for future work, that can be divided in two main directions.

The first potential avenue of research is a wider characterization, with different encoding parameters and even different encoders, and considering different applications, going beyond simple \gls{vr} games to include the industrial and commercial use cases we mentioned above, and a wider set of subjects. The traces should also integrate a record of the head movements of the users, as they correspond to shifts in the point of view of the \gls{xr} headset and are expected to be strongly correlated with frame size changes.

The other challenge is to actually design schemes and scheduling algorithms able to take into account the nature of the traffic and accommodate it, efficiently exploiting the prediction and adapting to the peculiarities of different communication technologies or even multiple independent links. In this sense, the allocation of resources over time and frequency, the prioritization of users and traffic types, and even the use of packet-level coding to protect the stream from link failures and deep fading events, can be promising avenues to design a solid framework to support \gls{xr} in mission-critical scenarios. The study of these techniques at all levels of the communication stack, simulating connection impairments in repeatable conditions through a full-stack network simulator, is our first priority in the ongoing work on this subject.

\bibliographystyle{IEEEtran}
\bibliography{./bibl} 

\end{document}

%% file: acronyms.tex
\newacronym{sla}{SLA}{Service Level Agreement}
\newacronym{urllc}{URLLC}{Ultra Reliable Low Latency Communication}
\newacronym{mmtc}{mMTC}{massive Machine Type Communication}
\newacronym{sdn}{SDN}{Software Defined Networking}
\newacronym{ns}{NS}{Network Slicing}
\newacronym{bbu}{BBU}{BaseBand Unit}
\newacronym{psc}{PSC}{Public Safety Communication}
\newacronym{lmr}{LMR}{Land Mobile Radio}
\newacronym{bb}{BB}{Broadband}
\newacronym{ps}{PS}{Public Safety}
\newacronym{uav}{UAV}{Unmaned Aerial Vehicle}

\newacronym{marl}{MARL}{Multi-Agent Reinforcement Learning}
\newacronym{drl}{DRL}{Deep Reinforcement Learning}
\newacronym{pomdp}{POMDP}{Partially Observable Markov Decision Process}
\newacronym{mdp}{MDP}{Markov Decision Process}
\newacronym{cnn}{CNN}{Convolutional Neural Network}
\newacronym{nn}{NN}{Neural Network}
\newacronym{radam}{RAdam}{Rectified Adam}
\newacronym{relu}{ReLU}{Rectified Linear Unit}
\newacronym{adam}{Adam}{Adaptive moment estimator}
\newacronym{ac}{AC}{Actor Critic}
\newacronym{a2c}{A2C}{Advance Actor Critic}
\newacronym{gnn}{GNN}{Graph Neural Network}
\newacronym{fl}{FL}{Federated Learning}
\newacronym{sarsa}{SARSA}{State-Action-Reward-State-Action}
\newacronym{mc}{MC}{Markov Chain}

\newacronym{fifo}{FIFO}{First In First Out}
\newacronym{cvo}{CVO}{Critical Voice}
\newacronym{cvi}{CVI}{Critical Video}
\newacronym{ncvo}{NCVO}{Non-Critical Voice}
\newacronym{ncvi}{NCVI}{Non-Critical Video}
\newacronym{ma}{MA}{Moving Average}

\newacronym{3dof}{3DoF}{3 Degrees of Freedom}
\newacronym{3gpp}{3GPP}{3rd Generation Partnership Project}
\newacronym{5g}{5G}{5\textsuperscript{th} Generation}
\newacronym{5gc}{5GC}{5G Core}
\newacronym{6dof}{6DoF}{6 Degrees of Freedom}
\newacronym{abft}{A-BFT}{Association-BeamForming Training}
\newacronym{adc}{ADC}{Analog to Digital Converter}
\newacronym{addts}{ADDTS}{Add Traffic Stream}
\newacronym{afbw}{AFBW}{Average Fading Bandwidth}
\newacronym{aid}{AID}{Association ID}
\newacronym{aifs}{AIFS}{Arbitration Inter-Frame Space}
\newacronym{aimd}{AIMD}{Additive Increase Multiplicative Decrease}
\newacronym{am}{AM}{Acknowledged Mode}
\newacronym{amc}{AMC}{Adaptive Modulation and Coding}
\newacronym{ampdu}{A-MPDU}{MAC Protocol Data Unit Aggregation}
\newacronym{aoa}{AoA}{Angle of Arrival}
\newacronym{aod}{AoD}{Angle of Departure}
\newacronym{ap}{AP}{Access Point}
\newacronym{api}{API}{Application Programming Interface}
\newacronym{app}{APP}{Application}
\newacronym{aqm}{AQM}{Active Queue Management}
\newacronym{ar}{AR}{Augmented Reality}
\newacronym{arf}{ARF}{Auto Rate Fallback}
\newacronym{arp}{ARP}{Address Resolution Protocol}
\newacronym{ati}{ATI}{Announcement Transmission Interval}
\newacronym{awgn}{AGWN}{Additive White Gaussian Noise}
\newacronym{awv}{AWV}{Antenna Weight Vector}
\newacronym{balia}{BALIA}{Balanced Link Adaptation}
\newacronym{bdp}{BDP}{Bandwidth-Delay Product}
\newacronym{ber}{BER}{Bit Error Rate}
\newacronym{bf}{BF}{Beamforming}
\newacronym{bframe}{B-frame}{Bipredictive-coded frame}
\newacronym{bhi}{BHI}{Beacon Header Interval}
\newacronym{bi}{BI}{Beacon Interval}
\newacronym{brp}{BRP}{Beam Refinement Protocol}
\newacronym{bs}{BS}{Base Station}
\newacronym{bss}{BSS}{Basic Service Set}
\newacronym{bti}{BTI}{Beacon Transmission Interval}
\newacronym{cad}{CAD}{Computer-aided Design}
\newacronym{cbap}{CBAP}{Contention-Based Access Period}
\newacronym{cbr}{CBR}{Constant Bit Rate}
\newacronym{cc}{CC}{Congestion Control}
\newacronym{cdf}{CDF}{Cumulative Distribution Function}
\newacronym{cir}{CIR}{Channel Impulse Response}
\newacronym{cn}{CN}{Core Network}
\newacronym{cp}{CP}{Control Plane}
\newacronym{cqi}{CQI}{Channel Quality Indicator}
\newacronym{crs}{CRS}{Cell Reference Signal}
\newacronym{csirs}{CSI-RS}{Channel State Information - Reference Signal}
\newacronym{csmaca}{CSMA/CA}{Carrier Sense Multiple Access with Collision Avoidance}
\newacronym{cts}{CTS}{Clear to Send}
\newacronym{dc}{DC}{Dual Connectivity}
\newacronym{dce}{DCE}{Direct Code Execution}
\newacronym{dcf}{DCF}{Distributed Coordination Function}
\newacronym{dci}{DCI}{Downlink Control Information}
\newacronym{delts}{DELTS}{Delete Traffic Stream}
\newacronym{dl}{DL}{Downlink}
\newacronym{dmg}{DMG}{Directional Multi-Gigabit}
\newacronym{dmr}{DMR}{Deadline Miss Ratio}
\newacronym{dmrs}{DMRS}{DeModulation Reference Signal}
\newacronym{dti}{DTI}{Data Transmission Interval}
\newacronym{e2e}{E2E}{End-to-End}
\newacronym{ecn}{ECN}{Explicit Congestion Notification}
\newacronym{edca}{EDCA}{Enhanced Distributed Channel Access}
\newacronym{edf}{EDF}{Earliest Deadline First}
\newacronym{embb}{eMBB}{Enhanced Mobile BroadBand}
\newacronym{enb}{eNB}{evolved Node Base}
\newacronym{endc}{EN-DC}{E-UTRAN-\gls{nr} \gls{dc}}
\newacronym{epc}{EPC}{Evolved Packet Core}
\newacronym{es}{ES}{Edge Server}
\newacronym{ese}{ESE}{Extended Schedule Element}
\newacronym{fdd}{FDD}{Frequency Division Duplexing}
\newacronym{fdma}{FDMA}{Frequency Division Multiple Access}
\newacronym{fec}{FEC}{Forward Error Correction}
\newacronym{fov}{FoV}{Field-of-View}
\newacronym{fps}{FPS}{Frames per Second}
\newacronym{fr2}{FR2}{Frequency Range 2}
\newacronym{ftp}{FTP}{File Transfer Protocol}
\newacronym{gmm}{GMM}{Gaussian Mixture Model}
\newacronym{gnb}{gNB}{Next Generation Node Base}
\newacronym[firstplural=Group of Pictures (GoPs)]{gop}{GoP}{Group of Pictures}
\newacronym{harq}{HARQ}{Hybrid Automatic Repeat reQuest}
\newacronym{hetnet}{HetNet}{Heterogeneous Network}
\newacronym{hh}{HH}{Hard Handover}
\newacronym{hmd}{HMD}{Head Mounted Device}
\newacronym{hol}{HOL}{Head-of-Line}
\newacronym{hqf}{HQF}{Highest-quality-first}
\newacronym{ia}{IA}{Initial Access}
\newacronym{iab}{IAB}{Integrated Access and Backhaul}
\newacronym{ibss}{IBSS}{Independent Basic Service Set}
\newacronym{id}{ID}{Identifier}
\newacronym{ifi}{IFI}{Inter-Frame Inter-arrival}
\newacronym{iframe}{I-frame}{Intra-coded frame}
\newacronym{imt}{IMT}{International Mobile Telecommunication}
\newacronym{imt2020}{IMT-2020}{International Mobile Te\-le\-com\-mu\-ni\-ca\-tion-2020}
\newacronym{inr}{INR}{Interference to Noise Ratio}
\newacronym{iot}{IoT}{Internet of Things}
\newacronym{ipa}{IPA}{Inter-Packet Arrival}
\newacronym{ism}{ISM}{Industrial, Scientific, and Medical}
\newacronym{itu}{ITU}{International Telecommunication Union}
\newacronym{kpi}{KPI}{Key Performance Indicator}
\newacronym{ks}{KS}{Kolmogorov-Smirnov}
\newacronym{lcf}{LCF}{Level Crossing Frequency}
\newacronym{lcm}{lcm}{least common multiple}
\newacronym{lcr}{LCR}{Level Crossing Rate}
\newacronym{los}{LoS}{Line-of-Sight}
\newacronym{lp}{LP}{Low Power}
\newacronym{lte}{LTE}{Long Term Evolution}
\newacronym{m2m}{M2M}{Machine to Machine}
\newacronym{ols}{OLS}{Ordinary Least Squares}
\newacronym{mac}{MAC}{Medium Access Control}
\newacronym{mcs}{MCS}{Modulation and Coding Scheme}
\newacronym{mec}{MEC}{Mobile Edge Cloud}
\newacronym{mi}{MI}{Mutual Information}
\newacronym{mib}{MIB}{Master Information Block}
\newacronym{mimo}{MIMO}{Multiple Input, Multiple Output}
\newacronym{ml}{ML}{Machine Learning}
\newacronym{mlr}{MLR}{Maximum-local-rate}
\newacronym[plural=\gls{mme}s,firstplural=Mobility Management Entities (MMEs)]{mme}{MME}{Mobility Management Entity}
\newacronym{mmw}{mmW}{Millimeter Wave}
\newacronym{moi}{MoI}{Method of Images}
\newacronym{mpc}{MPC}{Multi Path Component}
\newacronym{mpdu}{MPDU}{MAC Protocol Data Unit}
\newacronym{mptcp}{MPTCP}{Multipath TCP}
\newacronym{mr}{MR}{Mixed Reality}
\newacronym{mrdc}{MR-DC}{Multi \gls{rat} \gls{dc}}
\newacronym{msdu}{MSDU}{MAC Service Data Unit}
\newacronym{mss}{MSS}{Maximum Segment Size}
\newacronym{mtd}{MTD}{Machine-Type Device}
\newacronym{mtu}{MTU}{Maximum Transmission Unit}
\newacronym{mumimo}{MU-MIMO}{Multi-User Multiple Input, Multiple Output}
\newacronym{nav}{NAV}{Network Allocation Vector}
\newacronym{ncbr}{NCBR}{Non-Constant Bit Rate}
\newacronym{nfv}{NFV}{Network Function Virtualization}
\newacronym{nlos}{NLoS}{Non-Line-of-Sight}
\newacronym{nr}{NR}{New Radio}
\newacronym{mcmc}{MCMC}{Markov Chain Monte Carlo}
\newacronym{mse}{MSE}{Mean Square Error}
\newacronym{nrmse}{NRMSE}{Normalized Root Mean Square Error}
\newacronym{ns2}{ns-2}{Network Simulator 2}
\newacronym{ns3}{ns-3}{Network Simulator 3}
\newacronym{nsa}{NSA}{Non Stand Alone}
\newacronym{o2i}{O2I}{Outdoor-to-Indoor}
\newacronym{ofdm}{OFDM}{Orthogonal Frequency Division Multiplexing}
\newacronym{pa}{PA}{Position-aware}
\newacronym{pan}{PAN}{Personal Area Network}
\newacronym{pbch}{PBCH}{Physical Broadcast Channel}
\newacronym{pbss}{PBSS}{Personal Basic Service Set}
\newacronym{pcf}{PCF}{Point Coordinator Function}
\newacronym{pcp}{PCP}{\gls{pbss} Central Point}
\newacronym{pcpap}{PCP/AP}{\acrlong{pcp}/\acrlong{ap}}
\newacronym{pdcch}{PDCCH}{Physical Downlonk Control Channel}
\newacronym{pdcp}{PDCP}{Packet Data Convergence Protocol}
\newacronym{pdf}{PDF}{Probability Density Function}
\newacronym{pdsch}{PDSCH}{Physical Downlink Shared Channel}
\newacronym{pdu}{PDU}{Packet Data Unit}
\newacronym{per}{PER}{Packet Error Rate}
\newacronym{pf}{PF}{Proportional Fair}
\newacronym{pframe}{P-frame}{Predictive-coded frame}
\newacronym{pgw}{PGW}{Packet Gateway}
\newacronym{phy}{PHY}{Physical Layer}
\newacronym{ppdu}{PPDU}{PHY Protocol Data Unit}
\newacronym{ppp}{PPP}{Poisson Point Process}
\newacronym{prb}{PRB}{Physical Resource Block}
\newacronym{pss}{PSS}{Primary Synchronization Signal}
\newacronym{pucch}{PUCCH}{Physical Uplink Control Channel}
\newacronym{pusch}{PUSCH}{Physical Uplink Shared Channel}
\newacronym{qd}{QD}{Quasi Deterministic}
\newacronym{qoe}{QoE}{Quality of Experience}
\newacronym{qos}{QoS}{Quality of Service}
\newacronym{rach}{RACH}{Random Access Channel}
\newacronym{ran}{RAN}{Radio Access Network}
\newacronym[firstplural=Radio Access Technologies (RATs)]{rat}{RAT}{Radio Access Technology}
\newacronym{red}{RED}{Random Early Detection}
\newacronym{rf}{RF}{Radio Frequency}
\newacronym{rl}{RL}{Reinforcement Learning}
\newacronym{rlc}{RLC}{Radio Link Control}
\newacronym{rlf}{RLF}{Radio Link Failure}
\newacronym{rr}{RR}{Round Robin}
\newacronym{rrc}{RRC}{Radio Resource Control}
\newacronym{rrm}{RRM}{Radio Resource Management}
\newacronym{rs}{RS}{Remote Server}
\newacronym{rsrp}{RSRP}{Reference Signal Received Power}
\newacronym{rsrq}{RSRQ}{Reference Signal Received Quality}
\newacronym{rss}{RSS}{Received Signal Strength}
\newacronym{rssi}{RSSI}{Received Signal Strength Indicator}
\newacronym{rt}{RT}{Ray Tracer}
\newacronym{rts}{RTS}{Request to Send}
\newacronym{rtt}{RTT}{Round Trip Time}
\newacronym{rw}{RW}{Receive Window}
\newacronym{rx}{RX}{Receiver}
\newacronym{sa}{SA}{standalone}
\newacronym{sack}{SACK}{Selective Acknowledgment}
\newacronym{sap}{SAP}{Service Access Point}
\newacronym{sc}{SC}{Single Carrier}
\newacronym{sch}{SCH}{Secondary Cell Handover}
\newacronym{scm}{SCM}{Spatial Channel Model}
\newacronym{scoot}{SCOOT}{Split Cycle Offset Optimization Technique}
\newacronym{sdma}{SDMA}{Spatial Division Multiple Access}
\newacronym{sdr}{SDR}{Software Defined Radio}
\newacronym{semm}{SEMM}{SPCA-EDCA Mixed Mode}
\newacronym{si}{SI}{Study Item}
\newacronym{sib}{SIB}{Secondary Information Block}
\newacronym{sinr}{SINR}{Signal-to-Interference-plus-Noise Ratio}
\newacronym{sir}{SIR}{Signal-to-Interference Ratio}
\newacronym{sls}{SLS}{Sector-Level Sweep}
\newacronym{sm}{SM}{Saturation Mode}
\newacronym{snr}{SNR}{Signal-to-Noise Ratio}
\newacronym{son}{SON}{Self-Organizing Network}
\newacronym{sp}{SP}{Service Period}
\newacronym{spr}{SPR}{Service Period Request}
\newacronym{srs}{SRS}{Sounding Reference Signal}
\newacronym{ss}{SS}{Synchronization Signal}
\newacronym{sss}{SSS}{Secondary Synchronization Signal}
\newacronym{ssw}{SSW}{Sector Sweep}
\newacronym{sta}{STA}{Station}
\newacronym{stb}{STB}{Set Top Box}
\newacronym{tb}{TB}{Transport Block}
\newacronym{tbtt}{TBTT}{Target Beacon Transmission Time}
\newacronym[firstplural=Traffic Categories (TCs)]{tc}{TC}{Traffic Category}
\newacronym{tcp}{TCP}{Transmission Control Protocol}
\newacronym{tdd}{TDD}{Time Division Duplexing}
\newacronym{tdma}{TDMA}{Time Division Multiple Access}
\newacronym{tfl}{TfL}{Transport for London}
\newacronym{tgad}{TGad}{Task Group ad}
\newacronym{tgay}{TGay}{Task Group ay}
\newacronym{tm}{TM}{Transparent Mode}
\newacronym{trp}{TRP}{Transmitter Receiver Pair}
\newacronym{ts}{TS}{Traffic Stream}
\newacronym{tsconst}{TSCONST}{Traffic Scheduling Constraint}
\newacronym{tsf}{TSF}{Timing Synchronization Function}
\newacronym{tspec}{TSPEC}{Traffic Specification}
\newacronym{tti}{TTI}{Transmission Time Interval}
\newacronym{ttt}{TTT}{Time-to-Trigger}
\newacronym{tx}{TX}{Transmitter}
\newacronym[firstplural=Transmission Opportunities (TXOPs)]{txop}{TXOP}{Transmission Opportunity}
\newacronym{udp}{UDP}{User Datagram Protocol}
\newacronym{ue}{UE}{User Equipment}
\newacronym{ul}{UL}{Uplink}
\newacronym{um}{UM}{Unacknowledged Mode}
\newacronym{uma}{UMa}{Urban Macro}
\newacronym{uml}{UML}{Unified Modeling Language}
\newacronym{up}{UP}{User Priority}
\newacronym{utc}{UTC}{Urban Traffic Control}
\newacronym{vbr}{VBR}{Variable Bit Rate}
\newacronym{vm}{VM}{Virtual Machine}
\newacronym{vr}{VR}{Virtual Reality}
\newacronym{wbf}{WBF}{Wired Bias Function}
\newacronym{wf}{WF}{Wired-first}
\newacronym{wifi}{Wi-Fi}{Wireless Fidelity}
\newacronym{wigig}{WiGig}{Wireless Gigabit}
\newacronym{wlan}{WLAN}{Wireless Local Area Network}
\newacronym{xr}{XR}{eXtended Reality}

\newacronym{nvenc}{NVENC}{Nvidia Encoder}

\newacronym{gm}{GM}{\emph{general} model}
\newacronym{cm}{CM}{\emph{content-dependent} model}
\newacronym{crm}{CRM}{\emph{content- and rate-dependent} model}
\newacronym{fs}{FS}{\emph{Frame-by-frame} scheduling}
\newacronym{cs}{CS}{\emph{Constant} scheduling}

%% file: myTikz.tex
\pgfplotsset{compat=newest}
\pgfplotsset{plot coordinates/math parser=false}
\pgfplotsset{every axis/.append style={
                    label style={font=\scriptsize},
                    tick label style={font=\scriptsize},
                    legend style={font=\scriptsize}
                    }}
\usetikzlibrary{plotmarks,shapes,patterns,decorations.pathreplacing,backgrounds,calc,arrows,arrows.meta,spy,matrix,shadows,trees,positioning,fit}
\usepgfplotslibrary{patchplots,groupplots}

\tikzstyle{startstop} = [rectangle, rounded corners, minimum width=2cm, minimum height=0.5cm,text centered, draw=black]
\tikzstyle{io} = [trapezium, trapezium left angle=70, trapezium right angle=110, minimum width=3cm, minimum height=1cm, text centered, draw=black]
\tikzstyle{process} = [rectangle, minimum width=2cm, minimum height=0.5cm, text centered, draw=black, alignb=center]
\tikzstyle{decision} = [ellipse, minimum width=2cm, minimum height=1cm, text centered, draw=black]
\tikzstyle{arrow} = [thick,<->,>=stealth]
\tikzstyle{line} = [thick,>=stealth]
\tikzstyle{darrow} = [thick,<->,>=stealth,dashed]
\tikzstyle{sarrow} = [thick,->,>=stealth]
\tikzstyle{larrow} = [line width=0.1mm,dashdotted,->,>=stealth]

\pgfkeys{/pgf/number format/.cd,1000 sep={\,}} 

\makeatletter
\def\grd@save@target#1{%
  \def\grd@target{#1}}
\def\grd@save@start#1{%
  \def\grd@start{#1}}
\tikzset{
  grid with coordinates/.style={
    to path={%
      \pgfextra{%
        \edef\grd@@target{(\tikztotarget)}%
        \tikz@scan@one@point\grd@save@target\grd@@target\relax
        \edef\grd@@start{(\tikztostart)}%
        \tikz@scan@one@point\grd@save@start\grd@@start\relax
        \draw[minor help lines] (\tikztostart) grid (\tikztotarget);
        \draw[major help lines] (\tikztostart) grid (\tikztotarget);
        \grd@start
        \pgfmathsetmacro{\grd@xa}{\the\pgf@x/1cm}
        \pgfmathsetmacro{\grd@ya}{\the\pgf@y/1cm}
        \grd@target
        \pgfmathsetmacro{\grd@xb}{\the\pgf@x/1cm}
        \pgfmathsetmacro{\grd@yb}{\the\pgf@y/1cm}
        \pgfmathsetmacro{\grd@xc}{\grd@xa + \pgfkeysvalueof{/tikz/grid with coordinates/major step x}}
        \pgfmathsetmacro{\grd@yc}{\grd@ya + \pgfkeysvalueof{/tikz/grid with coordinates/major step y}}
        \foreach \x in {\grd@xa,\grd@xc,...,\grd@xb}
        \node[anchor=north] at (\x,\grd@ya) {\pgfmathprintnumber{\x}};
        \foreach \y in {\grd@ya,\grd@yc,...,\grd@yb}
        \node[anchor=east] at (\grd@xa,\y) {\pgfmathprintnumber{\y}};
      }
    }
  },
  minor help lines/.style={
    help lines,
    gray,
    line cap =round,
    xstep=\pgfkeysvalueof{/tikz/grid with coordinates/minor step x},
    ystep=\pgfkeysvalueof{/tikz/grid with coordinates/minor step y}
  },
  major help lines/.style={
    help lines,
    line cap =round,
    line width=\pgfkeysvalueof{/tikz/grid with coordinates/major line width},
    xstep=\pgfkeysvalueof{/tikz/grid with coordinates/major step x},
    ystep=\pgfkeysvalueof{/tikz/grid with coordinates/major step y}
  },
  grid with coordinates/.cd,
  minor step x/.initial=.5,
  minor step y/.initial=.2,
  major step x/.initial=1,
  major step y/.initial=1,
  major line width/.initial=1pt,
}
\makeatother

\newlength\fheight
\newlength\fwidth

%% file: img/stem_legend.tex
%
%

\definecolor{color0}{rgb}{1,0.498039215686275,0.0549019607843137}
\definecolor{color1}{rgb}{0.172549019607843,0.627450980392157,0.172549019607843}
\definecolor{color2}{rgb}{0.580392156862745,0.403921568627451,0.741176470588235}
\definecolor{color3}{rgb}{0.549019607843137,0.337254901960784,0.294117647058824}
\definecolor{color4}{rgb}{0.12156862745098,0.466666666666667,0.705882352941177}
\definecolor{color5}{rgb}{0.83921568627451,0.152941176470588,0.156862745098039}







\begin{tikzpicture}
\pgfplotsset{every tick label/.append style={font=\scriptsize}}

\begin{axis}[%
width=0,
height=0,
at={(0,0)},
scale only axis,
xmin=0,
xmax=0,
xtick={},
ymin=0,
ymax=0,
ytick={},
axis background/.style={fill=white},
legend style={legend cell align=center, align=center, draw=white!15!black, font=\scriptsize, at={(0, 0)}, anchor=center, /tikz/every even column/.append style={column sep=2em}},
legend columns=4,
]

\addplot [thick, color1, mark=triangle*, mark size=3, mark options={solid,rotate=180}, only marks]
table {%
0 0
};
\addlegendentry{Video Frame (DL)}

\addplot [thick, color4, mark=triangle*, mark size=3, mark options={solid,rotate=180}, only marks]
table {%
0 0
};
\addlegendentry{Frame Feedback (DL)}

\addplot [thick, color0, mark=triangle*, mark size=3, mark options={solid}, only marks]
table {%
0 0
};
\addlegendentry{Frame Feedback (UL)}

\addplot [thick, color5, mark=triangle*, mark size=3, mark options={solid}, only marks]
table {%
0 0
};
\addlegendentry{Head Tracking (UL)}

\end{axis}
\end{tikzpicture}%

%% file: img/window_deviation_vp_30_60.tex
\begin{tikzpicture}

\pgfplotsset{every tick label/.append style={font=\scriptsize}}

\begin{axis}[
width=\fwidth,
height=\fheight,
legend cell align={left},
legend style={font=\tiny, at={(0.99,0.99)}, anchor=north east, legend cell align=left, align=left,fill opacity=0.8, draw opacity=1, text opacity=1, draw=white!80!black},
tick align=outside,
tick pos=left,
x grid style={white!69.0196078431373!black},
xlabel={Window (frames)},
xmin=0, xmax=300,
xmajorgrids,
y grid style={white!69.0196078431373!black},
ylabel={Rate (Mb/s)},
ymin=0, ymax=12,
ymajorgrids,
ytick style={color=black}
]

\addplot [thick, color2,mark=o,mark repeat=10]
table {%
1 5.00902666100751
2 4.36391956614027
3 4.05556307731122
4 3.84247263193696
5 3.69478852382225
6 3.5929201537428
7 3.51703495689458
8 3.45521554293712
9 3.40187877845132
10 3.35295788296274
11 3.30794046870069
12 3.26681474261399
13 3.22990053689358
14 3.19623290727957
15 3.16484303651213
16 3.13513673208693
17 3.10722139887122
18 3.0812055804598
19 3.05678923048056
20 3.03390106310579
21 3.01256342022792
22 2.99270796354647
23 2.97412584341535
24 2.95665503542409
25 2.94007903058197
26 2.92447757017274
27 2.90960643438522
28 2.8954587040002
29 2.88205129168994
30 2.86936254494485
31 2.85728325471014
32 2.84570439129952
33 2.83461032885064
34 2.82398935826612
35 2.81379568600954
36 2.80395407168165
37 2.7944547076178
38 2.78534687396478
39 2.77655482154103
40 2.76802184596197
41 2.7597283902418
42 2.7516781481541
43 2.74386763072479
44 2.73624404045819
45 2.72882553973491
46 2.721611496171
47 2.71462592294805
48 2.70784552932416
49 2.70124096314229
50 2.69477911183199
51 2.68846815361907
52 2.68228617830618
53 2.67622036056978
54 2.67027946919437
55 2.66447643754034
56 2.65877589436982
57 2.65317000294093
58 2.6476485328791
59 2.64222529226311
60 2.63690973164567
61 2.63171077074027
62 2.62659728568922
63 2.62156270691922
64 2.61659233205332
65 2.61168569290097
66 2.60685002592331
67 2.60208850387723
68 2.59740618974862
69 2.59279957980711
70 2.58826341820214
71 2.5837962639282
72 2.5794014925867
73 2.57506895946782
74 2.57080140457142
75 2.56659517775026
76 2.56245837277296
77 2.55838570067995
78 2.55437664370719
79 2.55043360641146
80 2.54655388431072
81 2.54273856588537
82 2.53897060240079
83 2.53524116694896
84 2.53156672557092
85 2.52794169509452
86 2.52436643028007
87 2.52083567785404
88 2.51735385675713
89 2.51391723498251
90 2.51052589336222
91 2.50717168797606
92 2.50385194592967
93 2.50056849468366
94 2.49732322869159
95 2.4941143848959
96 2.49094138782288
97 2.48780587841255
98 2.48471533579151
99 2.48165934804907
100 2.47864031438313
101 2.47565725322757
102 2.47270365839636
103 2.46978285873502
104 2.46688820073159
105 2.46402592535468
106 2.46119375273038
107 2.45839123604102
108 2.45561237277373
109 2.45286478577121
110 2.45014287614483
111 2.44744213278508
112 2.44475832607284
113 2.44208677937187
114 2.43943440652665
115 2.43680039719532
116 2.43418731451847
117 2.43159314218463
118 2.42901971450719
119 2.42646479390257
120 2.42393004418425
121 2.42141910760274
122 2.4189367051411
123 2.41648102716505
124 2.41404511388225
125 2.41162333956549
126 2.40921682860711
127 2.40683296277415
128 2.4044694601837
129 2.40212825260536
130 2.39980473914204
131 2.39749773446355
132 2.39520738007832
133 2.3929316782305
134 2.39066989294771
135 2.38842022379664
136 2.3861839547134
137 2.38396202607728
138 2.38175269608344
139 2.37956047766042
140 2.37738370771358
141 2.37522103900792
142 2.37307175206106
143 2.37093541128851
144 2.36880952521006
145 2.36669646865908
146 2.36459710042473
147 2.36251217705489
148 2.36043855947565
149 2.35837357305557
150 2.3563163338487
151 2.35426866878285
152 2.35223010440329
153 2.35019957897065
154 2.34817181360167
155 2.34614892041024
156 2.34413070970674
157 2.34211623381966
158 2.34010659719847
159 2.33810174281951
160 2.33609861426352
161 2.33409871116407
162 2.33210256525377
163 2.33011068813952
164 2.32812166791821
165 2.32613566579735
166 2.3241538621216
167 2.32217945121128
168 2.32021267318563
169 2.31825183264119
170 2.31629641402185
171 2.314345101588
172 2.31239658732739
173 2.31045418434182
174 2.30851617971831
175 2.30658245516683
176 2.30465162104178
177 2.3027237160156
178 2.30079697669857
179 2.29887219250397
180 2.2969502192699
181 2.2950301470657
182 2.29311092123975
183 2.29119067731949
184 2.28927038709196
185 2.28735048768302
186 2.28543354985781
187 2.28351989715294
188 2.28160990494731
189 2.27970455057858
190 2.27780417288507
191 2.2759085685105
192 2.27401679658337
193 2.27212937541805
194 2.27024779200527
195 2.2683718785501
196 2.26650226255599
197 2.26463896052264
198 2.26278088607004
199 2.26092602355552
200 2.25907388974742
201 2.2572257225114
202 2.25538170712024
203 2.25354024446215
204 2.25170149318565
205 2.2498642433013
206 2.24802964534872
207 2.24619723205667
208 2.24436713244839
209 2.24253958438087
210 2.24071547934471
211 2.23889381772867
212 2.23707432996358
213 2.23525794717762
214 2.23344412842973
215 2.23163362598225
216 2.22982742365917
217 2.22802362220954
218 2.22622141854245
219 2.22442048073361
220 2.22262077010025
221 2.22082363800592
222 2.21902844914625
223 2.21723505142452
224 2.21544280983682
225 2.21365212645183
226 2.21186419517462
227 2.21007725375424
228 2.20829056927377
229 2.20650441998952
230 2.20472061175479
231 2.2029393467267
232 2.20115913299086
233 2.19938099009139
234 2.19760550228272
235 2.19583249336771
236 2.19406231384538
237 2.19229551256137
238 2.19053304098415
239 2.18877356767543
240 2.18701678904918
241 2.1852630888493
242 2.18351400821891
243 2.18176780321531
244 2.18002416175787
245 2.17828121781131
246 2.17653888514419
247 2.1747978401626
248 2.17305899687129
249 2.17132193322218
250 2.16958494245379
251 2.16784745295778
252 2.1661099218089
253 2.16437258009844
254 2.16263675586074
255 2.16090250279111
256 2.15916954921027
257 2.15743691315405
258 2.155705818194
259 2.15397685686949
260 2.15225013997392
261 2.15052570416173
262 2.14880252998395
263 2.14708147908411
264 2.14536192283662
265 2.14364412135527
266 2.14192708898646
267 2.14021150164782
268 2.13849795197404
269 2.13678612103464
270 2.13507574919383
271 2.13336754080068
272 2.13166196662404
273 2.12995928026783
274 2.12825913123816
275 2.12656166334772
276 2.12486675705062
277 2.12317343970266
278 2.12148067343197
279 2.11978897472427
280 2.11809888268614
281 2.11641066739336
282 2.11472421157424
283 2.11303885518077
284 2.11135519716988
285 2.10967326601717
286 2.10799258130856
287 2.10631301480696
288 2.10463456137468
289 2.10295699143846
290 2.10128066451023
291 2.09960526307326
292 2.09793205370011
293 2.09626104579068
294 2.09459200309238
295 2.09292538574471
296 2.09126143144201
297 2.08960089580192
298 2.08794266610941
299 2.08628621899916
300 2.08463145988984
};
\addlegendentry{Standard deviation}
\addplot [thick, color4,mark=triangle,mark repeat=10]
table {%
1 7.03152
2 5.92344
3 5.3584
4 4.97904
5 4.760544
6 4.61184
7 4.45467428571428
8 4.34622
9 4.26698666666667
10 4.19232
11 4.08183272727273
12 3.98251999999999
13 3.9020676923077
14 3.8364
15 3.764832
16 3.6921
17 3.62871529411764
18 3.54704
19 3.46380631578948
20 3.402816
21 3.35008
22 3.30333818181818
23 3.25550608695652
24 3.20976
25 3.16867199999999
26 3.13685538461539
27 3.10746666666667
28 3.07606285714285
29 3.03305379310345
30 3.00432
31 2.97733161290322
32 2.95533
33 2.92538181818182
34 2.89572705882352
35 2.86660114285714
36 2.83957333333333
37 2.81846918918919
38 2.79659368421052
39 2.78907076923077
40 2.769
41 2.75672195121951
42 2.74109714285714
43 2.72343069767442
44 2.7107890909091
45 2.69051733333333
46 2.6766052173913
47 2.66479659574468
48 2.65425999999999
49 2.65172571428571
50 2.628624
51 2.62055529411764
52 2.61138461538462
53 2.59860226415094
54 2.58761777777777
55 2.56745890909091
56 2.55324
57 2.53621052631579
58 2.52222620689655
59 2.5063240677966
60 2.489592
61 2.47116590163934
62 2.44902193548386
63 2.43231238095238
64 2.4088875
65 2.38687015384615
66 2.37601454545455
67 2.36415044776119
68 2.35174588235294
69 2.32726956521739
70 2.31812571428571
71 2.31187605633803
72 2.29421333333333
73 2.27602191780822
74 2.26350486486486
75 2.23980160000001
76 2.21858526315789
77 2.21606025974026
78 2.20926153846154
79 2.20448202531646
80 2.200614
81 2.19793185185185
82 2.19206634146342
83 2.18646361445784
84 2.17280571428571
85 2.15835105882353
86 2.14775441860465
87 2.13715862068965
88 2.13501818181818
89 2.12373033707865
90 2.12121066666667
91 2.10372395604396
92 2.09013391304347
93 2.076
94 2.06737531914893
95 2.05962442105263
96 2.04648
97 2.03852041237113
98 2.03332408163265
99 2.02871757575758
100 2.019696
101 2.00953188118812
102 2.00351529411765
103 1.98784310679611
104 1.98195230769231
105 1.97722971428572
106 1.97457509433962
107 1.96894654205608
108 1.96820888888889
109 1.96642348623853
110 1.97064
111 1.97411891891892
112 1.97748
113 1.97211185840708
114 1.96680842105263
115 1.95862539130435
116 1.95398068965518
117 1.94432
118 1.93852881355932
119 1.93390386554622
120 1.921864
121 1.91820694214876
122 1.91239868852459
123 1.91091512195123
124 1.9105935483871
125 1.91006592
126 1.90456380952381
127 1.9000894488189
128 1.89192
129 1.88372837209302
130 1.87756430769231
131 1.87562381679389
132 1.87433454545454
133 1.87134676691729
134 1.86742208955224
135 1.86505955555556
136 1.86040941176471
137 1.85671708029197
138 1.85605217391305
139 1.85800402877697
140 1.84890171428571
141 1.84871489361702
142 1.84224
143 1.83651692307692
144 1.83115999999999
145 1.82864110344827
146 1.83157150684931
147 1.82871836734693
148 1.82768756756757
149 1.82483758389262
150 1.821584
151 1.81910463576159
152 1.81740315789473
153 1.81708235294118
154 1.81190337662337
155 1.80707303225807
156 1.80228307692308
157 1.79471388535032
158 1.78741367088607
159 1.78368301886793
160 1.77787200000001
161 1.77054708074535
162 1.76431111111111
163 1.7573418404908
164 1.75174243902439
165 1.747008
166 1.74350168674699
167 1.73927856287425
168 1.73000857142857
169 1.72168047337278
170 1.71545505882353
171 1.71046736842105
172 1.70185395348837
173 1.69666404624277
174 1.68835034482758
175 1.68129188571429
176 1.67486454545454
177 1.66535593220339
178 1.66061662921348
179 1.65594368715084
180 1.65203733333333
181 1.64691712707182
182 1.64411340659341
183 1.63889049180328
184 1.63305913043478
185 1.62519437837838
186 1.61866838709678
187 1.61227893048128
188 1.60444085106383
189 1.60010158730158
190 1.59266526315789
191 1.58988816753926
192 1.58806999999999
193 1.58446756476684
194 1.57678762886598
195 1.57042461538462
196 1.56822612244898
197 1.5648633502538
198 1.5626496969697
199 1.56097447236181
200 1.55986800000001
201 1.55823522388059
202 1.55660198019802
203 1.54944236453202
204 1.54522823529412
205 1.54083746341463
206 1.5368481553398
207 1.53275826086956
208 1.52618307692308
209 1.52708669856458
210 1.52538514285715
211 1.51541004739337
212 1.50837509433962
213 1.50407661971831
214 1.50171140186916
215 1.49814251162791
216 1.49314222222222
217 1.48632552995392
218 1.48254385321101
219 1.47936
220 1.47782181818182
221 1.47667330316742
222 1.46987027027027
223 1.46614170403588
224 1.46339357142857
225 1.45868373333334
226 1.45278796460177
227 1.44884511013216
228 1.44558315789473
229 1.44442270742358
230 1.440864
231 1.43777246753247
232 1.43562620689655
233 1.43145888412017
234 1.43088820512821
235 1.42908459574468
236 1.42597220338983
237 1.42347949367089
238 1.42292369747899
239 1.42062928870292
240 1.417348
241 1.41406605809129
242 1.41001785123967
243 1.4052562962963
244 1.40505639344263
245 1.39985240816327
246 1.39363317073171
247 1.38868858299595
248 1.38696580645161
249 1.38472674698795
250 1.38297024
251 1.38107474103585
252 1.3801180952381
253 1.37449422924901
254 1.37109921259843
255 1.36924611764706
256 1.366996875
257 1.36494070038911
258 1.36367813953488
259 1.36260324324325
260 1.36083692307692
261 1.35430252873563
262 1.35183389312977
263 1.34604410646388
264 1.34109818181818
265 1.33789947169812
266 1.33278676691729
267 1.33078831460674
268 1.32752597014925
269 1.3222643866171
270 1.32035911111112
271 1.31794361623616
272 1.31615117647058
273 1.30877010989011
274 1.30673518248175
275 1.30161163636364
276 1.29952695652174
277 1.29477487364621
278 1.29042302158273
279 1.28556215053764
280 1.28097085714286
281 1.27648740213523
282 1.27554042553192
283 1.26913017667844
284 1.2634276056338
285 1.25697178947369
286 1.25150937062937
287 1.24563512195122
288 1.24167833333333
289 1.23721577854671
290 1.23621020689655
291 1.23184164948454
292 1.22963506849315
293 1.22566279863481
294 1.22095346938776
295 1.2179259661017
296 1.21378216216216
297 1.21121616161616
298 1.20878657718121
299 1.20715023411371
300 1.2046128
};
\addlegendentry{95th percentile overflow}
\addplot [thick, color6,mark=square,mark repeat=10]
table {%
1 11.61504
2 9.86904
3 9.26623999999999
4 8.73384
5 8.352192
6 8.10408
7 7.74041142857143
8 7.47396
9 7.2744
10 7.07116800000001
11 6.87072
12 6.71856
13 6.54705230769231
14 6.32252571428572
15 6.127424
16 5.95344
17 5.86198588235293
18 5.76658666666666
19 5.60574315789473
20 5.47723200000001
21 5.36781714285714
22 5.27424
23 5.15267478260869
24 5.07412
25 4.98875520000001
26 4.90872
27 4.81969777777778
28 4.80810857142857
29 4.77826758620689
30 4.737584
31 4.65564387096774
32 4.605195
33 4.56850909090909
34 4.53251294117648
35 4.497792
36 4.46962666666666
37 4.43281297297298
38 4.41731368421052
39 4.40152615384616
40 4.372764
41 4.33660097560976
42 4.30965714285714
43 4.28648930232558
44 4.25763272727273
45 4.22331733333334
46 4.18455652173913
47 4.14687319148936
48 4.12475
49 4.09000163265306
50 4.0644
51 4.03266823529412
52 4.00976307692308
53 3.98712452830188
54 3.96248
55 3.94911709090908
56 3.93998571428571
57 3.93356631578948
58 3.91142896551724
59 3.89957694915255
60 3.868712
61 3.85414032786885
62 3.84277161290322
63 3.80978285714286
64 3.78528
65 3.7633476923077
66 3.7448
67 3.73781014925373
68 3.73872
69 3.72197565217392
70 3.71695542857143
71 3.70326084507042
72 3.69273333333333
73 3.68249424657534
74 3.66696648648649
75 3.66376320000001
76 3.64415368421052
77 3.61267948051948
78 3.61897230769231
79 3.59967189873417
80 3.602616
81 3.59377777777778
82 3.58886048780487
83 3.56276819277109
84 3.55055428571428
85 3.5288865882353
86 3.50705860465116
87 3.5054951724138
88 3.50297454545454
89 3.49222112359551
90 3.46324266666667
91 3.45681758241759
92 3.44333217391304
93 3.43186064516129
94 3.42217531914893
95 3.4399427368421
96 3.42957499999999
97 3.4120775257732
98 3.40402775510204
99 3.38425212121212
100 3.3804768
101 3.36356910891089
102 3.36879529411765
103 3.35713398058252
104 3.34182461538462
105 3.33018057142857
106 3.31312301886792
107 3.28889271028038
108 3.27984888888889
109 3.2684609174312
110 3.25273745454545
111 3.23258378378378
112 3.22476857142857
113 3.20832
114 3.19602526315789
115 3.17991652173913
116 3.16848413793104
117 3.15664410256411
118 3.13587254237288
119 3.13319394957983
120 3.122176
121 3.11395834710743
122 3.10757508196722
123 3.08908487804878
124 3.07416774193548
125 3.06404736
126 3.0497219047619
127 3.03292346456693
128 3.02491125
129 3.01373023255815
130 2.99359384615385
131 2.98215938931298
132 2.96575636363636
133 2.94678857142857
134 2.93746029850746
135 2.93220977777778
136 2.92069764705883
137 2.91300788321168
138 2.90030956521738
139 2.88907165467626
140 2.87294742857143
141 2.85373617021276
142 2.85022309859155
143 2.84280503496504
144 2.83812666666667
145 2.82096
146 2.81020273972602
147 2.80031346938776
148 2.79508216216217
149 2.80317100671141
150 2.79255680000001
151 2.79182304635761
152 2.78111368421052
153 2.78228392156863
154 2.77589610389611
155 2.76611922580645
156 2.75214461538461
157 2.74351490445859
158 2.73048607594936
159 2.72809358490566
160 2.719593
161 2.70483875776397
162 2.69776296296296
163 2.69267337423313
164 2.69785463414634
165 2.68206254545455
166 2.66960385542168
167 2.66908455089821
168 2.66782571428571
169 2.66114556213017
170 2.65793788235294
171 2.66215859649122
172 2.66261860465116
173 2.65927768786126
174 2.65521655172413
175 2.65101257142857
176 2.64682363636364
177 2.64758779661017
178 2.64530966292135
179 2.64200044692738
180 2.64257066666667
181 2.63651270718232
182 2.63992087912088
183 2.64066360655737
184 2.63924608695652
185 2.64164237837838
186 2.64171096774194
187 2.64419165775401
188 2.64518553191489
189 2.64625015873016
190 2.64804884210526
191 2.64397068062828
192 2.6426275
193 2.6366549222798
194 2.63733030927835
195 2.63241353846153
196 2.63090204081632
197 2.63258558375634
198 2.62508848484848
199 2.6243191959799
200 2.620092
201 2.61668298507463
202 2.61007841584158
203 2.60812374384236
204 2.60619058823529
205 2.6016327804878
206 2.59709592233009
207 2.59657275362319
208 2.59717846153847
209 2.59938373205741
210 2.601088
211 2.59015507109005
212 2.57489886792452
213 2.5602185915493
214 2.55422355140186
215 2.54651386046512
216 2.54520222222222
217 2.54366599078341
218 2.54024366972477
219 2.52545534246575
220 2.52289963636363
221 2.51812995475114
222 2.51044756756757
223 2.49427156950673
224 2.48696571428571
225 2.47902293333333
226 2.47559575221239
227 2.45546220264318
228 2.45418736842105
229 2.44846742358079
230 2.45151443478261
231 2.44854649350649
232 2.43198827586207
233 2.42922025751073
234 2.41842256410256
235 2.41085412765958
236 2.40202576271187
237 2.3914146835443
238 2.38062857142857
239 2.37374661087866
240 2.358234
241 2.3564574273859
242 2.34604363636364
243 2.34070518518519
244 2.33927213114754
245 2.32851526530612
246 2.33572292682927
247 2.3256971659919
248 2.32191096774193
249 2.32100048192771
250 2.3252352
251 2.33136764940239
252 2.32213333333333
253 2.32136727272727
254 2.31886299212599
255 2.31487058823529
256 2.30666062500001
257 2.30903159533074
258 2.30274232558139
259 2.29622362934363
260 2.28783323076924
261 2.28415632183908
262 2.28864183206107
263 2.28916927756654
264 2.28562181818182
265 2.27993660377358
266 2.2784679699248
267 2.27896269662921
268 2.27750149253731
269 2.27500014869888
270 2.27097422222222
271 2.27078258302583
272 2.26786588235294
273 2.26230153846154
274 2.26248875912409
275 2.25496494545455
276 2.25281043478262
277 2.24271422382672
278 2.23741467625899
279 2.23253161290322
280 2.23138628571428
281 2.23697935943061
282 2.22905021276596
283 2.22138402826855
284 2.22040225352113
285 2.22280084210526
286 2.22330293706294
287 2.22238327526132
288 2.21271333333333
289 2.21376498269896
290 2.2104364137931
291 2.20618226804124
292 2.19981863013698
293 2.19966307167236
294 2.20005224489795
295 2.19867986440678
296 2.20012540540541
297 2.19695030303031
298 2.19623194630872
299 2.19119518394648
300 2.19051200000001
};
\addlegendentry{99th percentile overflow}

\end{axis}

\begin{axis}[
width=\fwidth,
height=\fheight,
tick align=outside,
tick pos=right,
xlabel={Window (s)},
xmin=0, xmax=5,
ymin=0, ymax=1,
ymajorticks = false,
]
    
\end{axis}
    
\end{tikzpicture}

%% file: img/size_autocorr_vp_30_60.tex
\begin{tikzpicture}

\pgfplotsset{every tick label/.append style={font=\scriptsize}}

\begin{axis}[
width=\sfwidth,
height=\sfheight,
legend cell align={left},
legend style={font=\tiny, at={(0.98,0.99)}, anchor=north east, legend cell align=left, align=left,fill opacity=0.8, draw opacity=1, text opacity=1, draw=white!80!black},
tick align=outside,
tick pos=left,
x grid style={white!69.0196078431373!black},
xlabel={Lag (frames)},
xmin=0, xmax=50,
xmajorgrids,
y grid style={white!69.0196078431373!black},
ylabel={Autocorrelation},
ymin=-0.5, ymax=1,
ymajorgrids,
ytick style={color=black}
]

\addplot [thick, color2,mark=o]
table {%
0 1
1 0.517836309350073
2 0.413433745231653
3 0.325157432676376
4 0.334787919362688
5 0.36519833379723
6 0.356017247272571
7 0.328556538997123
8 0.304253705535198
9 0.266788043173347
10 0.255763516076102
11 0.2548017444313
12 0.26646456353622
13 0.255464803188973
14 0.236750594796484
15 0.220307147882125
16 0.223062177290464
17 0.229831506934101
18 0.222898388658022
19 0.223897545031599
20 0.230686767907829
21 0.234647400834151
22 0.230260534822164
23 0.226602491583914
24 0.217071994617214
25 0.227452603733461
26 0.213386068697139
27 0.215787871258549
28 0.221592055961187
29 0.224567490702634
30 0.218419964178907
31 0.210120189862408
32 0.211239130223141
33 0.212972097197209
34 0.210054525646951
35 0.202210908472013
36 0.202675305478668
37 0.212528836200849
38 0.203010795535743
39 0.195270971362609
40 0.192884017775452
41 0.194220521732189
42 0.194832131188118
43 0.184805020749906
44 0.190071612671639
45 0.191038955662475
46 0.197135402897756
47 0.193320784862826
48 0.186847058531481
49 0.180483883732332
50 0.183084442475377
};
\addlegendentry{Autocorr. of $F(t)$}
\addplot [thick, color6,mark=triangle]
table {%
0 1
1 -0.391773918934805
2 -0.0167388583966961
3 -0.101523244932691
4 -0.021531004237653
5 0.0410247990022447
6 0.0189691616296346
7 -0.00329634684125523
8 0.0136480089590587
9 -0.0274076688622721
10 -0.0104279178233368
11 -0.013082646070798
12 0.0234735500234375
13 0.0079971119581818
14 -0.00235404027385207
15 -0.0198941317471495
16 -0.00415761941443526
17 0.014192500934362
18 -0.00822239184843067
19 -0.00599209756683808
20 0.0029341319357817
21 0.00863783139347187
22 -0.000749942477685081
23 0.00607456771605123
24 -0.0206171505779058
25 0.0253087640063166
26 -0.0170456500865272
27 -0.00352288647866168
28 0.00291153497114848
29 0.00944091159789572
30 0.00223142835361686
31 -0.00973955903802881
32 -0.000639128580590911
33 0.00481533889499011
34 0.00508919076522155
35 -0.00859679376871189
36 -0.00970065563295453
37 0.0200334177820948
38 -0.0018446008745478
39 -0.0055280800472064
40 -0.00385233849658026
41 0.000753202118417935
42 0.0109999169994948
43 -0.0158194972698275
44 0.00444128984054233
45 -0.00528620020506456
46 0.0102407456333972
47 0.00274593261622197
48 -0.000119883283715285
49 -0.00925807406492241
50 0.00821833954332399
};
\addlegendentry{Autocorr. of $\Delta F(t)$}

\addplot [thick, gray, dashed, forget plot]
table {%
0 0.05
1 0.05
2 0.05
3 0.05
4 0.05
5 0.05
6 0.05
7 0.05
8 0.05
9 0.05
10 0.05
11 0.05
12 0.05
13 0.05
14 0.05
15 0.05
16 0.05
17 0.05
18 0.05
19 0.05
20 0.05
21 0.05
22 0.05
23 0.05
24 0.05
25 0.05
26 0.05
27 0.05
28 0.05
29 0.05
30 0.05
31 0.05
32 0.05
33 0.05
34 0.05
35 0.05
36 0.05
37 0.05
38 0.05
39 0.05
40 0.05
41 0.05
42 0.05
43 0.05
44 0.05
45 0.05
46 0.05
47 0.05
48 0.05
49 0.05
50 0.05
};
\addplot [thick, gray, dashed, forget plot]
table {%
0 -0.05
1 -0.05
2 -0.05
3 -0.05
4 -0.05
5 -0.05
6 -0.05
7 -0.05
8 -0.05
9 -0.05
10 -0.05
11 -0.05
12 -0.05
13 -0.05
14 -0.05
15 -0.05
16 -0.05
17 -0.05
18 -0.05
19 -0.05
20 -0.05
21 -0.05
22 -0.05
23 -0.05
24 -0.05
25 -0.05
26 -0.05
27 -0.05
28 -0.05
29 -0.05
30 -0.05
31 -0.05
32 -0.05
33 -0.05
34 -0.05
35 -0.05
36 -0.05
37 -0.05
38 -0.05
39 -0.05
40 -0.05
41 -0.05
42 -0.05
43 -0.05
44 -0.05
45 -0.05
46 -0.05
47 -0.05
48 -0.05
49 -0.05
50 -0.05
};

\end{axis}

\begin{axis}[
width=\sfwidth,
height=\sfheight,
tick align=outside,
tick pos=right,
xlabel={Lag (ms)},
xmin=0, xmax=833.33333333,
ymin=0, ymax=1,
ymajorticks = false,
]
    
\end{axis}

\end{tikzpicture}

%% file: img/size_rolling_autocorr_vp_30_60.tex
\begin{tikzpicture}

\begin{axis}[
width=\fwidth,
height=\fheight,
colormap={mymap}{[1pt]
 rgb(0pt)=(0.699611712320639,0.155783736277259,0.200466931944463);
  rgb(1pt)=(0.741098023510142,0.285713892997613,0.323031245695321);
  rgb(2pt)=(0.782584334699646,0.415644049717968,0.445595559446178);
  rgb(3pt)=(0.827030095357373,0.554842847640379,0.576903067825654);
  rgb(4pt)=(0.868516406546877,0.684773004360733,0.699467381576512);
  rgb(5pt)=(0.91000271773638,0.814703161081087,0.822031695327369);
  rgb(6pt)=(0.944495395711727,0.947045513158855,0.95014114060491);
  rgb(7pt)=(0.816034334821363,0.854298101102617,0.883181238459061);
  rgb(8pt)=(0.690058891103904,0.762308988051814,0.814457809630203);
  rgb(9pt)=(0.555835055238277,0.66429677831316,0.741234632485169);
  rgb(10pt)=(0.429121474155285,0.571768666553075,0.672108527315517);
  rgb(11pt)=(0.303146030437825,0.479779553502272,0.603385098486659);
  rgb(12pt)=(0.177170586720365,0.387790440451469,0.534661669657801)
  rgb(13pt)=(0.177170586720365,0.387790440451469,0.534661669657801)
},
colorbar sampled,
colormap access=piecewise constant,
colorbar right,
point meta max=1,
point meta min=-1,
colorbar style={ytick={-1,-0.75,-0.5,-0.25,0,0.25,0.5,0.75,1},
samples=14,
yticklabels={
  \ensuremath{-}1,
  \ensuremath{-}0.75,
  \ensuremath{-}0.50,
  \ensuremath{-}0.25,
  0,
  0.25,
  0.50,
  0.75,
  1
},ylabel={}},
tick align=outside,
tick pos=left,
x grid style={white!69.0196078431373!black},
xlabel={Lag (frames)},
xmin=0, xmax=61,
xtick={0.5,5.5,10.5,15.5,20.5,25.5,30.5,35.5,40.5,45.5,50.5,55.5,60.5},
xticklabels={0,5,10,15,20,25,30,35,40,45,50,55,60},
y dir=reverse,
y grid style={white!69.0196078431373!black},
ylabel={Time (s)},
ymin=0, ymax=566,
ytick style={color=black},
ytick={0.5,60.5,120.5,180.5,240.5,300.5,360.5,420.5,480.5,540.5},
yticklabels={
  0,60,120,180,240,300,360,420,480,540}
]
\addplot graphics [includegraphics cmd=\pgfimage,xmin=0, xmax=61, ymin=566, ymax=0] {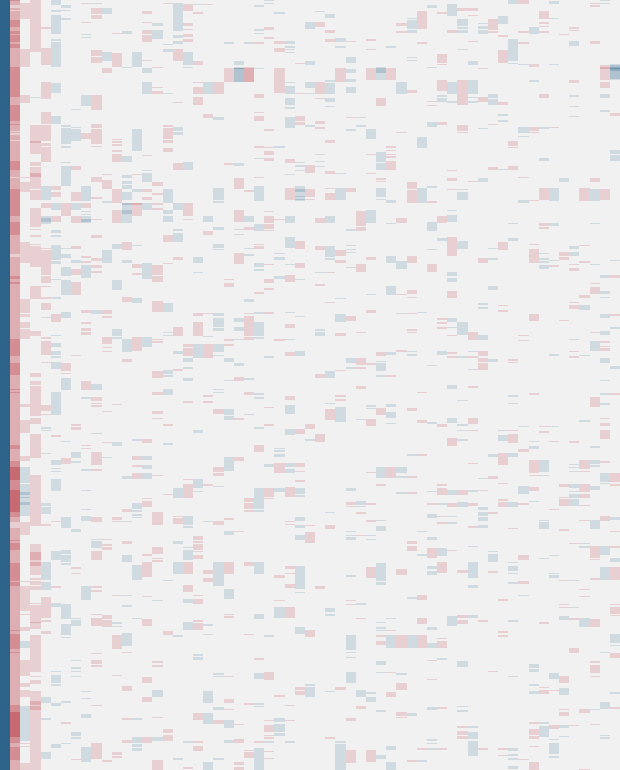};
\end{axis}

\begin{axis}[
  width=\fwidth,
  height=\fheight,
  tick align=outside,
  tick pos=right,
  xlabel={Lag (ms)},
  xmin=0, xmax=1016.667,
  ymin=0, ymax=1,
  ymajorticks = false,
  ]

  \end{axis}

\end{tikzpicture}

%% file: img/legend_n.tex
\begin{tikzpicture}

\begin{axis}[
    width=0,
    height=0,
    at={(0,0)},
    scale only axis,
    xmin=0,
    xmax=0,
    xtick={},
    ymin=0,
    ymax=0,
    ytick={},
    axis background/.style={fill=white},
    legend style={legend cell align=center, align=center, draw=white!15!black, font=\scriptsize, at={(0, 0)}, anchor=center, /tikz/every even column/.append style={column sep=2em}},
    legend columns=10,
]
\addplot [thick, color0]
table {%
0 1
};
\addlegendentry{$N=0$}
\addplot [thick, color1]
table {%
0 1
};
\addlegendentry{$N=1$}
\addplot [thick, color2]
table {%
0 1
};
\addlegendentry{$N=2$}
\addplot [thick, color3]
table {%
0 1
};
\addlegendentry{$N=3$}
\addplot [thick, color4]
table {%
0 1
};
\addlegendentry{$N=4$}
\addplot [thick, color5]
table {%
0 1
};
\addlegendentry{$N=5$}
\addplot [thick, color6]
table {%
0 1
};
\addlegendentry{$N=6$}
\end{axis}

\end{tikzpicture}

%% file: img/autocorr_Linear_1_frames.tex
\begin{tikzpicture}

\begin{axis}[
width=\sfwidth,
height=\sfheight,
legend style={at={(0.98,0.98)}, anchor=north east, fill opacity=0.8, legend columns=4, draw opacity=1, text opacity=1,
draw=white!80!black, /tikz/every even column/.append style={column sep=0.5em}, font=\tiny},
tick align=outside,
tick pos=left,
x grid style={white!69.0196078431373!black},
xlabel={Lag (frames)},
xmin=-1, xmax=21,
xmajorgrids,
y grid style={white!69.0196078431373!black},
ylabel={Autocorrelation},
ymin=-0.153089121643665, ymax=1.05490900579256,
ymajorgrids,
ytick style={color=black}
]
\addplot [thick, color0]
table {%
0 1
1 0.467627024563451
2 0.386678461671627
3 0.289915198879574
4 0.275341597285425
5 0.270901758454847
6 0.248515400368811
7 0.221993838594262
8 0.174933283065541
9 0.13787191459132
10 0.108290797337813
11 0.0907287081702729
12 0.103473392129106
13 0.0753888838219881
14 0.0669010114992973
15 0.0489823718882972
16 0.0515623035432974
17 0.0498877306298421
18 0.0485075634479317
19 0.0468378237299123
20 0.0463075565134515
};
\addplot [thick, color1]
table {%
0 1
1 -0.0981801158511094
2 0.150727876249507
3 0.0569536333138058
4 0.0945702071523353
5 0.109681763589611
6 0.093295851119966
7 0.0933436899686838
8 0.0579502860841136
9 0.0458793165918818
10 0.0324117660134167
11 0.0151054264356603
12 0.0621405907530121
13 0.0158851546791096
14 0.0300677691084331
15 0.00567374053361614
16 0.021370106517629
17 0.018034155019033
18 0.0179125819417568
19 0.0164247763242153
20 0.0159557097716162
};
\addplot [thick, color2]
table {%
0 1
1 -0.0123542659112769
2 -0.0459725029274762
3 -0.00838855384686502
4 0.0372228213004587
5 0.0798062695607738
6 0.0740336677715648
7 0.0707189477427878
8 0.0356486651473089
9 0.0191725256450366
10 0.00334588427917296
11 0.000745876565648191
12 0.0494437424866286
13 0.013536039878729
14 0.0134577681478106
15 -0.00411508966893306
16 0.0103456153901497
17 0.0128301491257589
18 0.0113924843772721
19 0.00962478497343389
20 0.0123376736821672
};
\addplot [thick, color3]
table {%
0 1
1 -0.00478337892363173
2 -0.023980827173394
3 -0.0622114442950089
4 0.0230155588767323
5 0.0684642148078619
6 0.0701085434370355
7 0.0671786968903481
8 0.0303720475889648
9 0.011212729310628
10 -0.0026147916839249
11 -0.00445928944217541
12 0.0456362305740469
13 0.0103573984825846
14 0.0121285864505771
15 -0.0080539497632333
16 0.00794901444093941
17 0.0100643863447718
18 0.0101337393670986
19 0.00896884738430489
20 0.0108747084526406
};
\addplot [thick, color4]
table {%
0 1
1 -0.00767622651644151
2 -0.00911913219876571
3 -0.0243293418515051
4 -0.050330227437615
5 0.0521471524304755
6 0.0544249216265398
7 0.0613449419629661
8 0.0217760181946534
9 0.00515678478885353
10 -0.0101410214492291
11 -0.0130000163398945
12 0.0373749631472741
13 0.00339960442326097
14 0.00820049224484625
15 -0.00946794358294854
16 0.00236726905841527
17 0.00650022868050762
18 0.00751618487076797
19 0.00754877690226283
20 0.00746307120497348
};
\addplot [thick, color5]
table {%
0 1
1 -0.00402143435353587
2 -0.00597279283027728
3 -0.0036070508727193
4 -0.00810269655930091
5 -0.0221992134729614
6 0.0376649368990387
7 0.0419847152896756
8 0.0174759162699821
9 0.000454529344881876
10 -0.0163585001222423
11 -0.0216056169160254
12 0.0272698965109604
13 -0.0044252513513018
14 0.00158802585705492
15 -0.0134132496141505
16 0.00079019301278003
17 0.00220890701502576
18 0.0042895687587392
19 0.00236457847961217
20 0.0058932188343839
};
\addplot [thick, color6]
table {%
0 1
1 -0.000759254633870117
2 -0.000853197018060813
3 0.000448584907611879
4 0.00333930346158908
5 -0.000139496948214762
6 -0.00470407465790527
7 0.0334446501055545
8 0.00905558940152654
9 -0.00197130860538306
10 -0.0194313680077943
11 -0.0257626732512656
12 0.0227010693515244
13 -0.00971851568058316
14 -0.00269143150622811
15 -0.0170476495799799
16 -0.000632848764992775
17 0.00150593700937746
18 0.000554235200813429
19 0.000489356756473908
20 0.00419233471715153
};
\addplot [thick, black, dashed, forget plot]
table {%
0 0.05
1 0.05
2 0.05
3 0.05
4 0.05
5 0.05
6 0.05
7 0.05
8 0.05
9 0.05
10 0.05
11 0.05
12 0.05
13 0.05
14 0.05
15 0.05
16 0.05
17 0.05
18 0.05
19 0.05
20 0.05
};
\addplot [thick, black, dashed, forget plot]
table {%
0 -0.05
1 -0.05
2 -0.05
3 -0.05
4 -0.05
5 -0.05
6 -0.05
7 -0.05
8 -0.05
9 -0.05
10 -0.05
11 -0.05
12 -0.05
13 -0.05
14 -0.05
15 -0.05
16 -0.05
17 -0.05
18 -0.05
19 -0.05
20 -0.05
};



\end{axis}

\begin{axis}[
width=\sfwidth,
height=\sfheight,
tick align=outside,
tick pos=right,
xlabel={Lag (ms)},
xmin=-16.6666, xmax=350,
ymin=0, ymax=1,
ymajorticks = false,
]
    
\end{axis}

\end{tikzpicture}

%% file: img/autocorr_Quantile_1_frames.tex
\begin{tikzpicture}

\begin{axis}[
width=\sfwidth,
height=\sfheight,
legend style={at={(0.98,0.98)}, anchor=north east, fill opacity=0.8, legend columns=4, draw opacity=1, text opacity=1,
draw=white!80!black, /tikz/every even column/.append style={column sep=0.5em}, font=\tiny},
tick align=outside,
tick pos=left,
x grid style={white!69.0196078431373!black},
xlabel={Lag (frames)},
xmin=-1, xmax=21,
xmajorgrids,
y grid style={white!69.0196078431373!black},
ylabel={Autocorrelation},
ymin=-0.1025, ymax=1.0525,
ymajorgrids,
ytick style={color=black}
]
\addplot [thick, color0]
table {%
0 1
1 0.467627024563451
2 0.386678461671627
3 0.289915198879574
4 0.275341597285425
5 0.270901758454847
6 0.248515400368811
7 0.221993838594262
8 0.174933283065541
9 0.13787191459132
10 0.108290797337813
11 0.0907287081702729
12 0.103473392129106
13 0.0753888838219881
14 0.0669010114992973
15 0.0489823718882972
16 0.0515623035432974
17 0.0498877306298421
18 0.0485075634479317
19 0.0468378237299123
20 0.0463075565134515
};
\addplot [thick, color1]
table {%
0 1
1 0.11185971871572
2 0.238317020566131
3 0.143432481945961
4 0.161675722595726
5 0.169530017466258
6 0.150913296625593
7 0.141098357843621
8 0.101372039057096
9 0.0800297389797738
10 0.0605786526595123
11 0.0431762006619358
12 0.0774847018057679
13 0.0379718738943583
14 0.0437391953861639
15 0.021748034004522
16 0.0325757555063692
17 0.0298597477040893
18 0.0292694013075912
19 0.0277163440116886
20 0.027224178574118
};
\addplot [thick, color2]
table {%
0 1
1 0.130716678672006
2 0.0521744616693844
3 0.0671832300271736
4 0.0969660351526297
5 0.129566599805081
6 0.120363435609184
7 0.110219527695984
8 0.0718538070756958
9 0.049168733548824
10 0.0292502901586177
11 0.0241436002676665
12 0.0630366816871386
13 0.0304631840422267
14 0.02644213288973
15 0.00953966755844923
16 0.0205611009803969
17 0.0225482128839955
18 0.0209776184070289
19 0.0192094579069057
20 0.0213786643399552
};
\addplot [thick, color3]
table {%
0 1
1 0.12796552716561
2 0.0541645156260542
3 -0.00711914451253354
4 0.0703331552555953
5 0.109021302105063
6 0.109343812279918
7 0.100231143061951
8 0.060278799609932
9 0.0349775543986403
10 0.0176797993938263
11 0.014611457524724
12 0.0563396614609958
13 0.02472157112382
14 0.0224176839489758
15 0.00286047177719979
16 0.0159718253266767
17 0.0179445024546402
18 0.0180354250405873
19 0.0169416637744654
20 0.0184405355930229
};
\addplot [thick, color4]
table {%
0 1
1 0.12783148979721
2 0.0601079926304476
3 0.00924952628325991
4 -0.016922614179137
5 0.0846953382532696
6 0.0887697814940289
7 0.0901403405638285
8 0.0468926982594005
9 0.0233695400868274
10 0.00512705693340227
11 0.0022784062159266
12 0.0451842755832299
13 0.0156887399056038
14 0.0163361194148314
15 -0.000954710119289624
16 0.00829132812481772
17 0.0126714535180997
18 0.0141355704391201
19 0.0144651878416944
20 0.0138534701947519
};
\addplot [thick, color5]
table {%
0 1
1 0.123701157101479
2 0.0536447749898354
3 0.0176001774622242
4 0.00700222893911009
5 -0.0177053802018401
6 0.0588654337966085
7 0.0596984201444377
8 0.0339846158843728
9 0.0101972776281435
10 -0.00937262295232793
11 -0.0141742138886515
12 0.0283759528680201
13 0.0022294884565295
14 0.00514201925398502
15 -0.00882931985587758
16 0.00310983093788083
17 0.00484913928538233
18 0.00813126445607092
19 0.0064792479513757
20 0.00982937324989914
};
\addplot [thick, color6]
table {%
0 1
1 0.121686208292072
2 0.0553768086562304
3 0.0243079745939452
4 0.0174688233218054
5 -0.00218919932627444
6 -0.000863150983135271
7 0.0439628528749094
8 0.0201273257392078
9 0.00443485933169177
10 -0.0160839034781266
11 -0.0224038854611669
12 0.0199508097986374
13 -0.00683705769588252
14 -0.00216797375647116
15 -0.0147706793723008
16 9.17449698461189e-05
17 0.00266124490489729
18 0.00226018784294586
19 0.00284471017179433
20 0.00673860883475947
};
\addplot [thick, black, dashed, forget plot]
table {%
0 0.05
1 0.05
2 0.05
3 0.05
4 0.05
5 0.05
6 0.05
7 0.05
8 0.05
9 0.05
10 0.05
11 0.05
12 0.05
13 0.05
14 0.05
15 0.05
16 0.05
17 0.05
18 0.05
19 0.05
20 0.05
};
\addplot [thick, black, dashed, forget plot]
table {%
0 -0.05
1 -0.05
2 -0.05
3 -0.05
4 -0.05
5 -0.05
6 -0.05
7 -0.05
8 -0.05
9 -0.05
10 -0.05
11 -0.05
12 -0.05
13 -0.05
14 -0.05
15 -0.05
16 -0.05
17 -0.05
18 -0.05
19 -0.05
20 -0.05
};



\end{axis}

\begin{axis}[
width=\sfwidth,
height=\sfheight,
tick align=outside,
tick pos=right,
xlabel={Lag (ms)},
xmin=-16.6666, xmax=350,
ymin=0, ymax=1,
ymajorticks = false,
]
    
\end{axis}

\end{tikzpicture}

%% file: img/heatmap_Linear_std.tex
\begin{tikzpicture} 
    \begin{axis}[
    width=\bpheight,
    height=\bpheight,
    xlabel=$N$,
    ylabel=$\tau$,
    mesh/cols=11,
    mesh/rows=11,
    xmin=-0.5,
    xmax=10.5,
    ymin=-0.5,
    ymax=9.5,
    point meta min=8.5,
    point meta max=10.5,
colormap={mymap}{[1pt]
 rgb(0pt)=(0.946403,0.937159,0.458592);
  rgb(1pt)=(0.962517,0.851476,0.285546);
  rgb(2pt)=(0.981173,0.759135,0.156863);
  rgb(3pt)=(0.987945,0.667748,0.058329);
  rgb(4pt)=(0.981895,0.579392,0.02625);
  rgb(5pt)=(0.961293,0.488716,0.084289);
  rgb(6pt)=(0.929644,0.411479,0.145367);
  rgb(7pt)=(0.886302,0.342586,0.202968);
  rgb(8pt)=(0.832299,0.283913,0.257383);
  rgb(9pt)=(0.769556,0.236077,0.307485);
  rgb(10pt)=(0.694627,0.195021,0.354388);
  rgb(11pt)=(0.621685,0.164184,0.388781);
  rgb(12pt)=(0.547157,0.136929,0.413511);
  rgb(13pt)=(0.472328,0.110547,0.428334);
  rgb(14pt)=(0.397674,0.083257,0.433183);
  rgb(15pt)=(0.316282,0.05349,0.425116);
  rgb(16pt)=(0.238273,0.036621,0.396353);
  rgb(17pt)=(0.15585,0.044559,0.325338);
  rgb(18pt)=(0.081962,0.043328,0.215289);
  rgb(19pt)=(0.025793,0.019331,0.10593);
},
    ytick={1,3,5,7,9},
    yticklabels={2,4,6,8,10}
]
    \addplot[matrix plot*, point meta=explicit] file [meta=index 2] {./img/imgdata/Linear_std.csv};
\end{axis}
\end{tikzpicture}

%% file: img/heatmap_Quantile_std.tex
\begin{tikzpicture} 
    \begin{axis}[
    width=\bpheight,
    height=\bpheight,
    xlabel=$N$,
    ylabel=$\tau$,
    mesh/cols=11,
    mesh/rows=11,
    xmin=-0.5,
    xmax=10.5,
    ymin=-0.5,
    ymax=9.5,
    point meta min=8.5,
    point meta max=10.5,
colormap={mymap}{[1pt]
 rgb(0pt)=(0.946403,0.937159,0.458592);
  rgb(1pt)=(0.962517,0.851476,0.285546);
  rgb(2pt)=(0.981173,0.759135,0.156863);
  rgb(3pt)=(0.987945,0.667748,0.058329);
  rgb(4pt)=(0.981895,0.579392,0.02625);
  rgb(5pt)=(0.961293,0.488716,0.084289);
  rgb(6pt)=(0.929644,0.411479,0.145367);
  rgb(7pt)=(0.886302,0.342586,0.202968);
  rgb(8pt)=(0.832299,0.283913,0.257383);
  rgb(9pt)=(0.769556,0.236077,0.307485);
  rgb(10pt)=(0.694627,0.195021,0.354388);
  rgb(11pt)=(0.621685,0.164184,0.388781);
  rgb(12pt)=(0.547157,0.136929,0.413511);
  rgb(13pt)=(0.472328,0.110547,0.428334);
  rgb(14pt)=(0.397674,0.083257,0.433183);
  rgb(15pt)=(0.316282,0.05349,0.425116);
  rgb(16pt)=(0.238273,0.036621,0.396353);
  rgb(17pt)=(0.15585,0.044559,0.325338);
  rgb(18pt)=(0.081962,0.043328,0.215289);
  rgb(19pt)=(0.025793,0.019331,0.10593);
},
    colorbar sampled,
    colormap access=piecewise constant,
    colorbar right,
    colorbar style={samples=21,
                    ylabel={$\sigma_w$ (kB)},
                    ytick={7, 7.5, ..., 15}},
    ytick={1,3,5,7,9},
    yticklabels={2,4,6,8,10}
]
    \addplot[matrix plot*, point meta=explicit] file [meta=index 2] {./img/imgdata/Quantile_std.csv};
\end{axis}
\end{tikzpicture}

%% file: img/legend_gen.tex
\begin{tikzpicture}

\begin{axis}[
    width=0,
    height=0,
    at={(0,0)},
    scale only axis,
    xmin=0,
    xmax=0,
    xtick={},
    ymin=0,
    ymax=0,
    ytick={},
    axis background/.style={fill=white},
    legend style={legend cell align=center, align=center, draw=white!15!black, font=\scriptsize, at={(0, 0)}, anchor=center, /tikz/every even column/.append style={column sep=2em}},
    legend columns=10,
]
\addplot [thick, color2]
table {%
0 1
};
\addlegendentry{General model}
\addplot [thick, color4]
table {%
0 1
};
\addlegendentry{Content-dependent model}
\addplot [thick, color6]
table {%
0 1
};
\addlegendentry{Content- and rate-dependent model}

\end{axis}

\end{tikzpicture}

%% file: img/gen_boxplot_Linear_1.tex
\begin{tikzpicture}

\begin{axis}[
width=\sfwidth,
height=\sfheight,
legend cell align={left},
legend style={
  fill opacity=0.8,
  draw opacity=1,
  text opacity=1,
  at={(0.03,0.97)},
  anchor=north west,
  draw=white!80!black
},
tick align=outside,
tick pos=left,
ylabel={Relative error},
xlabel={Rate (Mb/s)},
x grid style={white!69.0196078431373!black},
xmin=-3, xmax=60,
xtick={0,3,6,9,12,15,18,21,24,27,30,33,36,39,42,45,48,51,54,57},
xticklabels={
  10,
  20,
  30,
  40,
  50,
  10,
  20,
  30,
  40,
  50,
  10,
  20,
  30,
  40,
  50,
  10,
  20,
  30,
  40,
  50
},
y grid style={white!69.0196078431373!black},
ymin=-1, ymax=0.5,
ytick style={color=black},
ymajorgrids,
ytick={-1,-0.75,-0.5,-0.25,0,0.25,0.5,0.75,1}
]

\addplot [semithick, black, forget plot]
table {%
-3 0
-2 0
-1 0
0 0
1 0
2 0
3 0
4 0
5 0
6 0
7 0
8 0
9 0
10 0
11 0
12 0
13 0
14 0
15 0
16 0
17 0
18 0
19 0
20 0
21 0
22 0
23 0
24 0
25 0
26 0
27 0
28 0
29 0
30 0
31 0
32 0
33 0
34 0
35 0
36 0
37 0
38 0
39 0
40 0
41 0
42 0
43 0
44 0
45 0
46 0
47 0
48 0
49 0
50 0
51 0
52 0
53 0
54 0
55 0
56 0
57 0
58 0
59 0
60 0
61 0
62 0
};
\addplot [semithick, color2]
table {%
-1.1 -0.105062729079818
-0.5 -0.105062729079818
-0.5 0.0761368181490279
-1.1 0.0761368181490279
-1.1 -0.105062729079818
};
\addplot [semithick, color2]
table {%
-0.8 -0.105062729079818
-0.8 -0.376653070446306
};
\addplot [semithick, color2]
table {%
-0.8 0.0761368181490279
-0.8 0.347800648356077
};
\addplot [semithick, color2]
table {%
-0.95 -0.376653070446306
-0.65 -0.376653070446306
};
\addplot [semithick, color2]
table {%
-0.95 0.347800648356077
-0.65 0.347800648356077
};
\addplot [semithick, color2]
table {%
1.9 -0.0727742446471767
2.5 -0.0727742446471767
2.5 0.0726008701653586
1.9 0.0726008701653586
1.9 -0.0727742446471767
};
\addplot [semithick, color2]
table {%
2.2 -0.0727742446471767
2.2 -0.290790487648007
};
\addplot [semithick, color2]
table {%
2.2 0.0726008701653586
2.2 0.290659760342465
};
\addplot [semithick, color2]
table {%
2.05 -0.290790487648007
2.35 -0.290790487648007
};
\addplot [semithick, color2]
table {%
2.05 0.290659760342465
2.35 0.290659760342465
};
\addplot [semithick, color2]
table {%
4.9 -0.079350927378063
5.5 -0.079350927378063
5.5 0.0522593013915245
4.9 0.0522593013915245
4.9 -0.079350927378063
};
\addplot [semithick, color2]
table {%
5.2 -0.079350927378063
5.2 -0.276736175380704
};
\addplot [semithick, color2]
table {%
5.2 0.0522593013915245
5.2 0.24963119508318
};
\addplot [semithick, color2]
table {%
5.05 -0.276736175380704
5.35 -0.276736175380704
};
\addplot [semithick, color2]
table {%
5.05 0.24963119508318
5.35 0.24963119508318
};
\addplot [semithick, color2]
table {%
7.9 -0.101605867758823
8.5 -0.101605867758823
8.5 0.0312509692280472
7.9 0.0312509692280472
7.9 -0.101605867758823
};
\addplot [semithick, color2]
table {%
8.2 -0.101605867758823
8.2 -0.300759622937054
};
\addplot [semithick, color2]
table {%
8.2 0.0312509692280472
8.2 0.2305011646969
};
\addplot [semithick, color2]
table {%
8.05 -0.300759622937054
8.35 -0.300759622937054
};
\addplot [semithick, color2]
table {%
8.05 0.2305011646969
8.35 0.2305011646969
};
\addplot [semithick, color2]
table {%
10.9 -0.0753534598706181
11.5 -0.0753534598706181
11.5 0.0394115070172826
10.9 0.0394115070172826
10.9 -0.0753534598706181
};
\addplot [semithick, color2]
table {%
11.2 -0.0753534598706181
11.2 -0.247428573271288
};
\addplot [semithick, color2]
table {%
11.2 0.0394115070172826
11.2 0.211551594504241
};
\addplot [semithick, color2]
table {%
11.05 -0.247428573271288
11.35 -0.247428573271288
};
\addplot [semithick, color2]
table {%
11.05 0.211551594504241
11.35 0.211551594504241
};
\addplot [semithick, color2]
table {%
13.9 -0.136482817590915
14.5 -0.136482817590915
14.5 0.157570881316689
13.9 0.157570881316689
13.9 -0.136482817590915
};
\addplot [semithick, color2]
table {%
14.2 -0.136482817590915
14.2 -0.577471682250312
};
\addplot [semithick, color2]
table {%
14.2 0.157570881316689
14.2 0.598491707330203
};
\addplot [semithick, color2]
table {%
14.05 -0.577471682250312
14.35 -0.577471682250312
};
\addplot [semithick, color2]
table {%
14.05 0.598491707330203
14.35 0.598491707330203
};
\addplot [semithick, color2]
table {%
16.9 -0.0764755712794797
17.5 -0.0764755712794797
17.5 0.0824990062326616
16.9 0.0824990062326616
16.9 -0.0764755712794797
};
\addplot [semithick, color2]
table {%
17.2 -0.0764755712794797
17.2 -0.314818182055628
};
\addplot [semithick, color2]
table {%
17.2 0.0824990062326616
17.2 0.320768817866117
};
\addplot [semithick, color2]
table {%
17.05 -0.314818182055628
17.35 -0.314818182055628
};
\addplot [semithick, color2]
table {%
17.05 0.320768817866117
17.35 0.320768817866117
};
\addplot [semithick, color2]
table {%
19.9 -0.0850677196326138
20.5 -0.0850677196326138
20.5 0.0990852908517967
19.9 0.0990852908517967
19.9 -0.0850677196326138
};
\addplot [semithick, color2]
table {%
20.2 -0.0850677196326138
20.2 -0.36125769792726
};
\addplot [semithick, color2]
table {%
20.2 0.0990852908517967
20.2 0.375300205638469
};
\addplot [semithick, color2]
table {%
20.05 -0.36125769792726
20.35 -0.36125769792726
};
\addplot [semithick, color2]
table {%
20.05 0.375300205638469
20.35 0.375300205638469
};
\addplot [semithick, color2]
table {%
22.9 -0.0627577089295605
23.5 -0.0627577089295605
23.5 0.079916362154732
22.9 0.079916362154732
22.9 -0.0627577089295605
};
\addplot [semithick, color2]
table {%
23.2 -0.0627577089295605
23.2 -0.276641912434393
};
\addplot [semithick, color2]
table {%
23.2 0.079916362154732
23.2 0.293386640150933
};
\addplot [semithick, color2]
table {%
23.05 -0.276641912434393
23.35 -0.276641912434393
};
\addplot [semithick, color2]
table {%
23.05 0.293386640150933
23.35 0.293386640150933
};
\addplot [semithick, color2]
table {%
25.9 -0.0588871283572223
26.5 -0.0588871283572223
26.5 0.0730760023611019
25.9 0.0730760023611019
25.9 -0.0588871283572223
};
\addplot [semithick, color2]
table {%
26.2 -0.0588871283572223
26.2 -0.256819383404528
};
\addplot [semithick, color2]
table {%
26.2 0.0730760023611019
26.2 0.270906178005579
};
\addplot [semithick, color2]
table {%
26.05 -0.256819383404528
26.35 -0.256819383404528
};
\addplot [semithick, color2]
table {%
26.05 0.270906178005579
26.35 0.270906178005579
};
\addplot [semithick, color2]
table {%
28.9 -0.12717220811693
29.5 -0.12717220811693
29.5 0.136771623796432
28.9 0.136771623796432
28.9 -0.12717220811693
};
\addplot [semithick, color2]
table {%
29.2 -0.12717220811693
29.2 -0.522727730957445
};
\addplot [semithick, color2]
table {%
29.2 0.136771623796432
29.2 0.532579713188845
};
\addplot [semithick, color2]
table {%
29.05 -0.522727730957445
29.35 -0.522727730957445
};
\addplot [semithick, color2]
table {%
29.05 0.532579713188845
29.35 0.532579713188845
};
\addplot [semithick, color2]
table {%
31.9 -0.102138691897775
32.5 -0.102138691897775
32.5 0.107944772262808
31.9 0.107944772262808
31.9 -0.102138691897775
};
\addplot [semithick, color2]
table {%
32.2 -0.102138691897775
32.2 -0.417150457850738
};
\addplot [semithick, color2]
table {%
32.2 0.107944772262808
32.2 0.423055227190453
};
\addplot [semithick, color2]
table {%
32.05 -0.417150457850738
32.35 -0.417150457850738
};
\addplot [semithick, color2]
table {%
32.05 0.423055227190453
32.35 0.423055227190453
};
\addplot [semithick, color2]
table {%
34.9 -0.081900275965834
35.5 -0.081900275965834
35.5 0.0855145982064358
34.9 0.0855145982064358
34.9 -0.081900275965834
};
\addplot [semithick, color2]
table {%
35.2 -0.081900275965834
35.2 -0.332997447796389
};
\addplot [semithick, color2]
table {%
35.2 0.0855145982064358
35.2 0.336572230777542
};
\addplot [semithick, color2]
table {%
35.05 -0.332997447796389
35.35 -0.332997447796389
};
\addplot [semithick, color2]
table {%
35.05 0.336572230777542
35.35 0.336572230777542
};
\addplot [semithick, color2]
table {%
37.9 -0.0838475401899211
38.5 -0.0838475401899211
38.5 0.0857883753538068
37.9 0.0857883753538068
37.9 -0.0838475401899211
};
\addplot [semithick, color2]
table {%
38.2 -0.0838475401899211
38.2 -0.338280126918848
};
\addplot [semithick, color2]
table {%
38.2 0.0857883753538068
38.2 0.340194999708612
};
\addplot [semithick, color2]
table {%
38.05 -0.338280126918848
38.35 -0.338280126918848
};
\addplot [semithick, color2]
table {%
38.05 0.340194999708612
38.35 0.340194999708612
};
\addplot [semithick, color2]
table {%
40.9 -0.0690176120416685
41.5 -0.0690176120416685
41.5 0.0747629941168072
40.9 0.0747629941168072
40.9 -0.0690176120416685
};
\addplot [semithick, color2]
table {%
41.2 -0.0690176120416685
41.2 -0.284400326631291
};
\addplot [semithick, color2]
table {%
41.2 0.0747629941168072
41.2 0.289693283756052
};
\addplot [semithick, color2]
table {%
41.05 -0.284400326631291
41.35 -0.284400326631291
};
\addplot [semithick, color2]
table {%
41.05 0.289693283756052
41.35 0.289693283756052
};
\addplot [semithick, color2]
table {%
43.9 -0.0661635027617125
44.5 -0.0661635027617125
44.5 0.0745858204643166
43.9 0.0745858204643166
43.9 -0.0661635027617125
};
\addplot [semithick, color2]
table {%
44.2 -0.0661635027617125
44.2 -0.277189718152137
};
\addplot [semithick, color2]
table {%
44.2 0.0745858204643166
44.2 0.285486210709381
};
\addplot [semithick, color2]
table {%
44.05 -0.277189718152137
44.35 -0.277189718152137
};
\addplot [semithick, color2]
table {%
44.05 0.285486210709381
44.35 0.285486210709381
};
\addplot [semithick, color2]
table {%
46.9 -0.0601655696774331
47.5 -0.0601655696774331
47.5 0.069870485736607
46.9 0.069870485736607
46.9 -0.0601655696774331
};
\addplot [semithick, color2]
table {%
47.2 -0.0601655696774331
47.2 -0.255201550806894
};
\addplot [semithick, color2]
table {%
47.2 0.069870485736607
47.2 0.26473378841603
};
\addplot [semithick, color2]
table {%
47.05 -0.255201550806894
47.35 -0.255201550806894
};
\addplot [semithick, color2]
table {%
47.05 0.26473378841603
47.35 0.26473378841603
};
\addplot [semithick, color2]
table {%
49.9 -0.0594293664254752
50.5 -0.0594293664254752
50.5 0.0665776280943636
49.9 0.0665776280943636
49.9 -0.0594293664254752
};
\addplot [semithick, color2]
table {%
50.2 -0.0594293664254752
50.2 -0.248434281377675
};
\addplot [semithick, color2]
table {%
50.2 0.0665776280943636
50.2 0.255517424866098
};
\addplot [semithick, color2]
table {%
50.05 -0.248434281377675
50.35 -0.248434281377675
};
\addplot [semithick, color2]
table {%
50.05 0.255517424866098
50.35 0.255517424866098
};
\addplot [semithick, color2]
table {%
52.9 -0.0495628591022651
53.5 -0.0495628591022651
53.5 0.0618744013645492
52.9 0.0618744013645492
52.9 -0.0495628591022651
};
\addplot [semithick, color2]
table {%
53.2 -0.0495628591022651
53.2 -0.216589759206651
};
\addplot [semithick, color2]
table {%
53.2 0.0618744013645492
53.2 0.228985276754349
};
\addplot [semithick, color2]
table {%
53.05 -0.216589759206651
53.35 -0.216589759206651
};
\addplot [semithick, color2]
table {%
53.05 0.228985276754349
53.35 0.228985276754349
};
\addplot [semithick, color2]
table {%
55.9 -0.0502191051833806
56.5 -0.0502191051833806
56.5 0.0588510095417407
55.9 0.0588510095417407
55.9 -0.0502191051833806
};
\addplot [semithick, color2]
table {%
56.2 -0.0502191051833806
56.2 -0.213823660503233
};
\addplot [semithick, color2]
table {%
56.2 0.0588510095417407
56.2 0.222362983709878
};
\addplot [semithick, color2]
table {%
56.05 -0.213823660503233
56.35 -0.213823660503233
};
\addplot [semithick, color2]
table {%
56.05 0.222362983709878
56.35 0.222362983709878
};
\addplot [semithick, color4]
table {%
-0.3 -0.092150609070505
0.3 -0.092150609070505
0.3 0.0834808676196756
-0.3 0.0834808676196756
-0.3 -0.092150609070505
};
\addplot [semithick, color4]
table {%
0 -0.092150609070505
0 -0.355575481351138
};
\addplot [semithick, color4]
table {%
0 0.0834808676196756
0 0.346894010015539
};
\addplot [semithick, color4]
table {%
-0.15 -0.355575481351138
0.15 -0.355575481351138
};
\addplot [semithick, color4]
table {%
-0.15 0.346894010015539
0.15 0.346894010015539
};
\addplot [semithick, color4]
table {%
2.7 -0.0673919233998065
3.3 -0.0673919233998065
3.3 0.0748701039474925
2.7 0.0748701039474925
2.7 -0.0673919233998065
};
\addplot [semithick, color4]
table {%
3 -0.0673919233998065
3 -0.280665265685393
};
\addplot [semithick, color4]
table {%
3 0.0748701039474925
3 0.288224137115969
};
\addplot [semithick, color4]
table {%
2.85 -0.280665265685393
3.15 -0.280665265685393
};
\addplot [semithick, color4]
table {%
2.85 0.288224137115969
3.15 0.288224137115969
};
\addplot [semithick, color4]
table {%
5.7 -0.0672313116589637
6.3 -0.0672313116589637
6.3 0.0600960997903217
5.7 0.0600960997903217
5.7 -0.0672313116589637
};
\addplot [semithick, color4]
table {%
6 -0.0672313116589637
6 -0.258183063493553
};
\addplot [semithick, color4]
table {%
6 0.0600960997903217
6 0.251072651508575
};
\addplot [semithick, color4]
table {%
5.85 -0.258183063493553
6.15 -0.258183063493553
};
\addplot [semithick, color4]
table {%
5.85 0.251072651508575
6.15 0.251072651508575
};
\addplot [semithick, color4]
table {%
8.7 -0.0583711839359467
9.3 -0.0583711839359467
9.3 0.0358820187041387
8.7 0.0358820187041387
8.7 -0.0583711839359467
};
\addplot [semithick, color4]
table {%
9 -0.0583711839359467
9 -0.199614821437832
};
\addplot [semithick, color4]
table {%
9 0.0358820187041387
9 0.177224739318838
};
\addplot [semithick, color4]
table {%
8.85 -0.199614821437832
9.15 -0.199614821437832
};
\addplot [semithick, color4]
table {%
8.85 0.177224739318838
9.15 0.177224739318838
};
\addplot [semithick, color4]
table {%
11.7 -0.0542705243776353
12.3 -0.0542705243776353
12.3 0.0458771839814049
11.7 0.0458771839814049
11.7 -0.0542705243776353
};
\addplot [semithick, color4]
table {%
12 -0.0542705243776353
12 -0.204318760175483
};
\addplot [semithick, color4]
table {%
12 0.0458771839814049
12 0.196085097832521
};
\addplot [semithick, color4]
table {%
11.85 -0.204318760175483
12.15 -0.204318760175483
};
\addplot [semithick, color4]
table {%
11.85 0.196085097832521
12.15 0.196085097832521
};
\addplot [semithick, color4]
table {%
14.7 -0.145201987657017
15.3 -0.145201987657017
15.3 0.146525819863302
14.7 0.146525819863302
14.7 -0.145201987657017
};
\addplot [semithick, color4]
table {%
15 -0.145201987657017
15 -0.581359884587916
};
\addplot [semithick, color4]
table {%
15 0.146525819863302
15 0.583778979814068
};
\addplot [semithick, color4]
table {%
14.85 -0.581359884587916
15.15 -0.581359884587916
};
\addplot [semithick, color4]
table {%
14.85 0.583778979814068
15.15 0.583778979814068
};
\addplot [semithick, color4]
table {%
17.7 -0.0884424563084793
18.3 -0.0884424563084793
18.3 0.079956568552195
17.7 0.079956568552195
17.7 -0.0884424563084793
};
\addplot [semithick, color4]
table {%
18 -0.0884424563084793
18 -0.341009043894346
};
\addplot [semithick, color4]
table {%
18 0.079956568552195
18 0.332337141582633
};
\addplot [semithick, color4]
table {%
17.85 -0.341009043894346
18.15 -0.341009043894346
};
\addplot [semithick, color4]
table {%
17.85 0.332337141582633
18.15 0.332337141582633
};
\addplot [semithick, color4]
table {%
20.7 -0.0964540515108308
21.3 -0.0964540515108308
21.3 0.0952142401506729
20.7 0.0952142401506729
20.7 -0.0964540515108308
};
\addplot [semithick, color4]
table {%
21 -0.0964540515108308
21 -0.383702738025014
};
\addplot [semithick, color4]
table {%
21 0.0952142401506729
21 0.382602106201315
};
\addplot [semithick, color4]
table {%
20.85 -0.383702738025014
21.15 -0.383702738025014
};
\addplot [semithick, color4]
table {%
20.85 0.382602106201315
21.15 0.382602106201315
};
\addplot [semithick, color4]
table {%
23.7 -0.0753616958632236
24.3 -0.0753616958632236
24.3 0.0745179323720098
23.7 0.0745179323720098
23.7 -0.0753616958632236
};
\addplot [semithick, color4]
table {%
24 -0.0753616958632236
24 -0.300167925043226
};
\addplot [semithick, color4]
table {%
24 0.0745179323720098
24 0.299096441691378
};
\addplot [semithick, color4]
table {%
23.85 -0.300167925043226
24.15 -0.300167925043226
};
\addplot [semithick, color4]
table {%
23.85 0.299096441691378
24.15 0.299096441691378
};
\addplot [semithick, color4]
table {%
26.7 -0.0682566586664053
27.3 -0.0682566586664053
27.3 0.069229907912704
26.7 0.069229907912704
26.7 -0.0682566586664053
};
\addplot [semithick, color4]
table {%
27 -0.0682566586664053
27 -0.274229082998969
};
\addplot [semithick, color4]
table {%
27 0.069229907912704
27 0.275439647770019
};
\addplot [semithick, color4]
table {%
26.85 -0.274229082998969
27.15 -0.274229082998969
};
\addplot [semithick, color4]
table {%
26.85 0.275439647770019
27.15 0.275439647770019
};
\addplot [semithick, color4]
table {%
29.7 -0.135331744962298
30.3 -0.135331744962298
30.3 0.140447020680566
29.7 0.140447020680566
29.7 -0.135331744962298
};
\addplot [semithick, color4]
table {%
30 -0.135331744962298
30 -0.548801668736331
};
\addplot [semithick, color4]
table {%
30 0.140447020680566
30 0.554077052064296
};
\addplot [semithick, color4]
table {%
29.85 -0.548801668736331
30.15 -0.548801668736331
};
\addplot [semithick, color4]
table {%
29.85 0.554077052064296
30.15 0.554077052064296
};
\addplot [semithick, color4]
table {%
32.7 -0.110031055516442
33.3 -0.110031055516442
33.3 0.109177942250541
32.7 0.109177942250541
32.7 -0.110031055516442
};
\addplot [semithick, color4]
table {%
33 -0.110031055516442
33 -0.438762560632515
};
\addplot [semithick, color4]
table {%
33 0.109177942250541
33 0.43796639492088
};
\addplot [semithick, color4]
table {%
32.85 -0.438762560632515
33.15 -0.438762560632515
};
\addplot [semithick, color4]
table {%
32.85 0.43796639492088
33.15 0.43796639492088
};
\addplot [semithick, color4]
table {%
35.7 -0.0872089790162356
36.3 -0.0872089790162356
36.3 0.0869218355994126
35.7 0.0869218355994126
35.7 -0.0872089790162356
};
\addplot [semithick, color4]
table {%
36 -0.0872089790162356
36 -0.348394361222264
};
\addplot [semithick, color4]
table {%
36 0.0869218355994126
36 0.348112703318457
};
\addplot [semithick, color4]
table {%
35.85 -0.348394361222264
36.15 -0.348394361222264
};
\addplot [semithick, color4]
table {%
35.85 0.348112703318457
36.15 0.348112703318457
};
\addplot [semithick, color4]
table {%
38.7 -0.090691600361321
39.3 -0.090691600361321
39.3 0.0877014444207784
38.7 0.0877014444207784
38.7 -0.090691600361321
};
\addplot [semithick, color4]
table {%
39 -0.090691600361321
39 -0.358105505616044
};
\addplot [semithick, color4]
table {%
39 0.0877014444207784
39 0.35495264964374
};
\addplot [semithick, color4]
table {%
38.85 -0.358105505616044
39.15 -0.358105505616044
};
\addplot [semithick, color4]
table {%
38.85 0.35495264964374
39.15 0.35495264964374
};
\addplot [semithick, color4]
table {%
41.7 -0.0769917088310674
42.3 -0.0769917088310674
42.3 0.0764484253022461
41.7 0.0764484253022461
41.7 -0.0769917088310674
};
\addplot [semithick, color4]
table {%
42 -0.0769917088310674
42 -0.307069341518069
};
\addplot [semithick, color4]
table {%
42 0.0764484253022461
42 0.306575072722165
};
\addplot [semithick, color4]
table {%
41.85 -0.307069341518069
42.15 -0.307069341518069
};
\addplot [semithick, color4]
table {%
41.85 0.306575072722165
42.15 0.306575072722165
};
\addplot [semithick, color4]
table {%
44.7 -0.0726540327429645
45.3 -0.0726540327429645
45.3 0.070258424806106
44.7 0.070258424806106
44.7 -0.0726540327429645
};
\addplot [semithick, color4]
table {%
45 -0.0726540327429645
45 -0.286939358580103
};
\addplot [semithick, color4]
table {%
45 0.070258424806106
45 0.284559681291769
};
\addplot [semithick, color4]
table {%
44.85 -0.286939358580103
45.15 -0.286939358580103
};
\addplot [semithick, color4]
table {%
44.85 0.284559681291769
45.15 0.284559681291769
};
\addplot [semithick, color4]
table {%
47.7 -0.0656167501590835
48.3 -0.0656167501590835
48.3 0.066500249591092
47.7 0.066500249591092
47.7 -0.0656167501590835
};
\addplot [semithick, color4]
table {%
48 -0.0656167501590835
48 -0.263774369719707
};
\addplot [semithick, color4]
table {%
48 0.066500249591092
48 0.264647400326503
};
\addplot [semithick, color4]
table {%
47.85 -0.263774369719707
48.15 -0.263774369719707
};
\addplot [semithick, color4]
table {%
47.85 0.264647400326503
48.15 0.264647400326503
};
\addplot [semithick, color4]
table {%
50.7 -0.0649284963162823
51.3 -0.0649284963162823
51.3 0.0619187693171194
50.7 0.0619187693171194
50.7 -0.0649284963162823
};
\addplot [semithick, color4]
table {%
51 -0.0649284963162823
51 -0.254839197385811
};
\addplot [semithick, color4]
table {%
51 0.0619187693171194
51 0.252124185869584
};
\addplot [semithick, color4]
table {%
50.85 -0.254839197385811
51.15 -0.254839197385811
};
\addplot [semithick, color4]
table {%
50.85 0.252124185869584
51.15 0.252124185869584
};
\addplot [semithick, color4]
table {%
53.7 -0.0545474084291705
54.3 -0.0545474084291705
54.3 0.0577165445868354
53.7 0.0577165445868354
53.7 -0.0545474084291705
};
\addplot [semithick, color4]
table {%
54 -0.0545474084291705
54 -0.222935665611319
};
\addplot [semithick, color4]
table {%
54 0.0577165445868354
54 0.226086593245831
};
\addplot [semithick, color4]
table {%
53.85 -0.222935665611319
54.15 -0.222935665611319
};
\addplot [semithick, color4]
table {%
53.85 0.226086593245831
54.15 0.226086593245831
};
\addplot [semithick, color4]
table {%
56.7 -0.0538808575200522
57.3 -0.0538808575200522
57.3 0.0535680276143302
56.7 0.0535680276143302
56.7 -0.0538808575200522
};
\addplot [semithick, color4]
table {%
57 -0.0538808575200522
57 -0.214742679950021
};
\addplot [semithick, color4]
table {%
57 0.0535680276143302
57 0.214684787114386
};
\addplot [semithick, color4]
table {%
56.85 -0.214742679950021
57.15 -0.214742679950021
};
\addplot [semithick, color4]
table {%
56.85 0.214684787114386
57.15 0.214684787114386
};
\addplot [semithick, color6]
table {%
0.5 -0.10173477531898
1.1 -0.10173477531898
1.1 0.0840219930871446
0.5 0.0840219930871446
0.5 -0.10173477531898
};
\addplot [semithick, color6]
table {%
0.8 -0.10173477531898
0.8 -0.380264376372473
};
\addplot [semithick, color6]
table {%
0.8 0.0840219930871446
0.8 0.362639434762832
};
\addplot [semithick, color6]
table {%
0.65 -0.380264376372473
0.95 -0.380264376372473
};
\addplot [semithick, color6]
table {%
0.65 0.362639434762832
0.95 0.362639434762832
};
\addplot [semithick, color6]
table {%
3.5 -0.0795881397485717
4.1 -0.0795881397485717
4.1 0.0684132399588336
3.5 0.0684132399588336
3.5 -0.0795881397485717
};
\addplot [semithick, color6]
table {%
3.8 -0.0795881397485717
3.8 -0.301428380712309
};
\addplot [semithick, color6]
table {%
3.8 0.0684132399588336
3.8 0.290387867941884
};
\addplot [semithick, color6]
table {%
3.65 -0.301428380712309
3.95 -0.301428380712309
};
\addplot [semithick, color6]
table {%
3.65 0.290387867941884
3.95 0.290387867941884
};
\addplot [semithick, color6]
table {%
6.5 -0.0744361399714504
7.1 -0.0744361399714504
7.1 0.0598026266363737
6.5 0.0598026266363737
6.5 -0.0744361399714504
};
\addplot [semithick, color6]
table {%
6.8 -0.0744361399714504
6.8 -0.275788401184036
};
\addplot [semithick, color6]
table {%
6.8 0.0598026266363737
6.8 0.261042849813383
};
\addplot [semithick, color6]
table {%
6.65 -0.275788401184036
6.95 -0.275788401184036
};
\addplot [semithick, color6]
table {%
6.65 0.261042849813383
6.95 0.261042849813383
};
\addplot [semithick, color6]
table {%
9.5 -0.0457408512661562
10.1 -0.0457408512661562
10.1 0.0418263336242468
9.5 0.0418263336242468
9.5 -0.0457408512661562
};
\addplot [semithick, color6]
table {%
9.8 -0.0457408512661562
9.8 -0.17696336060105
};
\addplot [semithick, color6]
table {%
9.8 0.0418263336242468
9.8 0.173154844383216
};
\addplot [semithick, color6]
table {%
9.65 -0.17696336060105
9.95 -0.17696336060105
};
\addplot [semithick, color6]
table {%
9.65 0.173154844383216
9.95 0.173154844383216
};
\addplot [semithick, color6]
table {%
12.5 -0.0524773514562209
13.1 -0.0524773514562209
13.1 0.044068598085479
12.5 0.044068598085479
12.5 -0.0524773514562209
};
\addplot [semithick, color6]
table {%
12.8 -0.0524773514562209
12.8 -0.197199613339235
};
\addplot [semithick, color6]
table {%
12.8 0.044068598085479
12.8 0.188850135769527
};
\addplot [semithick, color6]
table {%
12.65 -0.197199613339235
12.95 -0.197199613339235
};
\addplot [semithick, color6]
table {%
12.65 0.188850135769527
12.95 0.188850135769527
};
\addplot [semithick, color6]
table {%
15.5 -0.153673413698667
16.1 -0.153673413698667
16.1 0.138789776576667
15.5 0.138789776576667
15.5 -0.153673413698667
};
\addplot [semithick, color6]
table {%
15.8 -0.153673413698667
15.8 -0.591342219490219
};
\addplot [semithick, color6]
table {%
15.8 0.138789776576667
15.8 0.576927463293929
};
\addplot [semithick, color6]
table {%
15.65 -0.591342219490219
15.95 -0.591342219490219
};
\addplot [semithick, color6]
table {%
15.65 0.576927463293929
15.95 0.576927463293929
};
\addplot [semithick, color6]
table {%
18.5 -0.0876318319044805
19.1 -0.0876318319044805
19.1 0.0760596176529636
18.5 0.0760596176529636
18.5 -0.0876318319044805
};
\addplot [semithick, color6]
table {%
18.8 -0.0876318319044805
18.8 -0.332992526239564
};
\addplot [semithick, color6]
table {%
18.8 0.0760596176529636
18.8 0.32155016189652
};
\addplot [semithick, color6]
table {%
18.65 -0.332992526239564
18.95 -0.332992526239564
};
\addplot [semithick, color6]
table {%
18.65 0.32155016189652
18.95 0.32155016189652
};
\addplot [semithick, color6]
table {%
21.5 -0.101582895800561
22.1 -0.101582895800561
22.1 0.094558224170003
21.5 0.094558224170003
21.5 -0.101582895800561
};
\addplot [semithick, color6]
table {%
21.8 -0.101582895800561
21.8 -0.395610434921939
};
\addplot [semithick, color6]
table {%
21.8 0.094558224170003
21.8 0.388691415165704
};
\addplot [semithick, color6]
table {%
21.65 -0.395610434921939
21.95 -0.395610434921939
};
\addplot [semithick, color6]
table {%
21.65 0.388691415165704
21.95 0.388691415165704
};
\addplot [semithick, color6]
table {%
24.5 -0.0787760924127822
25.1 -0.0787760924127822
25.1 0.0728185336349476
24.5 0.0728185336349476
24.5 -0.0787760924127822
};
\addplot [semithick, color6]
table {%
24.8 -0.0787760924127822
24.8 -0.305772487690363
};
\addplot [semithick, color6]
table {%
24.8 0.0728185336349476
24.8 0.300151854775085
};
\addplot [semithick, color6]
table {%
24.65 -0.305772487690363
24.95 -0.305772487690363
};
\addplot [semithick, color6]
table {%
24.65 0.300151854775085
24.95 0.300151854775085
};
\addplot [semithick, color6]
table {%
27.5 -0.0663453310699755
28.1 -0.0663453310699755
28.1 0.0670745595284851
27.5 0.0670745595284851
27.5 -0.0663453310699755
};
\addplot [semithick, color6]
table {%
27.8 -0.0663453310699755
27.8 -0.266074888528599
};
\addplot [semithick, color6]
table {%
27.8 0.0670745595284851
27.8 0.266601948497243
};
\addplot [semithick, color6]
table {%
27.65 -0.266074888528599
27.95 -0.266074888528599
};
\addplot [semithick, color6]
table {%
27.65 0.266601948497243
27.95 0.266601948497243
};
\addplot [semithick, color6]
table {%
30.5 -0.154822621781034
31.1 -0.154822621781034
31.1 0.141089295541809
30.5 0.141089295541809
30.5 -0.154822621781034
};
\addplot [semithick, color6]
table {%
30.8 -0.154822621781034
30.8 -0.598594746763573
};
\addplot [semithick, color6]
table {%
30.8 0.141089295541809
30.8 0.584872229969847
};
\addplot [semithick, color6]
table {%
30.65 -0.598594746763573
30.95 -0.598594746763573
};
\addplot [semithick, color6]
table {%
30.65 0.584872229969847
30.95 0.584872229969847
};
\addplot [semithick, color6]
table {%
33.5 -0.119873406293718
34.1 -0.119873406293718
34.1 0.106507630613833
33.5 0.106507630613833
33.5 -0.119873406293718
};
\addplot [semithick, color6]
table {%
33.8 -0.119873406293718
33.8 -0.459391274626619
};
\addplot [semithick, color6]
table {%
33.8 0.106507630613833
33.8 0.446039457823008
};
\addplot [semithick, color6]
table {%
33.65 -0.459391274626619
33.95 -0.459391274626619
};
\addplot [semithick, color6]
table {%
33.65 0.446039457823008
33.95 0.446039457823008
};
\addplot [semithick, color6]
table {%
36.5 -0.0901901994580864
37.1 -0.0901901994580864
37.1 0.081046616226083
36.5 0.081046616226083
36.5 -0.0901901994580864
};
\addplot [semithick, color6]
table {%
36.8 -0.0901901994580864
36.8 -0.34702340695837
};
\addplot [semithick, color6]
table {%
36.8 0.081046616226083
36.8 0.337629271326574
};
\addplot [semithick, color6]
table {%
36.65 -0.34702340695837
36.95 -0.34702340695837
};
\addplot [semithick, color6]
table {%
36.65 0.337629271326574
36.95 0.337629271326574
};
\addplot [semithick, color6]
table {%
39.5 -0.0921276270543989
40.1 -0.0921276270543989
40.1 0.0841565752663468
39.5 0.0841565752663468
39.5 -0.0921276270543989
};
\addplot [semithick, color6]
table {%
39.8 -0.0921276270543989
39.8 -0.356343815405432
};
\addplot [semithick, color6]
table {%
39.8 0.0841565752663468
39.8 0.348566622523028
};
\addplot [semithick, color6]
table {%
39.65 -0.356343815405432
39.95 -0.356343815405432
};
\addplot [semithick, color6]
table {%
39.65 0.348566622523028
39.95 0.348566622523028
};
\addplot [semithick, color6]
table {%
42.5 -0.0787329673610492
43.1 -0.0787329673610492
43.1 0.0726903805591578
42.5 0.0726903805591578
42.5 -0.0787329673610492
};
\addplot [semithick, color6]
table {%
42.8 -0.0787329673610492
42.8 -0.305585399197842
};
\addplot [semithick, color6]
table {%
42.8 0.0726903805591578
42.8 0.299543970648234
};
\addplot [semithick, color6]
table {%
42.65 -0.305585399197842
42.95 -0.305585399197842
};
\addplot [semithick, color6]
table {%
42.65 0.299543970648234
42.95 0.299543970648234
};
\addplot [semithick, color6]
table {%
45.5 -0.0739939388351548
46.1 -0.0739939388351548
46.1 0.0660366782893659
45.5 0.0660366782893659
45.5 -0.0739939388351548
};
\addplot [semithick, color6]
table {%
45.8 -0.0739939388351548
45.8 -0.284022072524827
};
\addplot [semithick, color6]
table {%
45.8 0.0660366782893659
45.8 0.276000263018169
};
\addplot [semithick, color6]
table {%
45.65 -0.284022072524827
45.95 -0.284022072524827
};
\addplot [semithick, color6]
table {%
45.65 0.276000263018169
45.95 0.276000263018169
};
\addplot [semithick, color6]
table {%
48.5 -0.0727297717408128
49.1 -0.0727297717408128
49.1 0.0669980743718107
48.5 0.0669980743718107
48.5 -0.0727297717408128
};
\addplot [semithick, color6]
table {%
48.8 -0.0727297717408128
48.8 -0.282268832610484
};
\addplot [semithick, color6]
table {%
48.8 0.0669980743718107
48.8 0.276314002995614
};
\addplot [semithick, color6]
table {%
48.65 -0.282268832610484
48.95 -0.282268832610484
};
\addplot [semithick, color6]
table {%
48.65 0.276314002995614
48.95 0.276314002995614
};
\addplot [semithick, color6]
table {%
51.5 -0.0680772975388619
52.1 -0.0680772975388619
52.1 0.0632902252336239
51.5 0.0632902252336239
51.5 -0.0680772975388619
};
\addplot [semithick, color6]
table {%
51.8 -0.0680772975388619
51.8 -0.265100239450917
};
\addplot [semithick, color6]
table {%
51.8 0.0632902252336239
51.8 0.260335071881501
};
\addplot [semithick, color6]
table {%
51.65 -0.265100239450917
51.95 -0.265100239450917
};
\addplot [semithick, color6]
table {%
51.65 0.260335071881501
51.95 0.260335071881501
};
\addplot [semithick, color6]
table {%
54.5 -0.0601611272664319
55.1 -0.0601611272664319
55.1 0.0567236833935339
54.5 0.0567236833935339
54.5 -0.0601611272664319
};
\addplot [semithick, color6]
table {%
54.8 -0.0601611272664319
54.8 -0.2350025554062
};
\addplot [semithick, color6]
table {%
54.8 0.0567236833935339
54.8 0.232015048870137
};
\addplot [semithick, color6]
table {%
54.65 -0.2350025554062
54.95 -0.2350025554062
};
\addplot [semithick, color6]
table {%
54.65 0.232015048870137
54.95 0.232015048870137
};
\addplot [semithick, color6]
table {%
57.5 -0.0563341082464858
58.1 -0.0563341082464858
58.1 0.0540805943834459
57.5 0.0540805943834459
57.5 -0.0563341082464858
};
\addplot [semithick, color6]
table {%
57.8 -0.0563341082464858
57.8 -0.22192989608271
};
\addplot [semithick, color6]
table {%
57.8 0.0540805943834459
57.8 0.21949247696989
};
\addplot [semithick, color6]
table {%
57.65 -0.22192989608271
57.95 -0.22192989608271
};
\addplot [semithick, color6]
table {%
57.65 0.21949247696989
57.95 0.21949247696989
};
\addplot [semithick, color2]
table {%
-1.1 -0.0197804115642586
-0.5 -0.0197804115642586
};
\addplot [semithick, color2]
table {%
1.9 -0.00015431863573967
2.5 -0.00015431863573967
};
\addplot [semithick, color2]
table {%
4.9 -0.0157096313091663
5.5 -0.0157096313091663
};
\addplot [semithick, color2]
table {%
7.9 -0.0505749032537851
8.5 -0.0505749032537851
};
\addplot [semithick, color2]
table {%
10.9 -0.0224559103011975
11.5 -0.0224559103011975
};
\addplot [semithick, color2]
table {%
13.9 0.0048081983492903
14.5 0.0048081983492903
};
\addplot [semithick, color2]
table {%
16.9 -0.0023064090210793
17.5 -0.0023064090210793
};
\addplot [semithick, color2]
table {%
19.9 0.00328041162311782
20.5 0.00328041162311782
};
\addplot [semithick, color2]
table {%
22.9 0.00659674083924724
23.5 0.00659674083924724
};
\addplot [semithick, color2]
table {%
25.9 0.0061257715777811
26.5 0.0061257715777811
};
\addplot [semithick, color2]
table {%
28.9 -0.00169974482926487
29.5 -0.00169974482926487
};
\addplot [semithick, color2]
table {%
31.9 -0.00191920960870139
32.5 -0.00191920960870139
};
\addplot [semithick, color2]
table {%
34.9 -0.000204615774079653
35.5 -0.000204615774079653
};
\addplot [semithick, color2]
table {%
37.9 0.00337637528546771
38.5 0.00337637528546771
};
\addplot [semithick, color2]
table {%
40.9 0.00018235623022405
41.5 0.00018235623022405
};
\addplot [semithick, color2]
table {%
43.9 0.00372845292601712
44.5 0.00372845292601712
};
\addplot [semithick, color2]
table {%
46.9 0.00416323328692325
47.5 0.00416323328692325
};
\addplot [semithick, color2]
table {%
49.9 0.00359773738869407
50.5 0.00359773738869407
};
\addplot [semithick, color2]
table {%
52.9 0.00526006051538161
53.5 0.00526006051538161
};
\addplot [semithick, color2]
table {%
55.9 0.00410391517792921
56.5 0.00410391517792921
};
\addplot [semithick, color4]
table {%
-0.3 -0.00820763754191858
0.3 -0.00820763754191858
};
\addplot [semithick, color4]
table {%
2.7 0.00293558350773301
3.3 0.00293558350773301
};
\addplot [semithick, color4]
table {%
5.7 -0.00504112823507089
6.3 -0.00504112823507089
};
\addplot [semithick, color4]
table {%
8.7 -0.0220345975983421
9.3 -0.0220345975983421
};
\addplot [semithick, color4]
table {%
11.7 -0.0093759795668817
12.3 -0.0093759795668817
};
\addplot [semithick, color4]
table {%
14.7 -3.81936954744165e-06
15.3 -3.81936954744165e-06
};
\addplot [semithick, color4]
table {%
17.7 -0.00940674738202079
18.3 -0.00940674738202079
};
\addplot [semithick, color4]
table {%
20.7 -0.00470956096335349
21.3 -0.00470956096335349
};
\addplot [semithick, color4]
table {%
23.7 -0.000926487736353007
24.3 -0.000926487736353007
};
\addplot [semithick, color4]
table {%
26.7 -0.000737337385551166
27.3 -0.000737337385551166
};
\addplot [semithick, color4]
table {%
29.7 -0.00422161039861568
30.3 -0.00422161039861568
};
\addplot [semithick, color4]
table {%
32.7 -0.00260118213564688
33.3 -0.00260118213564688
};
\addplot [semithick, color4]
table {%
35.7 -0.00185605475023687
36.3 -0.00185605475023687
};
\addplot [semithick, color4]
table {%
38.7 0.00156315722137152
39.3 0.00156315722137152
};
\addplot [semithick, color4]
table {%
41.7 -0.00250085807339036
42.3 -0.00250085807339036
};
\addplot [semithick, color4]
table {%
44.7 -0.00222672744462605
45.3 -0.00222672744462605
};
\addplot [semithick, color4]
table {%
47.7 -0.000676368479594701
48.3 -0.000676368479594701
};
\addplot [semithick, color4]
table {%
50.7 -0.00126255402665501
51.3 -0.00126255402665501
};
\addplot [semithick, color4]
table {%
53.7 0.000345253023516269
54.3 0.000345253023516269
};
\addplot [semithick, color4]
table {%
56.7 -0.00133056699352504
57.3 -0.00133056699352504
};
\addplot [semithick, color6]
table {%
0.5 -0.0145879776150627
1.1 -0.0145879776150627
};
\addplot [semithick, color6]
table {%
3.5 -0.00567233613706727
4.1 -0.00567233613706727
};
\addplot [semithick, color6]
table {%
6.5 -0.00950644060287795
7.1 -0.00950644060287795
};
\addplot [semithick, color6]
table {%
9.5 -0.00702236151481288
10.1 -0.00702236151481288
};
\addplot [semithick, color6]
table {%
12.5 -0.00851412777147786
13.1 -0.00851412777147786
};
\addplot [semithick, color6]
table {%
15.5 -0.0035584011629143
16.1 -0.0035584011629143
};
\addplot [semithick, color6]
table {%
18.5 -0.0110913610799693
19.1 -0.0110913610799693
};
\addplot [semithick, color6]
table {%
21.5 -0.00862259269992935
22.1 -0.00862259269992935
};
\addplot [semithick, color6]
table {%
24.5 -0.00393115285607695
25.1 -0.00393115285607695
};
\addplot [semithick, color6]
table {%
27.5 -0.00139496765888573
28.1 -0.00139496765888573
};
\addplot [semithick, color6]
table {%
30.5 -0.0119699422282862
31.1 -0.0119699422282862
};
\addplot [semithick, color6]
table {%
33.5 -0.00707098338511447
34.1 -0.00707098338511447
};
\addplot [semithick, color6]
table {%
36.5 -0.0065458640360426
37.1 -0.0065458640360426
};
\addplot [semithick, color6]
table {%
39.5 0.000697354300848763
40.1 0.000697354300848763
};
\addplot [semithick, color6]
table {%
42.5 -0.00421037013806839
43.1 -0.00421037013806839
};
\addplot [semithick, color6]
table {%
45.5 -0.00490904026428164
46.1 -0.00490904026428164
};
\addplot [semithick, color6]
table {%
48.5 -0.00432321929543622
49.1 -0.00432321929543622
};
\addplot [semithick, color6]
table {%
51.5 -0.00242493017914305
52.1 -0.00242493017914305
};
\addplot [semithick, color6]
table {%
54.5 -0.00257508738899613
55.1 -0.00257508738899613
};
\addplot [semithick, color6]
table {%
57.5 -0.00289010499546601
58.1 -0.00289010499546601
};
\end{axis}

\begin{axis}[
width=\sfwidth,
height=\sfheight,
tick align=outside,
tick pos=right,
xlabel={Video content},
xmin=-3, xmax=60,
xtick={6,21,36,51},
xticklabels={Minecraft,Cities,Tour,Virus popper},
extra x ticks = {13.5, 28.5, 43.5},
extra x tick labels = {},
extra tick style={grid=major, tick align=inside},
ymin=-30, ymax=30,
ymajorticks = false,
]
\end{axis}

\end{tikzpicture}

%% file: img/gen_boxplot_Quantile_1.tex
\begin{tikzpicture}

\begin{axis}[
width=\sfwidth,
height=\sfheight,
legend cell align={left},
legend style={
  fill opacity=0.8,
  draw opacity=1,
  text opacity=1,
  at={(0.03,0.97)},
  anchor=north west,
  draw=white!80!black
},
tick align=outside,
tick pos=left,
ylabel={Relative error},
xlabel={Rate (Mb/s)},
x grid style={white!69.0196078431373!black},
xmin=-3, xmax=60,
xtick={0,3,6,9,12,15,18,21,24,27,30,33,36,39,42,45,48,51,54,57},
xticklabels={
  10,
  20,
  30,
  40,
  50,
  10,
  20,
  30,
  40,
  50,
  10,
  20,
  30,
  40,
  50,
  10,
  20,
  30,
  40,
  50
},
y grid style={white!69.0196078431373!black},
ymin=-1, ymax=0.5,
ytick style={color=black},
ymajorgrids,
ytick={-1,-0.75,-0.5,-0.25,0,0.25,0.5,0.75,1}
]

\addplot [thick, black, forget plot]
table {%
-3 0
-2 0
-1 0
0 0
1 0
2 0
3 0
4 0
5 0
6 0
7 0
8 0
9 0
10 0
11 0
12 0
13 0
14 0
15 0
16 0
17 0
18 0
19 0
20 0
21 0
22 0
23 0
24 0
25 0
26 0
27 0
28 0
29 0
30 0
31 0
32 0
33 0
34 0
35 0
36 0
37 0
38 0
39 0
40 0
41 0
42 0
43 0
44 0
45 0
46 0
47 0
48 0
49 0
50 0
51 0
52 0
53 0
54 0
55 0
56 0
57 0
58 0
59 0
60 0
61 0
62 0
};
\addplot [semithick, color2]
table {%
-1.1 -0.37479389480433
-0.5 -0.37479389480433
-0.5 -0.193468186792113
-1.1 -0.193468186792113
-1.1 -0.37479389480433
};
\addplot [semithick, color2]
table {%
-0.8 -0.37479389480433
-0.8 -0.646713016043388
};
\addplot [semithick, color2]
table {%
-0.8 -0.193468186792113
-0.8 0.0784134421070117
};
\addplot [semithick, color2]
table {%
-0.95 -0.646713016043388
-0.65 -0.646713016043388
};
\addplot [semithick, color2]
table {%
-0.95 0.0784134421070117
-0.65 0.0784134421070117
};
\addplot [semithick, color2]
table {%
1.9 -0.343142724454444
2.5 -0.343142724454444
2.5 -0.197486114600046
1.9 -0.197486114600046
1.9 -0.343142724454444
};
\addplot [semithick, color2]
table {%
2.2 -0.343142724454444
2.2 -0.561557823285426
};
\addplot [semithick, color2]
table {%
2.2 -0.197486114600046
2.2 0.0208437064762537
};
\addplot [semithick, color2]
table {%
2.05 -0.561557823285426
2.35 -0.561557823285426
};
\addplot [semithick, color2]
table {%
2.05 0.0208437064762537
2.35 0.0208437064762537
};
\addplot [semithick, color2]
table {%
4.9 -0.349338916065414
5.5 -0.349338916065414
5.5 -0.21818419667769
4.9 -0.21818419667769
4.9 -0.349338916065414
};
\addplot [semithick, color2]
table {%
5.2 -0.349338916065414
5.2 -0.546069522217039
};
\addplot [semithick, color2]
table {%
5.2 -0.21818419667769
5.2 -0.0216017469589519
};
\addplot [semithick, color2]
table {%
5.05 -0.546069522217039
5.35 -0.546069522217039
};
\addplot [semithick, color2]
table {%
5.05 -0.0216017469589519
5.35 -0.0216017469589519
};
\addplot [semithick, color2]
table {%
7.9 -0.369086390276717
8.5 -0.369086390276717
8.5 -0.23827700831948
7.9 -0.23827700831948
7.9 -0.369086390276717
};
\addplot [semithick, color2]
table {%
8.2 -0.369086390276717
8.2 -0.565108990295029
};
\addplot [semithick, color2]
table {%
8.2 -0.23827700831948
8.2 -0.042156869582989
};
\addplot [semithick, color2]
table {%
8.05 -0.565108990295029
8.35 -0.565108990295029
};
\addplot [semithick, color2]
table {%
8.05 -0.042156869582989
8.35 -0.042156869582989
};
\addplot [semithick, color2]
table {%
10.9 -0.344760788701446
11.5 -0.344760788701446
11.5 -0.230177442671936
10.9 -0.230177442671936
10.9 -0.344760788701446
};
\addplot [semithick, color2]
table {%
11.2 -0.344760788701446
11.2 -0.516564495219509
};
\addplot [semithick, color2]
table {%
11.2 -0.230177442671936
11.2 -0.058379270198138
};
\addplot [semithick, color2]
table {%
11.05 -0.516564495219509
11.35 -0.516564495219509
};
\addplot [semithick, color2]
table {%
11.05 -0.058379270198138
11.35 -0.058379270198138
};
\addplot [semithick, color2]
table {%
13.9 -0.405636140747284
14.5 -0.405636140747284
14.5 -0.115983850449255
13.9 -0.115983850449255
13.9 -0.405636140747284
};
\addplot [semithick, color2]
table {%
14.2 -0.405636140747284
14.2 -0.840104052187016
};
\addplot [semithick, color2]
table {%
14.2 -0.115983850449255
14.2 0.317153124027107
};
\addplot [semithick, color2]
table {%
14.05 -0.840104052187016
14.35 -0.840104052187016
};
\addplot [semithick, color2]
table {%
14.05 0.317153124027107
14.35 0.317153124027107
};
\addplot [semithick, color2]
table {%
16.9 -0.34677367935808
17.5 -0.34677367935808
17.5 -0.188630458055251
16.9 -0.188630458055251
16.9 -0.34677367935808
};
\addplot [semithick, color2]
table {%
17.2 -0.34677367935808
17.2 -0.583460570360169
};
\addplot [semithick, color2]
table {%
17.2 -0.188630458055251
17.2 0.048169336494136
};
\addplot [semithick, color2]
table {%
17.05 -0.583460570360169
17.35 -0.583460570360169
};
\addplot [semithick, color2]
table {%
17.05 0.048169336494136
17.35 0.048169336494136
};
\addplot [semithick, color2]
table {%
19.9 -0.355860660614776
20.5 -0.355860660614776
20.5 -0.17072338919474
19.9 -0.17072338919474
19.9 -0.355860660614776
};
\addplot [semithick, color2]
table {%
20.2 -0.355860660614776
20.2 -0.63346543289759
};
\addplot [semithick, color2]
table {%
20.2 -0.17072338919474
20.2 0.10687828171069
};
\addplot [semithick, color2]
table {%
20.05 -0.63346543289759
20.35 -0.63346543289759
};
\addplot [semithick, color2]
table {%
20.05 0.10687828171069
20.35 0.10687828171069
};
\addplot [semithick, color2]
table {%
22.9 -0.333004106107936
23.5 -0.333004106107936
23.5 -0.190693081532132
22.9 -0.190693081532132
22.9 -0.333004106107936
};
\addplot [semithick, color2]
table {%
23.2 -0.333004106107936
23.2 -0.546461206934254
};
\addplot [semithick, color2]
table {%
23.2 -0.190693081532132
23.2 0.0225595666009532
};
\addplot [semithick, color2]
table {%
23.05 -0.546461206934254
23.35 -0.546461206934254
};
\addplot [semithick, color2]
table {%
23.05 0.0225595666009532
23.35 0.0225595666009532
};
\addplot [semithick, color2]
table {%
25.9 -0.328957894830295
26.5 -0.328957894830295
26.5 -0.197266384328433
25.9 -0.197266384328433
25.9 -0.328957894830295
};
\addplot [semithick, color2]
table {%
26.2 -0.328957894830295
26.2 -0.525980897083731
};
\addplot [semithick, color2]
table {%
26.2 -0.197266384328433
26.2 2.6599603174568e-05
};
\addplot [semithick, color2]
table {%
26.05 -0.525980897083731
26.35 -0.525980897083731
};
\addplot [semithick, color2]
table {%
26.05 2.6599603174568e-05
26.35 2.6599603174568e-05
};
\addplot [semithick, color2]
table {%
28.9 -0.39588406240358
29.5 -0.39588406240358
29.5 -0.134601885119418
28.9 -0.134601885119418
28.9 -0.39588406240358
};
\addplot [semithick, color2]
table {%
29.2 -0.39588406240358
29.2 -0.787540195064975
};
\addplot [semithick, color2]
table {%
29.2 -0.134601885119418
29.2 0.256688792318988
};
\addplot [semithick, color2]
table {%
29.05 -0.787540195064975
29.35 -0.787540195064975
};
\addplot [semithick, color2]
table {%
29.05 0.256688792318988
29.35 0.256688792318988
};
\addplot [semithick, color2]
table {%
31.9 -0.372746039912096
32.5 -0.372746039912096
32.5 -0.162938912827823
31.9 -0.162938912827823
31.9 -0.372746039912096
};
\addplot [semithick, color2]
table {%
32.2 -0.372746039912096
32.2 -0.687240174371104
};
\addplot [semithick, color2]
table {%
32.2 -0.162938912827823
32.2 0.151644662706372
};
\addplot [semithick, color2]
table {%
32.05 -0.687240174371104
32.35 -0.687240174371104
};
\addplot [semithick, color2]
table {%
32.05 0.151644662706372
32.35 0.151644662706372
};
\addplot [semithick, color2]
table {%
34.9 -0.351463742956726
35.5 -0.351463742956726
35.5 -0.185161532358366
34.9 -0.185161532358366
34.9 -0.351463742956726
};
\addplot [semithick, color2]
table {%
35.2 -0.351463742956726
35.2 -0.600683015463343
};
\addplot [semithick, color2]
table {%
35.2 -0.185161532358366
35.2 0.0641827100952105
};
\addplot [semithick, color2]
table {%
35.05 -0.600683015463343
35.35 -0.600683015463343
};
\addplot [semithick, color2]
table {%
35.05 0.0641827100952105
35.35 0.0641827100952105
};
\addplot [semithick, color2]
table {%
37.9 -0.353893311025699
38.5 -0.353893311025699
38.5 -0.18410082892122
37.9 -0.18410082892122
37.9 -0.353893311025699
};
\addplot [semithick, color2]
table {%
38.2 -0.353893311025699
38.2 -0.608552605567032
};
\addplot [semithick, color2]
table {%
38.2 -0.18410082892122
38.2 0.0702610228469705
};
\addplot [semithick, color2]
table {%
38.05 -0.608552605567032
38.35 -0.608552605567032
};
\addplot [semithick, color2]
table {%
38.05 0.0702610228469705
38.35 0.0702610228469705
};
\addplot [semithick, color2]
table {%
40.9 -0.33864545400146
41.5 -0.33864545400146
41.5 -0.194923167754415
40.9 -0.194923167754415
40.9 -0.33864545400146
};
\addplot [semithick, color2]
table {%
41.2 -0.33864545400146
41.2 -0.553234752390781
};
\addplot [semithick, color2]
table {%
41.2 -0.194923167754415
41.2 0.019078862618532
};
\addplot [semithick, color2]
table {%
41.05 -0.553234752390781
41.35 -0.553234752390781
};
\addplot [semithick, color2]
table {%
41.05 0.019078862618532
41.35 0.019078862618532
};
\addplot [semithick, color2]
table {%
43.9 -0.336544818142926
44.5 -0.336544818142926
44.5 -0.196460883600771
43.9 -0.196460883600771
43.9 -0.336544818142926
};
\addplot [semithick, color2]
table {%
44.2 -0.336544818142926
44.2 -0.546656419022993
};
\addplot [semithick, color2]
table {%
44.2 -0.196460883600771
44.2 0.0134254396253451
};
\addplot [semithick, color2]
table {%
44.05 -0.546656419022993
44.35 -0.546656419022993
};
\addplot [semithick, color2]
table {%
44.05 0.0134254396253451
44.35 0.0134254396253451
};
\addplot [semithick, color2]
table {%
46.9 -0.330784385062131
47.5 -0.330784385062131
47.5 -0.200223997112359
46.9 -0.200223997112359
46.9 -0.330784385062131
};
\addplot [semithick, color2]
table {%
47.2 -0.330784385062131
47.2 -0.526199637973946
};
\addplot [semithick, color2]
table {%
47.2 -0.200223997112359
47.2 -0.00440057421561522
};
\addplot [semithick, color2]
table {%
47.05 -0.526199637973946
47.35 -0.526199637973946
};
\addplot [semithick, color2]
table {%
47.05 -0.00440057421561522
47.35 -0.00440057421561522
};
\addplot [semithick, color2]
table {%
49.9 -0.330313964884948
50.5 -0.330313964884948
50.5 -0.204037353468936
49.9 -0.204037353468936
49.9 -0.330313964884948
};
\addplot [semithick, color2]
table {%
50.2 -0.330313964884948
50.2 -0.519640523320114
};
\addplot [semithick, color2]
table {%
50.2 -0.204037353468936
50.2 -0.0146328638965783
};
\addplot [semithick, color2]
table {%
50.05 -0.519640523320114
50.35 -0.519640523320114
};
\addplot [semithick, color2]
table {%
50.05 -0.0146328638965783
50.35 -0.0146328638965783
};
\addplot [semithick, color2]
table {%
52.9 -0.319801611564975
53.5 -0.319801611564975
53.5 -0.208390304315384
52.9 -0.208390304315384
52.9 -0.319801611564975
};
\addplot [semithick, color2]
table {%
53.2 -0.319801611564975
53.2 -0.486800395657818
};
\addplot [semithick, color2]
table {%
53.2 -0.208390304315384
53.2 -0.0416367787990245
};
\addplot [semithick, color2]
table {%
53.05 -0.486800395657818
53.35 -0.486800395657818
};
\addplot [semithick, color2]
table {%
53.05 -0.0416367787990245
53.35 -0.0416367787990245
};
\addplot [semithick, color2]
table {%
55.9 -0.319988653233499
56.5 -0.319988653233499
56.5 -0.21135103475138
55.9 -0.21135103475138
55.9 -0.319988653233499
};
\addplot [semithick, color2]
table {%
56.2 -0.319988653233499
56.2 -0.482928945476707
};
\addplot [semithick, color2]
table {%
56.2 -0.21135103475138
56.2 -0.0484022335145042
};
\addplot [semithick, color2]
table {%
56.05 -0.482928945476707
56.35 -0.482928945476707
};
\addplot [semithick, color2]
table {%
56.05 -0.0484022335145042
56.35 -0.0484022335145042
};
\addplot [semithick, color4]
table {%
-0.3 -0.346969272149431
0.3 -0.346969272149431
0.3 -0.154695450276181
-0.3 -0.154695450276181
-0.3 -0.346969272149431
};
\addplot [semithick, color4]
table {%
0 -0.346969272149431
0 -0.635189853227798
};
\addplot [semithick, color4]
table {%
0 -0.154695450276181
0 0.133592099776505
};
\addplot [semithick, color4]
table {%
-0.15 -0.635189853227798
0.15 -0.635189853227798
};
\addplot [semithick, color4]
table {%
-0.15 0.133592099776505
0.15 0.133592099776505
};
\addplot [semithick, color4]
table {%
2.7 -0.333568803310931
3.3 -0.333568803310931
3.3 -0.184035099028026
2.7 -0.184035099028026
2.7 -0.333568803310931
};
\addplot [semithick, color4]
table {%
3 -0.333568803310931
3 -0.557794210271248
};
\addplot [semithick, color4]
table {%
3 -0.184035099028026
3 0.0402078538074668
};
\addplot [semithick, color4]
table {%
2.85 -0.557794210271248
3.15 -0.557794210271248
};
\addplot [semithick, color4]
table {%
2.85 0.0402078538074668
3.15 0.0402078538074668
};
\addplot [semithick, color4]
table {%
5.7 -0.318936581227873
6.3 -0.318936581227873
6.3 -0.182123460429616
5.7 -0.182123460429616
5.7 -0.318936581227873
};
\addplot [semithick, color4]
table {%
6 -0.318936581227873
6 -0.523794552605719
};
\addplot [semithick, color4]
table {%
6 -0.182123460429616
6 0.0228839860718535
};
\addplot [semithick, color4]
table {%
5.85 -0.523794552605719
6.15 -0.523794552605719
};
\addplot [semithick, color4]
table {%
5.85 0.0228839860718535
6.15 0.0228839860718535
};
\addplot [semithick, color4]
table {%
8.7 -0.285354596298007
9.3 -0.285354596298007
9.3 -0.166708715305928
8.7 -0.166708715305928
8.7 -0.285354596298007
};
\addplot [semithick, color4]
table {%
9 -0.285354596298007
9 -0.463295935463189
};
\addplot [semithick, color4]
table {%
9 -0.166708715305928
9 0.0111670373516476
};
\addplot [semithick, color4]
table {%
8.85 -0.463295935463189
9.15 -0.463295935463189
};
\addplot [semithick, color4]
table {%
8.85 0.0111670373516476
9.15 0.0111670373516476
};
\addplot [semithick, color4]
table {%
11.7 -0.303029545008247
12.3 -0.303029545008247
12.3 -0.184938730749802
11.7 -0.184938730749802
11.7 -0.303029545008247
};
\addplot [semithick, color4]
table {%
12 -0.303029545008247
12 -0.480126580288933
};
\addplot [semithick, color4]
table {%
12 -0.184938730749802
12 -0.00793435226165773
};
\addplot [semithick, color4]
table {%
11.85 -0.480126580288933
12.15 -0.480126580288933
};
\addplot [semithick, color4]
table {%
11.85 -0.00793435226165773
12.15 -0.00793435226165773
};
\addplot [semithick, color4]
table {%
14.7 -0.444318782160088
15.3 -0.444318782160088
15.3 -0.148185981470148
14.7 -0.148185981470148
14.7 -0.444318782160088
};
\addplot [semithick, color4]
table {%
15 -0.444318782160088
15 -0.888473060995665
};
\addplot [semithick, color4]
table {%
15 -0.148185981470148
15 0.295803313281823
};
\addplot [semithick, color4]
table {%
14.85 -0.888473060995665
15.15 -0.888473060995665
};
\addplot [semithick, color4]
table {%
14.85 0.295803313281823
15.15 0.295803313281823
};
\addplot [semithick, color4]
table {%
17.7 -0.388434619445616
18.3 -0.388434619445616
18.3 -0.219279946180877
17.7 -0.219279946180877
17.7 -0.388434619445616
};
\addplot [semithick, color4]
table {%
18 -0.388434619445616
18 -0.641986290635873
};
\addplot [semithick, color4]
table {%
18 -0.219279946180877
18 0.0344325352618801
};
\addplot [semithick, color4]
table {%
17.85 -0.641986290635873
18.15 -0.641986290635873
};
\addplot [semithick, color4]
table {%
17.85 0.0344325352618801
18.15 0.0344325352618801
};
\addplot [semithick, color4]
table {%
20.7 -0.395399207099567
21.3 -0.395399207099567
21.3 -0.201750281919559
20.7 -0.201750281919559
20.7 -0.395399207099567
};
\addplot [semithick, color4]
table {%
21 -0.395399207099567
21 -0.685509961699013
};
\addplot [semithick, color4]
table {%
21 -0.201750281919559
21 0.0885508341843963
};
\addplot [semithick, color4]
table {%
20.85 -0.685509961699013
21.15 -0.685509961699013
};
\addplot [semithick, color4]
table {%
20.85 0.0885508341843963
21.15 0.0885508341843963
};
\addplot [semithick, color4]
table {%
23.7 -0.375302214321359
24.3 -0.375302214321359
24.3 -0.222828960001934
23.7 -0.222828960001934
23.7 -0.375302214321359
};
\addplot [semithick, color4]
table {%
24 -0.375302214321359
24 -0.603990761560403
};
\addplot [semithick, color4]
table {%
24 -0.222828960001934
24 0.00573347555052596
};
\addplot [semithick, color4]
table {%
23.85 -0.603990761560403
24.15 -0.603990761560403
};
\addplot [semithick, color4]
table {%
23.85 0.00573347555052596
24.15 0.00573347555052596
};
\addplot [semithick, color4]
table {%
26.7 -0.367157305252649
27.3 -0.367157305252649
27.3 -0.230035594349362
26.7 -0.230035594349362
26.7 -0.367157305252649
};
\addplot [semithick, color4]
table {%
27 -0.367157305252649
27 -0.572747717896342
};
\addplot [semithick, color4]
table {%
27 -0.230035594349362
27 -0.0249278587108286
};
\addplot [semithick, color4]
table {%
26.85 -0.572747717896342
27.15 -0.572747717896342
};
\addplot [semithick, color4]
table {%
26.85 -0.0249278587108286
27.15 -0.0249278587108286
};
\addplot [semithick, color4]
table {%
29.7 -0.465427553558305
30.3 -0.465427553558305
30.3 -0.18760039343787
29.7 -0.18760039343787
29.7 -0.465427553558305
};
\addplot [semithick, color4]
table {%
30 -0.465427553558305
30 -0.882115604051783
};
\addplot [semithick, color4]
table {%
30 -0.18760039343787
30 0.228734717970457
};
\addplot [semithick, color4]
table {%
29.85 -0.882115604051783
30.15 -0.882115604051783
};
\addplot [semithick, color4]
table {%
29.85 0.228734717970457
30.15 0.228734717970457
};
\addplot [semithick, color4]
table {%
32.7 -0.440026109209839
33.3 -0.440026109209839
33.3 -0.217815409627836
32.7 -0.217815409627836
32.7 -0.440026109209839
};
\addplot [semithick, color4]
table {%
33 -0.440026109209839
33 -0.773166419397605
};
\addplot [semithick, color4]
table {%
33 -0.217815409627836
33 0.115384936279141
};
\addplot [semithick, color4]
table {%
32.85 -0.773166419397605
33.15 -0.773166419397605
};
\addplot [semithick, color4]
table {%
32.85 0.115384936279141
33.15 0.115384936279141
};
\addplot [semithick, color4]
table {%
35.7 -0.415099934176455
36.3 -0.415099934176455
36.3 -0.242570341267745
35.7 -0.242570341267745
35.7 -0.415099934176455
};
\addplot [semithick, color4]
table {%
36 -0.415099934176455
36 -0.673765510845714
};
\addplot [semithick, color4]
table {%
36 -0.242570341267745
36 0.0161282969407171
};
\addplot [semithick, color4]
table {%
35.85 -0.673765510845714
36.15 -0.673765510845714
};
\addplot [semithick, color4]
table {%
35.85 0.0161282969407171
36.15 0.0161282969407171
};
\addplot [semithick, color4]
table {%
38.7 -0.419854926807695
39.3 -0.419854926807695
39.3 -0.242262866609292
38.7 -0.242262866609292
38.7 -0.419854926807695
};
\addplot [semithick, color4]
table {%
39 -0.419854926807695
39 -0.685477011073925
};
\addplot [semithick, color4]
table {%
39 -0.242262866609292
39 0.0241000134830339
};
\addplot [semithick, color4]
table {%
38.85 -0.685477011073925
39.15 -0.685477011073925
};
\addplot [semithick, color4]
table {%
38.85 0.0241000134830339
39.15 0.0241000134830339
};
\addplot [semithick, color4]
table {%
41.7 -0.40659856607508
42.3 -0.40659856607508
42.3 -0.252642391682264
41.7 -0.252642391682264
41.7 -0.40659856607508
};
\addplot [semithick, color4]
table {%
42 -0.40659856607508
42 -0.637149713839708
};
\addplot [semithick, color4]
table {%
42 -0.252642391682264
42 -0.0219484699252094
};
\addplot [semithick, color4]
table {%
41.85 -0.637149713839708
42.15 -0.637149713839708
};
\addplot [semithick, color4]
table {%
41.85 -0.0219484699252094
42.15 -0.0219484699252094
};
\addplot [semithick, color4]
table {%
44.7 -0.285395042927795
45.3 -0.285395042927795
45.3 -0.138187042228734
44.7 -0.138187042228734
44.7 -0.285395042927795
};
\addplot [semithick, color4]
table {%
45 -0.285395042927795
45 -0.506176153674206
};
\addplot [semithick, color4]
table {%
45 -0.138187042228734
45 0.082396287933084
};
\addplot [semithick, color4]
table {%
44.85 -0.506176153674206
45.15 -0.506176153674206
};
\addplot [semithick, color4]
table {%
44.85 0.082396287933084
45.15 0.082396287933084
};
\addplot [semithick, color4]
table {%
47.7 -0.27802103179376
48.3 -0.27802103179376
48.3 -0.142597249566772
47.7 -0.142597249566772
47.7 -0.27802103179376
};
\addplot [semithick, color4]
table {%
48 -0.27802103179376
48 -0.481108451833191
};
\addplot [semithick, color4]
table {%
48 -0.142597249566772
48 0.0604213087498759
};
\addplot [semithick, color4]
table {%
47.85 -0.481108451833191
48.15 -0.481108451833191
};
\addplot [semithick, color4]
table {%
47.85 0.0604213087498759
48.15 0.0604213087498759
};
\addplot [semithick, color4]
table {%
50.7 -0.277373476491036
51.3 -0.277373476491036
51.3 -0.147116896846326
50.7 -0.147116896846326
50.7 -0.277373476491036
};
\addplot [semithick, color4]
table {%
51 -0.277373476491036
51 -0.472669151001219
};
\addplot [semithick, color4]
table {%
51 -0.147116896846326
51 0.0481136416612995
};
\addplot [semithick, color4]
table {%
50.85 -0.472669151001219
51.15 -0.472669151001219
};
\addplot [semithick, color4]
table {%
50.85 0.0481136416612995
51.15 0.0481136416612995
};
\addplot [semithick, color4]
table {%
53.7 -0.267213137548391
54.3 -0.267213137548391
54.3 -0.151122428291081
53.7 -0.151122428291081
53.7 -0.267213137548391
};
\addplot [semithick, color4]
table {%
54 -0.267213137548391
54 -0.441200576584291
};
\addplot [semithick, color4]
table {%
54 -0.151122428291081
54 0.0230094501406409
};
\addplot [semithick, color4]
table {%
53.85 -0.441200576584291
54.15 -0.441200576584291
};
\addplot [semithick, color4]
table {%
53.85 0.0230094501406409
54.15 0.0230094501406409
};
\addplot [semithick, color4]
table {%
56.7 -0.26492662861801
57.3 -0.26492662861801
57.3 -0.155878007909422
56.7 -0.155878007909422
56.7 -0.26492662861801
};
\addplot [semithick, color4]
table {%
57 -0.26492662861801
57 -0.428406805961727
};
\addplot [semithick, color4]
table {%
57 -0.155878007909422
57 0.00765127260010413
};
\addplot [semithick, color4]
table {%
56.85 -0.428406805961727
57.15 -0.428406805961727
};
\addplot [semithick, color4]
table {%
56.85 0.00765127260010413
57.15 0.00765127260010413
};
\addplot [semithick, color6]
table {%
0.5 -0.418922013074773
1.1 -0.418922013074773
1.1 -0.236735970855662
0.5 -0.236735970855662
0.5 -0.418922013074773
};
\addplot [semithick, color6]
table {%
0.8 -0.418922013074773
0.8 -0.692181549846063
};
\addplot [semithick, color6]
table {%
0.8 -0.236735970855662
0.8 0.0364641315635409
};
\addplot [semithick, color6]
table {%
0.65 -0.692181549846063
0.95 -0.692181549846063
};
\addplot [semithick, color6]
table {%
0.65 0.0364641315635409
0.95 0.0364641315635409
};
\addplot [semithick, color6]
table {%
3.5 -0.329007789911032
4.1 -0.329007789911032
4.1 -0.178999835700238
3.5 -0.178999835700238
3.5 -0.329007789911032
};
\addplot [semithick, color6]
table {%
3.8 -0.329007789911032
3.8 -0.553843283681754
};
\addplot [semithick, color6]
table {%
3.8 -0.178999835700238
3.8 0.0459177297815939
};
\addplot [semithick, color6]
table {%
3.65 -0.553843283681754
3.95 -0.553843283681754
};
\addplot [semithick, color6]
table {%
3.65 0.0459177297815939
3.95 0.0459177297815939
};
\addplot [semithick, color6]
table {%
6.5 -0.305384167563784
7.1 -0.305384167563784
7.1 -0.175283459821
6.5 -0.175283459821
6.5 -0.305384167563784
};
\addplot [semithick, color6]
table {%
6.8 -0.305384167563784
6.8 -0.500460775575825
};
\addplot [semithick, color6]
table {%
6.8 -0.175283459821
6.8 0.0198163680755441
};
\addplot [semithick, color6]
table {%
6.65 -0.500460775575825
6.95 -0.500460775575825
};
\addplot [semithick, color6]
table {%
6.65 0.0198163680755441
6.95 0.0198163680755441
};
\addplot [semithick, color6]
table {%
9.5 -0.263533848665294
10.1 -0.263533848665294
10.1 -0.0858216566857598
9.5 -0.0858216566857598
9.5 -0.263533848665294
};
\addplot [semithick, color6]
table {%
9.8 -0.263533848665294
9.8 -0.530038945390526
};
\addplot [semithick, color6]
table {%
9.8 -0.0858216566857598
9.8 0.180399091218799
};
\addplot [semithick, color6]
table {%
9.65 -0.530038945390526
9.95 -0.530038945390526
};
\addplot [semithick, color6]
table {%
9.65 0.180399091218799
9.95 0.180399091218799
};
\addplot [semithick, color6]
table {%
12.5 -0.256299406811325
13.1 -0.256299406811325
13.1 -0.148214289789167
12.5 -0.148214289789167
12.5 -0.256299406811325
};
\addplot [semithick, color6]
table {%
12.8 -0.256299406811325
12.8 -0.418317528324205
};
\addplot [semithick, color6]
table {%
12.8 -0.148214289789167
12.8 0.0137723953689482
};
\addplot [semithick, color6]
table {%
12.65 -0.418317528324205
12.95 -0.418317528324205
};
\addplot [semithick, color6]
table {%
12.65 0.0137723953689482
12.95 0.0137723953689482
};
\addplot [semithick, color6]
table {%
15.5 -0.577296623960893
16.1 -0.577296623960893
16.1 -0.27654523802156
15.5 -0.27654523802156
15.5 -0.577296623960893
};
\addplot [semithick, color6]
table {%
15.8 -0.577296623960893
15.8 -1.02791480161685
};
\addplot [semithick, color6]
table {%
15.8 -0.27654523802156
15.8 0.171095512425698
};
\addplot [semithick, color6]
table {%
15.65 -1.02791480161685
15.95 -1.02791480161685
};
\addplot [semithick, color6]
table {%
15.65 0.171095512425698
15.95 0.171095512425698
};
\addplot [semithick, color6]
table {%
18.5 -0.379966463731894
19.1 -0.379966463731894
19.1 -0.219369862425389
18.5 -0.219369862425389
18.5 -0.379966463731894
};
\addplot [semithick, color6]
table {%
18.8 -0.379966463731894
18.8 -0.620859286761183
};
\addplot [semithick, color6]
table {%
18.8 -0.219369862425389
18.8 0.0215124000924525
};
\addplot [semithick, color6]
table {%
18.65 -0.620859286761183
18.95 -0.620859286761183
};
\addplot [semithick, color6]
table {%
18.65 0.0215124000924525
18.95 0.0215124000924525
};
\addplot [semithick, color6]
table {%
21.5 -0.420328037212974
22.1 -0.420328037212974
22.1 -0.226152211153859
21.5 -0.226152211153859
21.5 -0.420328037212974
};
\addplot [semithick, color6]
table {%
21.8 -0.420328037212974
21.8 -0.711062593112632
};
\addplot [semithick, color6]
table {%
21.8 -0.226152211153859
21.8 0.0648301197453413
};
\addplot [semithick, color6]
table {%
21.65 -0.711062593112632
21.95 -0.711062593112632
};
\addplot [semithick, color6]
table {%
21.65 0.0648301197453413
21.95 0.0648301197453413
};
\addplot [semithick, color6]
table {%
24.5 -0.34021220196698
25.1 -0.34021220196698
25.1 -0.187211636029766
24.5 -0.187211636029766
24.5 -0.34021220196698
};
\addplot [semithick, color6]
table {%
24.8 -0.34021220196698
24.8 -0.569508975453248
};
\addplot [semithick, color6]
table {%
24.8 -0.187211636029766
24.8 0.0420801135219994
};
\addplot [semithick, color6]
table {%
24.65 -0.569508975453248
24.95 -0.569508975453248
};
\addplot [semithick, color6]
table {%
24.65 0.0420801135219994
24.95 0.0420801135219994
};
\addplot [semithick, color6]
table {%
27.5 -0.308082584758826
28.1 -0.308082584758826
28.1 -0.162210431102987
27.5 -0.162210431102987
27.5 -0.308082584758826
};
\addplot [semithick, color6]
table {%
27.8 -0.308082584758826
27.8 -0.526637281761782
};
\addplot [semithick, color6]
table {%
27.8 -0.162210431102987
27.8 0.0565276337737901
};
\addplot [semithick, color6]
table {%
27.65 -0.526637281761782
27.95 -0.526637281761782
};
\addplot [semithick, color6]
table {%
27.65 0.0565276337737901
27.95 0.0565276337737901
};
\addplot [semithick, color6]
table {%
30.5 -0.573197648448188
31.1 -0.573197648448188
31.1 -0.309970156569644
30.5 -0.309970156569644
30.5 -0.573197648448188
};
\addplot [semithick, color6]
table {%
30.8 -0.573197648448188
30.8 -0.967538477346715
};
\addplot [semithick, color6]
table {%
30.8 -0.309970156569644
30.8 0.0848276172093886
};
\addplot [semithick, color6]
table {%
30.65 -0.967538477346715
30.95 -0.967538477346715
};
\addplot [semithick, color6]
table {%
30.65 0.0848276172093886
30.95 0.0848276172093886
};
\addplot [semithick, color6]
table {%
33.5 -0.485810385577678
34.1 -0.485810385577678
34.1 -0.256051672111005
33.5 -0.256051672111005
33.5 -0.485810385577678
};
\addplot [semithick, color6]
table {%
33.8 -0.485810385577678
33.8 -0.830344197162562
};
\addplot [semithick, color6]
table {%
33.8 -0.256051672111005
33.8 0.0884108463667134
};
\addplot [semithick, color6]
table {%
33.65 -0.830344197162562
33.95 -0.830344197162562
};
\addplot [semithick, color6]
table {%
33.65 0.0884108463667134
33.95 0.0884108463667134
};
\addplot [semithick, color6]
table {%
36.5 -0.389087868462263
37.1 -0.389087868462263
37.1 -0.213739323348456
36.5 -0.213739323348456
36.5 -0.389087868462263
};
\addplot [semithick, color6]
table {%
36.8 -0.389087868462263
36.8 -0.65139319601635
};
\addplot [semithick, color6]
table {%
36.8 -0.213739323348456
36.8 0.0491705256082244
};
\addplot [semithick, color6]
table {%
36.65 -0.65139319601635
36.95 -0.65139319601635
};
\addplot [semithick, color6]
table {%
36.65 0.0491705256082244
36.95 0.0491705256082244
};
\addplot [semithick, color6]
table {%
39.5 -0.3816476079423
40.1 -0.3816476079423
40.1 -0.198000937861297
39.5 -0.198000937861297
39.5 -0.3816476079423
};
\addplot [semithick, color6]
table {%
39.8 -0.3816476079423
39.8 -0.657023000590669
};
\addplot [semithick, color6]
table {%
39.8 -0.198000937861297
39.8 0.077287637684414
};
\addplot [semithick, color6]
table {%
39.65 -0.657023000590669
39.95 -0.657023000590669
};
\addplot [semithick, color6]
table {%
39.65 0.077287637684414
39.95 0.077287637684414
};
\addplot [semithick, color6]
table {%
42.5 -0.372767380360126
43.1 -0.372767380360126
43.1 -0.207990491049083
42.5 -0.207990491049083
42.5 -0.372767380360126
};
\addplot [semithick, color6]
table {%
42.8 -0.372767380360126
42.8 -0.619851164019952
};
\addplot [semithick, color6]
table {%
42.8 -0.207990491049083
42.8 0.0386234964138311
};
\addplot [semithick, color6]
table {%
42.65 -0.619851164019952
42.95 -0.619851164019952
};
\addplot [semithick, color6]
table {%
42.65 0.0386234964138311
42.95 0.0386234964138311
};
\addplot [semithick, color6]
table {%
45.5 -0.298709769448805
46.1 -0.298709769448805
46.1 -0.159807635333233
45.5 -0.159807635333233
45.5 -0.298709769448805
};
\addplot [semithick, color6]
table {%
45.8 -0.298709769448805
45.8 -0.507037805418009
};
\addplot [semithick, color6]
table {%
45.8 -0.159807635333233
45.8 0.0485427631495898
};
\addplot [semithick, color6]
table {%
45.65 -0.507037805418009
45.95 -0.507037805418009
};
\addplot [semithick, color6]
table {%
45.65 0.0485427631495898
45.95 0.0485427631495898
};
\addplot [semithick, color6]
table {%
48.5 -0.290239812387956
49.1 -0.290239812387956
49.1 -0.144964267486741
48.5 -0.144964267486741
48.5 -0.290239812387956
};
\addplot [semithick, color6]
table {%
48.8 -0.290239812387956
48.8 -0.50801488226906
};
\addplot [semithick, color6]
table {%
48.8 -0.144964267486741
48.8 0.0728901503653805
};
\addplot [semithick, color6]
table {%
48.65 -0.50801488226906
48.95 -0.50801488226906
};
\addplot [semithick, color6]
table {%
48.65 0.0728901503653805
48.95 0.0728901503653805
};
\addplot [semithick, color6]
table {%
51.5 -0.285834708322571
52.1 -0.285834708322571
52.1 -0.145674643188293
51.5 -0.145674643188293
51.5 -0.285834708322571
};
\addplot [semithick, color6]
table {%
51.8 -0.285834708322571
51.8 -0.496057075912645
};
\addplot [semithick, color6]
table {%
51.8 -0.145674643188293
51.8 0.064514477940685
};
\addplot [semithick, color6]
table {%
51.65 -0.496057075912645
51.95 -0.496057075912645
};
\addplot [semithick, color6]
table {%
51.65 0.064514477940685
51.95 0.064514477940685
};
\addplot [semithick, color6]
table {%
54.5 -0.255235346898487
55.1 -0.255235346898487
55.1 -0.1306210220497
54.5 -0.1306210220497
54.5 -0.255235346898487
};
\addplot [semithick, color6]
table {%
54.8 -0.255235346898487
54.8 -0.442071129332339
};
\addplot [semithick, color6]
table {%
54.8 -0.1306210220497
54.8 0.056107429969041
};
\addplot [semithick, color6]
table {%
54.65 -0.442071129332339
54.95 -0.442071129332339
};
\addplot [semithick, color6]
table {%
54.65 0.056107429969041
54.95 0.056107429969041
};
\addplot [semithick, color6]
table {%
57.5 -0.228490140027836
58.1 -0.228490140027836
58.1 -0.112552496508074
57.5 -0.112552496508074
57.5 -0.228490140027836
};
\addplot [semithick, color6]
table {%
57.8 -0.228490140027836
57.8 -0.402352771528904
};
\addplot [semithick, color6]
table {%
57.8 -0.112552496508074
57.8 0.0613378744838862
};
\addplot [semithick, color6]
table {%
57.65 -0.402352771528904
57.95 -0.402352771528904
};
\addplot [semithick, color6]
table {%
57.65 0.0613378744838862
57.95 0.0613378744838862
};
\addplot [semithick, color2]
table {%
-1.1 -0.290198288008496
-0.5 -0.290198288008496
};
\addplot [semithick, color2]
table {%
1.9 -0.270691212593367
2.5 -0.270691212593367
};
\addplot [semithick, color2]
table {%
4.9 -0.285876421039037
5.5 -0.285876421039037
};
\addplot [semithick, color2]
table {%
7.9 -0.319408549732595
8.5 -0.319408549732595
};
\addplot [semithick, color2]
table {%
10.9 -0.29241528222112
11.5 -0.29241528222112
};
\addplot [semithick, color2]
table {%
13.9 -0.263575343643521
14.5 -0.263575343643521
};
\addplot [semithick, color2]
table {%
16.9 -0.272751805711441
17.5 -0.272751805711441
};
\addplot [semithick, color2]
table {%
19.9 -0.266713289328013
20.5 -0.266713289328013
};
\addplot [semithick, color2]
table {%
22.9 -0.263776462466308
23.5 -0.263776462466308
};
\addplot [semithick, color2]
table {%
25.9 -0.264086630598824
26.5 -0.264086630598824
};
\addplot [semithick, color2]
table {%
28.9 -0.270674891450916
29.5 -0.270674891450916
};
\addplot [semithick, color2]
table {%
31.9 -0.271782054911818
32.5 -0.271782054911818
};
\addplot [semithick, color2]
table {%
34.9 -0.270670066913261
35.5 -0.270670066913261
};
\addplot [semithick, color2]
table {%
37.9 -0.266779442537412
38.5 -0.266779442537412
};
\addplot [semithick, color2]
table {%
40.9 -0.270242021944875
41.5 -0.270242021944875
};
\addplot [semithick, color2]
table {%
43.9 -0.26620409656194
44.5 -0.26620409656194
};
\addplot [semithick, color2]
table {%
46.9 -0.266083679553809
47.5 -0.266083679553809
};
\addplot [semithick, color2]
table {%
49.9 -0.26676014979802
50.5 -0.26676014979802
};
\addplot [semithick, color2]
table {%
52.9 -0.264841949736565
53.5 -0.264841949736565
};
\addplot [semithick, color2]
table {%
55.9 -0.266472222384394
56.5 -0.266472222384394
};
\addplot [semithick, color4]
table {%
-0.3 -0.249699567954852
0.3 -0.249699567954852
};
\addplot [semithick, color4]
table {%
2.7 -0.257067728190229
3.3 -0.257067728190229
};
\addplot [semithick, color4]
table {%
5.7 -0.248424288097508
6.3 -0.248424288097508
};
\addplot [semithick, color4]
table {%
8.7 -0.213011591625433
9.3 -0.213011591625433
};
\addplot [semithick, color4]
table {%
11.7 -0.241346204719834
12.3 -0.241346204719834
};
\addplot [semithick, color4]
table {%
14.7 -0.297081348724595
15.3 -0.297081348724595
};
\addplot [semithick, color4]
table {%
17.7 -0.3080206015214
18.3 -0.3080206015214
};
\addplot [semithick, color4]
table {%
20.7 -0.303454738045169
21.3 -0.303454738045169
};
\addplot [semithick, color4]
table {%
23.7 -0.300335027724417
24.3 -0.300335027724417
};
\addplot [semithick, color4]
table {%
26.7 -0.300012926202118
27.3 -0.300012926202118
};
\addplot [semithick, color4]
table {%
29.7 -0.331384435936354
30.3 -0.331384435936354
};
\addplot [semithick, color4]
table {%
32.7 -0.331390851021521
33.3 -0.331390851021521
};
\addplot [semithick, color4]
table {%
35.7 -0.330621782415612
36.3 -0.330621782415612
};
\addplot [semithick, color4]
table {%
38.7 -0.326347623209421
39.3 -0.326347623209421
};
\addplot [semithick, color4]
table {%
41.7 -0.330818995102396
42.3 -0.330818995102396
};
\addplot [semithick, color4]
table {%
44.7 -0.213406105380071
45.3 -0.213406105380071
};
\addplot [semithick, color4]
table {%
47.7 -0.211570499974828
48.3 -0.211570499974828
};
\addplot [semithick, color4]
table {%
50.7 -0.212076129098749
51.3 -0.212076129098749
};
\addplot [semithick, color4]
table {%
53.7 -0.210099176856626
54.3 -0.210099176856626
};
\addplot [semithick, color4]
table {%
56.7 -0.212185275040659
57.3 -0.212185275040659
};
\addplot [semithick, color6]
table {%
0.5 -0.327788531100682
1.1 -0.327788531100682
};
\addplot [semithick, color6]
table {%
3.5 -0.253964875842319
4.1 -0.253964875842319
};
\addplot [semithick, color6]
table {%
6.5 -0.240727233300614
7.1 -0.240727233300614
};
\addplot [semithick, color6]
table {%
9.5 -0.155893326032583
10.1 -0.155893326032583
};
\addplot [semithick, color6]
table {%
12.5 -0.19919963037552
13.1 -0.19919963037552
};
\addplot [semithick, color6]
table {%
15.5 -0.424864321669639
16.1 -0.424864321669639
};
\addplot [semithick, color6]
table {%
18.5 -0.304921740365482
19.1 -0.304921740365482
};
\addplot [semithick, color6]
table {%
21.5 -0.329174997758687
22.1 -0.329174997758687
};
\addplot [semithick, color6]
table {%
24.5 -0.26579821168302
25.1 -0.26579821168302
};
\addplot [semithick, color6]
table {%
27.5 -0.236150911841601
28.1 -0.236150911841601
};
\addplot [semithick, color6]
table {%
30.5 -0.442804246026334
31.1 -0.442804246026334
};
\addplot [semithick, color6]
table {%
33.5 -0.371821847266187
34.1 -0.371821847266187
};
\addplot [semithick, color6]
table {%
36.5 -0.302864598987909
37.1 -0.302864598987909
};
\addplot [semithick, color6]
table {%
39.5 -0.284699633995435
40.1 -0.284699633995435
};
\addplot [semithick, color6]
table {%
42.5 -0.29007187459228
43.1 -0.29007187459228
};
\addplot [semithick, color6]
table {%
45.5 -0.229311773169159
46.1 -0.229311773169159
};
\addplot [semithick, color6]
table {%
48.5 -0.219440095672465
49.1 -0.219440095672465
};
\addplot [semithick, color6]
table {%
51.5 -0.216398533192849
52.1 -0.216398533192849
};
\addplot [semithick, color6]
table {%
54.5 -0.193322478754317
55.1 -0.193322478754317
};
\addplot [semithick, color6]
table {%
57.5 -0.171712488633295
58.1 -0.171712488633295
};
\end{axis}

\begin{axis}[
width=\sfwidth,
height=\sfheight,
tick align=outside,
tick pos=right,
xlabel={Video content},
xmin=-3, xmax=60,
xtick={6,21,36,51},
xticklabels={Minecraft,Cities,Tour,Virus popper},
extra x ticks = {13.5, 28.5, 43.5},
extra x tick labels = {},
extra tick style={grid=major, tick align=inside},
ymin=-30, ymax=30,
ymajorticks = false,
]
\end{axis}

\end{tikzpicture}

%% file: img/gen_boxplot_Linear_6.tex
\begin{tikzpicture}

\begin{axis}[
width=\sfwidth,
height=\sfheight,
legend cell align={left},
legend style={
  fill opacity=0.8,
  draw opacity=1,
  text opacity=1,
  at={(0.03,0.97)},
  anchor=north west,
  draw=white!80!black
},
tick align=outside,
tick pos=left,
ylabel={Relative error},
xlabel={Rate (Mb/s)},
x grid style={white!69.0196078431373!black},
xmin=-3, xmax=60,
xtick={0,3,6,9,12,15,18,21,24,27,30,33,36,39,42,45,48,51,54,57},
xticklabels={
  10,
  20,
  30,
  40,
  50,
  10,
  20,
  30,
  40,
  50,
  10,
  20,
  30,
  40,
  50,
  10,
  20,
  30,
  40,
  50
},
y grid style={white!69.0196078431373!black},
ymin=-1, ymax=0.5,
ytick style={color=black},
ymajorgrids,
ytick={-1,-0.75,-0.5,-0.25,0,0.25,0.5,0.75,1}
]

\addplot [thick, black, forget plot]
table {%
-3 0
-2 0
-1 0
0 0
1 0
2 0
3 0
4 0
5 0
6 0
7 0
8 0
9 0
10 0
11 0
12 0
13 0
14 0
15 0
16 0
17 0
18 0
19 0
20 0
21 0
22 0
23 0
24 0
25 0
26 0
27 0
28 0
29 0
30 0
31 0
32 0
33 0
34 0
35 0
36 0
37 0
38 0
39 0
40 0
41 0
42 0
43 0
44 0
45 0
46 0
47 0
48 0
49 0
50 0
51 0
52 0
53 0
54 0
55 0
56 0
57 0
58 0
59 0
60 0
61 0
62 0
};
\addplot [semithick, color4]
table {%
-1.1 -0.0939246415967927
-0.5 -0.0939246415967927
-0.5 0.0617913903640893
-1.1 0.0617913903640893
-1.1 -0.0939246415967927
};
\addplot [semithick, color4]
table {%
-0.8 -0.0939246415967927
-0.8 -0.325318689671142
};
\addplot [semithick, color4]
table {%
-0.8 0.0617913903640893
-0.8 0.294881299580876
};
\addplot [semithick, color4]
table {%
-0.95 -0.325318689671142
-0.65 -0.325318689671142
};
\addplot [semithick, color4]
table {%
-0.95 0.294881299580876
-0.65 0.294881299580876
};
\addplot [semithick, color4]
table {%
1.9 -0.0510471762546101
2.5 -0.0510471762546101
2.5 0.0649137100152593
1.9 0.0649137100152593
1.9 -0.0510471762546101
};
\addplot [semithick, color4]
table {%
2.2 -0.0510471762546101
2.2 -0.223340511204278
};
\addplot [semithick, color4]
table {%
2.2 0.0649137100152593
2.2 0.238747149067937
};
\addplot [semithick, color4]
table {%
2.05 -0.223340511204278
2.35 -0.223340511204278
};
\addplot [semithick, color4]
table {%
2.05 0.238747149067937
2.35 0.238747149067937
};
\addplot [semithick, color4]
table {%
4.9 -0.069681683781115
5.5 -0.069681683781115
5.5 0.0427916699077448
4.9 0.0427916699077448
4.9 -0.069681683781115
};
\addplot [semithick, color4]
table {%
5.2 -0.069681683781115
5.2 -0.23797034625187
};
\addplot [semithick, color4]
table {%
5.2 0.0427916699077448
5.2 0.210869606709369
};
\addplot [semithick, color4]
table {%
5.05 -0.23797034625187
5.35 -0.23797034625187
};
\addplot [semithick, color4]
table {%
5.05 0.210869606709369
5.35 0.210869606709369
};
\addplot [semithick, color4]
table {%
7.9 -0.145911983731284
8.5 -0.145911983731284
8.5 0.0320556140549908
7.9 0.0320556140549908
7.9 -0.145911983731284
};
\addplot [semithick, color4]
table {%
8.2 -0.145911983731284
8.2 -0.394008061794859
};
\addplot [semithick, color4]
table {%
8.2 0.0320556140549908
8.2 0.297018226275402
};
\addplot [semithick, color4]
table {%
8.05 -0.394008061794859
8.35 -0.394008061794859
};
\addplot [semithick, color4]
table {%
8.05 0.297018226275402
8.35 0.297018226275402
};
\addplot [semithick, color4]
table {%
10.9 -0.0948320820801244
11.5 -0.0948320820801244
11.5 0.0401854016545855
10.9 0.0401854016545855
10.9 -0.0948320820801244
};
\addplot [semithick, color4]
table {%
11.2 -0.0948320820801244
11.2 -0.296220060974539
};
\addplot [semithick, color4]
table {%
11.2 0.0401854016545855
11.2 0.242365970090137
};
\addplot [semithick, color4]
table {%
11.05 -0.296220060974539
11.35 -0.296220060974539
};
\addplot [semithick, color4]
table {%
11.05 0.242365970090137
11.35 0.242365970090137
};
\addplot [semithick, color4]
table {%
13.9 -0.0786986947823202
14.5 -0.0786986947823202
14.5 0.0950513428108475
13.9 0.0950513428108475
13.9 -0.0786986947823202
};
\addplot [semithick, color4]
table {%
14.2 -0.0786986947823202
14.2 -0.338785942912333
};
\addplot [semithick, color4]
table {%
14.2 0.0950513428108475
14.2 0.355163756763397
};
\addplot [semithick, color4]
table {%
14.05 -0.338785942912333
14.35 -0.338785942912333
};
\addplot [semithick, color4]
table {%
14.05 0.355163756763397
14.35 0.355163756763397
};
\addplot [semithick, color4]
table {%
16.9 -0.0534023819842844
17.5 -0.0534023819842844
17.5 0.0692773308051889
16.9 0.0692773308051889
16.9 -0.0534023819842844
};
\addplot [semithick, color4]
table {%
17.2 -0.0534023819842844
17.2 -0.236793751090389
};
\addplot [semithick, color4]
table {%
17.2 0.0692773308051889
17.2 0.252921081972311
};
\addplot [semithick, color4]
table {%
17.05 -0.236793751090389
17.35 -0.236793751090389
};
\addplot [semithick, color4]
table {%
17.05 0.252921081972311
17.35 0.252921081972311
};
\addplot [semithick, color4]
table {%
19.9 -0.0549547231696513
20.5 -0.0549547231696513
20.5 0.0805823999354582
19.9 0.0805823999354582
19.9 -0.0549547231696513
};
\addplot [semithick, color4]
table {%
20.2 -0.0549547231696513
20.2 -0.258073126241122
};
\addplot [semithick, color4]
table {%
20.2 0.0805823999354582
20.2 0.282547689251727
};
\addplot [semithick, color4]
table {%
20.05 -0.258073126241122
20.35 -0.258073126241122
};
\addplot [semithick, color4]
table {%
20.05 0.282547689251727
20.35 0.282547689251727
};
\addplot [semithick, color4]
table {%
22.9 -0.0508613503398645
23.5 -0.0508613503398645
23.5 0.0732125207776023
22.9 0.0732125207776023
22.9 -0.0508613503398645
};
\addplot [semithick, color4]
table {%
23.2 -0.0508613503398645
23.2 -0.236409766583877
};
\addplot [semithick, color4]
table {%
23.2 0.0732125207776023
23.2 0.259305108563906
};
\addplot [semithick, color4]
table {%
23.05 -0.236409766583877
23.35 -0.236409766583877
};
\addplot [semithick, color4]
table {%
23.05 0.259305108563906
23.35 0.259305108563906
};
\addplot [semithick, color4]
table {%
25.9 -0.0520446045414447
26.5 -0.0520446045414447
26.5 0.0738827579054597
25.9 0.0738827579054597
25.9 -0.0520446045414447
};
\addplot [semithick, color4]
table {%
26.2 -0.0520446045414447
26.2 -0.240826726497408
};
\addplot [semithick, color4]
table {%
26.2 0.0738827579054597
26.2 0.260898185671643
};
\addplot [semithick, color4]
table {%
26.05 -0.240826726497408
26.35 -0.240826726497408
};
\addplot [semithick, color4]
table {%
26.05 0.260898185671643
26.35 0.260898185671643
};
\addplot [semithick, color4]
table {%
28.9 -0.0760444015037661
29.5 -0.0760444015037661
29.5 0.087350138526474
28.9 0.087350138526474
28.9 -0.0760444015037661
};
\addplot [semithick, color4]
table {%
29.2 -0.0760444015037661
29.2 -0.320062962973704
};
\addplot [semithick, color4]
table {%
29.2 0.087350138526474
29.2 0.331407880862555
};
\addplot [semithick, color4]
table {%
29.05 -0.320062962973704
29.35 -0.320062962973704
};
\addplot [semithick, color4]
table {%
29.05 0.331407880862555
29.35 0.331407880862555
};
\addplot [semithick, color4]
table {%
31.9 -0.0631281669058243
32.5 -0.0631281669058243
32.5 0.075808958245037
31.9 0.075808958245037
31.9 -0.0631281669058243
};
\addplot [semithick, color4]
table {%
32.2 -0.0631281669058243
32.2 -0.271340967423875
};
\addplot [semithick, color4]
table {%
32.2 0.075808958245037
32.2 0.284207342659765
};
\addplot [semithick, color4]
table {%
32.05 -0.271340967423875
32.35 -0.271340967423875
};
\addplot [semithick, color4]
table {%
32.05 0.284207342659765
32.35 0.284207342659765
};
\addplot [semithick, color4]
table {%
34.9 -0.0532597997044164
35.5 -0.0532597997044164
35.5 0.067555677301311
34.9 0.067555677301311
34.9 -0.0532597997044164
};
\addplot [semithick, color4]
table {%
35.2 -0.0532597997044164
35.2 -0.232089705215211
};
\addplot [semithick, color4]
table {%
35.2 0.067555677301311
35.2 0.24800968969133
};
\addplot [semithick, color4]
table {%
35.05 -0.232089705215211
35.35 -0.232089705215211
};
\addplot [semithick, color4]
table {%
35.05 0.24800968969133
35.35 0.24800968969133
};
\addplot [semithick, color4]
table {%
37.9 -0.0532590819502743
38.5 -0.0532590819502743
38.5 0.0621639809952993
37.9 0.0621639809952993
37.9 -0.0532590819502743
};
\addplot [semithick, color4]
table {%
38.2 -0.0532590819502743
38.2 -0.226132954661159
};
\addplot [semithick, color4]
table {%
38.2 0.0621639809952993
38.2 0.234993415798565
};
\addplot [semithick, color4]
table {%
38.05 -0.226132954661159
38.35 -0.226132954661159
};
\addplot [semithick, color4]
table {%
38.05 0.234993415798565
38.35 0.234993415798565
};
\addplot [semithick, color4]
table {%
40.9 -0.0557292714655744
41.5 -0.0557292714655744
41.5 0.0675961612438983
40.9 0.0675961612438983
40.9 -0.0557292714655744
};
\addplot [semithick, color4]
table {%
41.2 -0.0557292714655744
41.2 -0.237683618693765
};
\addplot [semithick, color4]
table {%
41.2 0.0675961612438983
41.2 0.252037131143684
};
\addplot [semithick, color4]
table {%
41.05 -0.237683618693765
41.35 -0.237683618693765
};
\addplot [semithick, color4]
table {%
41.05 0.252037131143684
41.35 0.252037131143684
};
\addplot [semithick, color4]
table {%
43.9 -0.0503428869853248
44.5 -0.0503428869853248
44.5 0.0701376984477445
43.9 0.0701376984477445
43.9 -0.0503428869853248
};
\addplot [semithick, color4]
table {%
44.2 -0.0503428869853248
44.2 -0.228195284007147
};
\addplot [semithick, color4]
table {%
44.2 0.0701376984477445
44.2 0.250711615939622
};
\addplot [semithick, color4]
table {%
44.05 -0.228195284007147
44.35 -0.228195284007147
};
\addplot [semithick, color4]
table {%
44.05 0.250711615939622
44.35 0.250711615939622
};
\addplot [semithick, color4]
table {%
46.9 -0.0346407720940231
47.5 -0.0346407720940231
47.5 0.0593847198899347
46.9 0.0593847198899347
46.9 -0.0346407720940231
};
\addplot [semithick, color4]
table {%
47.2 -0.0346407720940231
47.2 -0.174655450416315
};
\addplot [semithick, color4]
table {%
47.2 0.0593847198899347
47.2 0.20039871546423
};
\addplot [semithick, color4]
table {%
47.05 -0.174655450416315
47.35 -0.174655450416315
};
\addplot [semithick, color4]
table {%
47.05 0.20039871546423
47.35 0.20039871546423
};
\addplot [semithick, color4]
table {%
49.9 -0.0350823505693294
50.5 -0.0350823505693294
50.5 0.0574680773646994
49.9 0.0574680773646994
49.9 -0.0350823505693294
};
\addplot [semithick, color4]
table {%
50.2 -0.0350823505693294
50.2 -0.173608219258721
};
\addplot [semithick, color4]
table {%
50.2 0.0574680773646994
50.2 0.195688341637035
};
\addplot [semithick, color4]
table {%
50.05 -0.173608219258721
50.35 -0.173608219258721
};
\addplot [semithick, color4]
table {%
50.05 0.195688341637035
50.35 0.195688341637035
};
\addplot [semithick, color4]
table {%
52.9 -0.0343272014945645
53.5 -0.0343272014945645
53.5 0.0577649808540542
52.9 0.0577649808540542
52.9 -0.0343272014945645
};
\addplot [semithick, color4]
table {%
53.2 -0.0343272014945645
53.2 -0.172409360105173
};
\addplot [semithick, color4]
table {%
53.2 0.0577649808540542
53.2 0.195465016030451
};
\addplot [semithick, color4]
table {%
53.05 -0.172409360105173
53.35 -0.172409360105173
};
\addplot [semithick, color4]
table {%
53.05 0.195465016030451
53.35 0.195465016030451
};
\addplot [semithick, color4]
table {%
55.9 -0.0228988867049045
56.5 -0.0228988867049045
56.5 0.050099502079219
55.9 0.050099502079219
55.9 -0.0228988867049045
};
\addplot [semithick, color4]
table {%
56.2 -0.0228988867049045
56.2 -0.131777396941793
};
\addplot [semithick, color4]
table {%
56.2 0.050099502079219
56.2 0.158622967067241
};
\addplot [semithick, color4]
table {%
56.05 -0.131777396941793
56.35 -0.131777396941793
};
\addplot [semithick, color4]
table {%
56.05 0.158622967067241
56.35 0.158622967067241
};
\addplot [semithick, color3]
table {%
-0.3 -0.0724508281671249
0.3 -0.0724508281671249
0.3 0.0726789559119333
-0.3 0.0726789559119333
-0.3 -0.0724508281671249
};
\addplot [semithick, color3]
table {%
0 -0.0724508281671249
0 -0.290116444846484
};
\addplot [semithick, color3]
table {%
0 0.0726789559119333
0 0.288275406988697
};
\addplot [semithick, color3]
table {%
-0.15 -0.290116444846484
0.15 -0.290116444846484
};
\addplot [semithick, color3]
table {%
-0.15 0.288275406988697
0.15 0.288275406988697
};
\addplot [semithick, color3]
table {%
2.7 -0.0482341775392704
3.3 -0.0482341775392704
3.3 0.070866990072131
2.7 0.070866990072131
2.7 -0.0482341775392704
};
\addplot [semithick, color3]
table {%
3 -0.0482341775392704
3 -0.226638288080448
};
\addplot [semithick, color3]
table {%
3 0.070866990072131
3 0.248598763095465
};
\addplot [semithick, color3]
table {%
2.85 -0.226638288080448
3.15 -0.226638288080448
};
\addplot [semithick, color3]
table {%
2.85 0.248598763095465
3.15 0.248598763095465
};
\addplot [semithick, color3]
table {%
5.7 -0.0498047373706457
6.3 -0.0498047373706457
6.3 0.0568525372474377
5.7 0.0568525372474377
5.7 -0.0498047373706457
};
\addplot [semithick, color3]
table {%
6 -0.0498047373706457
6 -0.209777932046418
};
\addplot [semithick, color3]
table {%
6 0.0568525372474377
6 0.216783406694379
};
\addplot [semithick, color3]
table {%
5.85 -0.209777932046418
6.15 -0.209777932046418
};
\addplot [semithick, color3]
table {%
5.85 0.216783406694379
6.15 0.216783406694379
};
\addplot [semithick, color3]
table {%
8.7 -0.0707254023170934
9.3 -0.0707254023170934
9.3 0.0370814094348482
8.7 0.0370814094348482
8.7 -0.0707254023170934
};
\addplot [semithick, color3]
table {%
9 -0.0707254023170934
9 -0.230456596856658
};
\addplot [semithick, color3]
table {%
9 0.0370814094348482
9 0.198643935239651
};
\addplot [semithick, color3]
table {%
8.85 -0.230456596856658
9.15 -0.230456596856658
};
\addplot [semithick, color3]
table {%
8.85 0.198643935239651
9.15 0.198643935239651
};
\addplot [semithick, color3]
table {%
11.7 -0.0602365232322419
12.3 -0.0602365232322419
12.3 0.0504221656399204
11.7 0.0504221656399204
11.7 -0.0602365232322419
};
\addplot [semithick, color3]
table {%
12 -0.0602365232322419
12 -0.225646315476736
};
\addplot [semithick, color3]
table {%
12 0.0504221656399204
12 0.215675912077503
};
\addplot [semithick, color3]
table {%
11.85 -0.225646315476736
12.15 -0.225646315476736
};
\addplot [semithick, color3]
table {%
11.85 0.215675912077503
12.15 0.215675912077503
};
\addplot [semithick, color3]
table {%
14.7 -0.0862260640787791
15.3 -0.0862260640787791
15.3 0.079217546847419
14.7 0.079217546847419
14.7 -0.0862260640787791
};
\addplot [semithick, color3]
table {%
15 -0.0862260640787791
15 -0.3317214671509
};
\addplot [semithick, color3]
table {%
15 0.079217546847419
15 0.327300545998761
};
\addplot [semithick, color3]
table {%
14.85 -0.3317214671509
15.15 -0.3317214671509
};
\addplot [semithick, color3]
table {%
14.85 0.327300545998761
15.15 0.327300545998761
};
\addplot [semithick, color3]
table {%
17.7 -0.0694553358165529
18.3 -0.0694553358165529
18.3 0.0626781496953002
17.7 0.0626781496953002
17.7 -0.0694553358165529
};
\addplot [semithick, color3]
table {%
18 -0.0694553358165529
18 -0.266148156606592
};
\addplot [semithick, color3]
table {%
18 0.0626781496953002
18 0.260753486717811
};
\addplot [semithick, color3]
table {%
17.85 -0.266148156606592
18.15 -0.266148156606592
};
\addplot [semithick, color3]
table {%
17.85 0.260753486717811
18.15 0.260753486717811
};
\addplot [semithick, color3]
table {%
20.7 -0.069245789156053
21.3 -0.069245789156053
21.3 0.0658539549216532
20.7 0.0658539549216532
20.7 -0.069245789156053
};
\addplot [semithick, color3]
table {%
21 -0.069245789156053
21 -0.27058230205541
};
\addplot [semithick, color3]
table {%
21 0.0658539549216532
21 0.267475017092561
};
\addplot [semithick, color3]
table {%
20.85 -0.27058230205541
21.15 -0.27058230205541
};
\addplot [semithick, color3]
table {%
20.85 0.267475017092561
21.15 0.267475017092561
};
\addplot [semithick, color3]
table {%
23.7 -0.0621424068009711
24.3 -0.0621424068009711
24.3 0.0595379910653287
23.7 0.0595379910653287
23.7 -0.0621424068009711
};
\addplot [semithick, color3]
table {%
24 -0.0621424068009711
24 -0.242759040365568
};
\addplot [semithick, color3]
table {%
24 0.0595379910653287
24 0.241292103757387
};
\addplot [semithick, color3]
table {%
23.85 -0.242759040365568
24.15 -0.242759040365568
};
\addplot [semithick, color3]
table {%
23.85 0.241292103757387
24.15 0.241292103757387
};
\addplot [semithick, color3]
table {%
26.7 -0.0642780266335657
27.3 -0.0642780266335657
27.3 0.0594837688617406
26.7 0.0594837688617406
26.7 -0.0642780266335657
};
\addplot [semithick, color3]
table {%
27 -0.0642780266335657
27 -0.249378440395176
};
\addplot [semithick, color3]
table {%
27 0.0594837688617406
27 0.243363213399898
};
\addplot [semithick, color3]
table {%
26.85 -0.249378440395176
27.15 -0.249378440395176
};
\addplot [semithick, color3]
table {%
26.85 0.243363213399898
27.15 0.243363213399898
};
\addplot [semithick, color3]
table {%
29.7 -0.0795837185458637
30.3 -0.0795837185458637
30.3 0.0842026470857949
29.7 0.0842026470857949
29.7 -0.0795837185458637
};
\addplot [semithick, color3]
table {%
30 -0.0795837185458637
30 -0.324935240269402
};
\addplot [semithick, color3]
table {%
30 0.0842026470857949
30 0.329199577208431
};
\addplot [semithick, color3]
table {%
29.85 -0.324935240269402
30.15 -0.324935240269402
};
\addplot [semithick, color3]
table {%
29.85 0.329199577208431
30.15 0.329199577208431
};
\addplot [semithick, color3]
table {%
32.7 -0.0597334865701577
33.3 -0.0597334865701577
33.3 0.0719960502865678
32.7 0.0719960502865678
32.7 -0.0597334865701577
};
\addplot [semithick, color3]
table {%
33 -0.0597334865701577
33 -0.257192500606071
};
\addplot [semithick, color3]
table {%
33 0.0719960502865678
33 0.269072804106087
};
\addplot [semithick, color3]
table {%
32.85 -0.257192500606071
33.15 -0.257192500606071
};
\addplot [semithick, color3]
table {%
32.85 0.269072804106087
33.15 0.269072804106087
};
\addplot [semithick, color3]
table {%
35.7 -0.0538396016106416
36.3 -0.0538396016106416
36.3 0.0654813099692614
35.7 0.0654813099692614
35.7 -0.0538396016106416
};
\addplot [semithick, color3]
table {%
36 -0.0538396016106416
36 -0.231945303543938
};
\addplot [semithick, color3]
table {%
36 0.0654813099692614
36 0.244031197676409
};
\addplot [semithick, color3]
table {%
35.85 -0.231945303543938
36.15 -0.231945303543938
};
\addplot [semithick, color3]
table {%
35.85 0.244031197676409
36.15 0.244031197676409
};
\addplot [semithick, color3]
table {%
38.7 -0.0533131807017401
39.3 -0.0533131807017401
39.3 0.0579338669378238
38.7 0.0579338669378238
38.7 -0.0533131807017401
};
\addplot [semithick, color3]
table {%
39 -0.0533131807017401
39 -0.220179030693249
};
\addplot [semithick, color3]
table {%
39 0.0579338669378238
39 0.224103430103831
};
\addplot [semithick, color3]
table {%
38.85 -0.220179030693249
39.15 -0.220179030693249
};
\addplot [semithick, color3]
table {%
38.85 0.224103430103831
39.15 0.224103430103831
};
\addplot [semithick, color3]
table {%
41.7 -0.0571198235958678
42.3 -0.0571198235958678
42.3 0.0649971478422028
41.7 0.0649971478422028
41.7 -0.0571198235958678
};
\addplot [semithick, color3]
table {%
42 -0.0571198235958678
42 -0.237269037387078
};
\addplot [semithick, color3]
table {%
42 0.0649971478422028
42 0.246270871440434
};
\addplot [semithick, color3]
table {%
41.85 -0.237269037387078
42.15 -0.237269037387078
};
\addplot [semithick, color3]
table {%
41.85 0.246270871440434
42.15 0.246270871440434
};
\addplot [semithick, color3]
table {%
44.7 -0.0611487143308532
45.3 -0.0611487143308532
45.3 0.0603836221811144
44.7 0.0603836221811144
44.7 -0.0611487143308532
};
\addplot [semithick, color3]
table {%
45 -0.0611487143308532
45 -0.243037917232454
};
\addplot [semithick, color3]
table {%
45 0.0603836221811144
45 0.240461865788729
};
\addplot [semithick, color3]
table {%
44.85 -0.243037917232454
45.15 -0.243037917232454
};
\addplot [semithick, color3]
table {%
44.85 0.240461865788729
45.15 0.240461865788729
};
\addplot [semithick, color3]
table {%
47.7 -0.0424955616924407
48.3 -0.0424955616924407
48.3 0.0512162319117103
47.7 0.0512162319117103
47.7 -0.0424955616924407
};
\addplot [semithick, color3]
table {%
48 -0.0424955616924407
48 -0.181809813201977
};
\addplot [semithick, color3]
table {%
48 0.0512162319117103
48 0.191661174973695
};
\addplot [semithick, color3]
table {%
47.85 -0.181809813201977
48.15 -0.181809813201977
};
\addplot [semithick, color3]
table {%
47.85 0.191661174973695
48.15 0.191661174973695
};
\addplot [semithick, color3]
table {%
50.7 -0.0424465847630331
51.3 -0.0424465847630331
51.3 0.0478268933718616
50.7 0.0478268933718616
50.7 -0.0424465847630331
};
\addplot [semithick, color3]
table {%
51 -0.0424465847630331
51 -0.177769194661308
};
\addplot [semithick, color3]
table {%
51 0.0478268933718616
51 0.182788443286279
};
\addplot [semithick, color3]
table {%
50.85 -0.177769194661308
51.15 -0.177769194661308
};
\addplot [semithick, color3]
table {%
50.85 0.182788443286279
51.15 0.182788443286279
};
\addplot [semithick, color3]
table {%
53.7 -0.0407217549617389
54.3 -0.0407217549617389
54.3 0.0488363081739351
53.7 0.0488363081739351
53.7 -0.0407217549617389
};
\addplot [semithick, color3]
table {%
54 -0.0407217549617389
54 -0.174142086370912
};
\addplot [semithick, color3]
table {%
54 0.0488363081739351
54 0.183134177721119
};
\addplot [semithick, color3]
table {%
53.85 -0.174142086370912
54.15 -0.174142086370912
};
\addplot [semithick, color3]
table {%
53.85 0.183134177721119
54.15 0.183134177721119
};
\addplot [semithick, color3]
table {%
56.7 -0.0291542135173569
57.3 -0.0291542135173569
57.3 0.0394952700581734
56.7 0.0394952700581734
56.7 -0.0291542135173569
};
\addplot [semithick, color3]
table {%
57 -0.0291542135173569
57 -0.131630716845605
};
\addplot [semithick, color3]
table {%
57 0.0394952700581734
57 0.140891773264484
};
\addplot [semithick, color3]
table {%
56.85 -0.131630716845605
57.15 -0.131630716845605
};
\addplot [semithick, color3]
table {%
56.85 0.140891773264484
57.15 0.140891773264484
};
\addplot [semithick, color6]
table {%
0.5 -0.0821061622447092
1.1 -0.0821061622447092
1.1 0.0700381256508618
0.5 0.0700381256508618
0.5 -0.0821061622447092
};
\addplot [semithick, color6]
table {%
0.8 -0.0821061622447092
0.8 -0.309524523439043
};
\addplot [semithick, color6]
table {%
0.8 0.0700381256508618
0.8 0.298118520354977
};
\addplot [semithick, color6]
table {%
0.65 -0.309524523439043
0.95 -0.309524523439043
};
\addplot [semithick, color6]
table {%
0.65 0.298118520354977
0.95 0.298118520354977
};
\addplot [semithick, color6]
table {%
3.5 -0.0584517689649558
4.1 -0.0584517689649558
4.1 0.0561943641768426
3.5 0.0561943641768426
3.5 -0.0584517689649558
};
\addplot [semithick, color6]
table {%
3.8 -0.0584517689649558
3.8 -0.230022414921267
};
\addplot [semithick, color6]
table {%
3.8 0.0561943641768426
3.8 0.228073642109188
};
\addplot [semithick, color6]
table {%
3.65 -0.230022414921267
3.95 -0.230022414921267
};
\addplot [semithick, color6]
table {%
3.65 0.228073642109188
3.95 0.228073642109188
};
\addplot [semithick, color6]
table {%
6.5 -0.0588593023735311
7.1 -0.0588593023735311
7.1 0.0543802714418838
6.5 0.0543802714418838
6.5 -0.0588593023735311
};
\addplot [semithick, color6]
table {%
6.8 -0.0588593023735311
6.8 -0.228175295501901
};
\addplot [semithick, color6]
table {%
6.8 0.0543802714418838
6.8 0.224228084982756
};
\addplot [semithick, color6]
table {%
6.65 -0.228175295501901
6.95 -0.228175295501901
};
\addplot [semithick, color6]
table {%
6.65 0.224228084982756
6.95 0.224228084982756
};
\addplot [semithick, color6]
table {%
9.5 -0.0418782264768914
10.1 -0.0418782264768914
10.1 0.03850830637587
9.5 0.03850830637587
9.5 -0.0418782264768914
};
\addplot [semithick, color6]
table {%
9.8 -0.0418782264768914
9.8 -0.162237387998247
};
\addplot [semithick, color6]
table {%
9.8 0.03850830637587
9.8 0.158859890217015
};
\addplot [semithick, color6]
table {%
9.65 -0.162237387998247
9.95 -0.162237387998247
};
\addplot [semithick, color6]
table {%
9.65 0.158859890217015
9.95 0.158859890217015
};
\addplot [semithick, color6]
table {%
12.5 -0.0587850082800096
13.1 -0.0587850082800096
13.1 0.0492516100864408
12.5 0.0492516100864408
12.5 -0.0587850082800096
};
\addplot [semithick, color6]
table {%
12.8 -0.0587850082800096
12.8 -0.220305836072063
};
\addplot [semithick, color6]
table {%
12.8 0.0492516100864408
12.8 0.210929361048869
};
\addplot [semithick, color6]
table {%
12.65 -0.220305836072063
12.95 -0.220305836072063
};
\addplot [semithick, color6]
table {%
12.65 0.210929361048869
12.95 0.210929361048869
};
\addplot [semithick, color6]
table {%
15.5 -0.0849928041276407
16.1 -0.0849928041276407
16.1 0.0727123531661402
15.5 0.0727123531661402
15.5 -0.0849928041276407
};
\addplot [semithick, color6]
table {%
15.8 -0.0849928041276407
15.8 -0.321030522758703
};
\addplot [semithick, color6]
table {%
15.8 0.0727123531661402
15.8 0.308156113418959
};
\addplot [semithick, color6]
table {%
15.65 -0.321030522758703
15.95 -0.321030522758703
};
\addplot [semithick, color6]
table {%
15.65 0.308156113418959
15.95 0.308156113418959
};
\addplot [semithick, color6]
table {%
18.5 -0.0683710967947519
19.1 -0.0683710967947519
19.1 0.056204309386298
18.5 0.056204309386298
18.5 -0.0683710967947519
};
\addplot [semithick, color6]
table {%
18.8 -0.0683710967947519
18.8 -0.254758670874398
};
\addplot [semithick, color6]
table {%
18.8 0.056204309386298
18.8 0.242358538858667
};
\addplot [semithick, color6]
table {%
18.65 -0.254758670874398
18.95 -0.254758670874398
};
\addplot [semithick, color6]
table {%
18.65 0.242358538858667
18.95 0.242358538858667
};
\addplot [semithick, color6]
table {%
21.5 -0.0719093009944017
22.1 -0.0719093009944017
22.1 0.0611249684781043
21.5 0.0611249684781043
21.5 -0.0719093009944017
};
\addplot [semithick, color6]
table {%
21.8 -0.0719093009944017
21.8 -0.268480848651119
};
\addplot [semithick, color6]
table {%
21.8 0.0611249684781043
21.8 0.260185638298264
};
\addplot [semithick, color6]
table {%
21.65 -0.268480848651119
21.95 -0.268480848651119
};
\addplot [semithick, color6]
table {%
21.65 0.260185638298264
21.95 0.260185638298264
};
\addplot [semithick, color6]
table {%
24.5 -0.0680257373650826
25.1 -0.0680257373650826
25.1 0.0568687103700731
24.5 0.0568687103700731
24.5 -0.0680257373650826
};
\addplot [semithick, color6]
table {%
24.8 -0.0680257373650826
24.8 -0.252605319212778
};
\addplot [semithick, color6]
table {%
24.8 0.0568687103700731
24.8 0.242312070781297
};
\addplot [semithick, color6]
table {%
24.65 -0.252605319212778
24.95 -0.252605319212778
};
\addplot [semithick, color6]
table {%
24.65 0.242312070781297
24.95 0.242312070781297
};
\addplot [semithick, color6]
table {%
27.5 -0.0652862390379675
28.1 -0.0652862390379675
28.1 0.0583286933372106
27.5 0.0583286933372106
27.5 -0.0652862390379675
};
\addplot [semithick, color6]
table {%
27.8 -0.0652862390379675
27.8 -0.250073295160828
};
\addplot [semithick, color6]
table {%
27.8 0.0583286933372106
27.8 0.243297235131617
};
\addplot [semithick, color6]
table {%
27.65 -0.250073295160828
27.95 -0.250073295160828
};
\addplot [semithick, color6]
table {%
27.65 0.243297235131617
27.95 0.243297235131617
};
\addplot [semithick, color6]
table {%
30.5 -0.0924296056125181
31.1 -0.0924296056125181
31.1 0.072790325349725
30.5 0.072790325349725
30.5 -0.0924296056125181
};
\addplot [semithick, color6]
table {%
30.8 -0.0924296056125181
30.8 -0.339457209047621
};
\addplot [semithick, color6]
table {%
30.8 0.072790325349725
30.8 0.316833555027471
};
\addplot [semithick, color6]
table {%
30.65 -0.339457209047621
30.95 -0.339457209047621
};
\addplot [semithick, color6]
table {%
30.65 0.316833555027471
30.95 0.316833555027471
};
\addplot [semithick, color6]
table {%
33.5 -0.065141769240703
34.1 -0.065141769240703
34.1 0.0637814985313738
33.5 0.0637814985313738
33.5 -0.065141769240703
};
\addplot [semithick, color6]
table {%
33.8 -0.065141769240703
33.8 -0.256784548329364
};
\addplot [semithick, color6]
table {%
33.8 0.0637814985313738
33.8 0.257126592021225
};
\addplot [semithick, color6]
table {%
33.65 -0.256784548329364
33.95 -0.256784548329364
};
\addplot [semithick, color6]
table {%
33.65 0.257126592021225
33.95 0.257126592021225
};
\addplot [semithick, color6]
table {%
36.5 -0.0613316535825392
37.1 -0.0613316535825392
37.1 0.0584509861843691
36.5 0.0584509861843691
36.5 -0.0613316535825392
};
\addplot [semithick, color6]
table {%
36.8 -0.0613316535825392
36.8 -0.240108384154971
};
\addplot [semithick, color6]
table {%
36.8 0.0584509861843691
36.8 0.236370484602809
};
\addplot [semithick, color6]
table {%
36.65 -0.240108384154971
36.95 -0.240108384154971
};
\addplot [semithick, color6]
table {%
36.65 0.236370484602809
36.95 0.236370484602809
};
\addplot [semithick, color6]
table {%
39.5 -0.0566177048361281
40.1 -0.0566177048361281
40.1 0.0547603273894162
39.5 0.0547603273894162
39.5 -0.0566177048361281
};
\addplot [semithick, color6]
table {%
39.8 -0.0566177048361281
39.8 -0.223383880109872
};
\addplot [semithick, color6]
table {%
39.8 0.0547603273894162
39.8 0.220875782961472
};
\addplot [semithick, color6]
table {%
39.65 -0.223383880109872
39.95 -0.223383880109872
};
\addplot [semithick, color6]
table {%
39.65 0.220875782961472
39.95 0.220875782961472
};
\addplot [semithick, color6]
table {%
42.5 -0.0611992216720572
43.1 -0.0611992216720572
43.1 0.0627043981772941
42.5 0.0627043981772941
42.5 -0.0611992216720572
};
\addplot [semithick, color6]
table {%
42.8 -0.0611992216720572
42.8 -0.242655243370165
};
\addplot [semithick, color6]
table {%
42.8 0.0627043981772941
42.8 0.247987556187362
};
\addplot [semithick, color6]
table {%
42.65 -0.242655243370165
42.95 -0.242655243370165
};
\addplot [semithick, color6]
table {%
42.65 0.247987556187362
42.95 0.247987556187362
};
\addplot [semithick, color6]
table {%
45.5 -0.0669851932684794
46.1 -0.0669851932684794
46.1 0.0542095018743899
45.5 0.0542095018743899
45.5 -0.0669851932684794
};
\addplot [semithick, color6]
table {%
45.8 -0.0669851932684794
45.8 -0.248442687203589
};
\addplot [semithick, color6]
table {%
45.8 0.0542095018743899
45.8 0.235864747512339
};
\addplot [semithick, color6]
table {%
45.65 -0.248442687203589
45.95 -0.248442687203589
};
\addplot [semithick, color6]
table {%
45.65 0.235864747512339
45.95 0.235864747512339
};
\addplot [semithick, color6]
table {%
48.5 -0.0465583437078482
49.1 -0.0465583437078482
49.1 0.044036967494251
48.5 0.044036967494251
48.5 -0.0465583437078482
};
\addplot [semithick, color6]
table {%
48.8 -0.0465583437078482
48.8 -0.180801373429123
};
\addplot [semithick, color6]
table {%
48.8 0.044036967494251
48.8 0.179244423130935
};
\addplot [semithick, color6]
table {%
48.65 -0.180801373429123
48.95 -0.180801373429123
};
\addplot [semithick, color6]
table {%
48.65 0.179244423130935
48.95 0.179244423130935
};
\addplot [semithick, color6]
table {%
51.5 -0.0448038843187301
52.1 -0.0448038843187301
52.1 0.0450837320954394
51.5 0.0450837320954394
51.5 -0.0448038843187301
};
\addplot [semithick, color6]
table {%
51.8 -0.0448038843187301
51.8 -0.179313891615717
};
\addplot [semithick, color6]
table {%
51.8 0.0450837320954394
51.8 0.17953718953366
};
\addplot [semithick, color6]
table {%
51.65 -0.179313891615717
51.95 -0.179313891615717
};
\addplot [semithick, color6]
table {%
51.65 0.17953718953366
51.95 0.17953718953366
};
\addplot [semithick, color6]
table {%
54.5 -0.0453841210476313
55.1 -0.0453841210476313
55.1 0.0429523806549188
54.5 0.0429523806549188
54.5 -0.0453841210476313
};
\addplot [semithick, color6]
table {%
54.8 -0.0453841210476313
54.8 -0.177707386004758
};
\addplot [semithick, color6]
table {%
54.8 0.0429523806549188
54.8 0.174958445660282
};
\addplot [semithick, color6]
table {%
54.65 -0.177707386004758
54.95 -0.177707386004758
};
\addplot [semithick, color6]
table {%
54.65 0.174958445660282
54.95 0.174958445660282
};
\addplot [semithick, color6]
table {%
57.5 -0.030794765855216
58.1 -0.030794765855216
58.1 0.0369392222605667
57.5 0.0369392222605667
57.5 -0.030794765855216
};
\addplot [semithick, color6]
table {%
57.8 -0.030794765855216
57.8 -0.13236531815099
};
\addplot [semithick, color6]
table {%
57.8 0.0369392222605667
57.8 0.137149394320923
};
\addplot [semithick, color6]
table {%
57.65 -0.13236531815099
57.95 -0.13236531815099
};
\addplot [semithick, color6]
table {%
57.65 0.137149394320923
57.95 0.137149394320923
};
\addplot [semithick, color4]
table {%
-1.1 -0.0142873002158352
-0.5 -0.0142873002158352
};
\addplot [semithick, color4]
table {%
1.9 0.00938642350704774
2.5 0.00938642350704774
};
\addplot [semithick, color4]
table {%
4.9 -0.0115641103824025
5.5 -0.0115641103824025
};
\addplot [semithick, color4]
table {%
7.9 -0.0519315950570356
8.5 -0.0519315950570356
};
\addplot [semithick, color4]
table {%
10.9 -0.0230376767682013
11.5 -0.0230376767682013
};
\addplot [semithick, color4]
table {%
13.9 0.00778669880437831
14.5 0.00778669880437831
};
\addplot [semithick, color4]
table {%
16.9 0.00504080098521097
17.5 0.00504080098521097
};
\addplot [semithick, color4]
table {%
19.9 0.0102153597877849
20.5 0.0102153597877849
};
\addplot [semithick, color4]
table {%
22.9 0.0114578068318258
23.5 0.0114578068318258
};
\addplot [semithick, color4]
table {%
25.9 0.00688311363990286
26.5 0.00688311363990286
};
\addplot [semithick, color4]
table {%
28.9 0.00376388083148998
29.5 0.00376388083148998
};
\addplot [semithick, color4]
table {%
31.9 0.00745183696073442
32.5 0.00745183696073442
};
\addplot [semithick, color4]
table {%
34.9 0.00688584260990843
35.5 0.00688584260990843
};
\addplot [semithick, color4]
table {%
37.9 0.00807708260589902
38.5 0.00807708260589902
};
\addplot [semithick, color4]
table {%
40.9 0.0027453449428099
41.5 0.0027453449428099
};
\addplot [semithick, color4]
table {%
43.9 0.00839445526625342
44.5 0.00839445526625342
};
\addplot [semithick, color4]
table {%
46.9 0.0121300901530485
47.5 0.0121300901530485
};
\addplot [semithick, color4]
table {%
49.9 0.0114068870077418
50.5 0.0114068870077418
};
\addplot [semithick, color4]
table {%
52.9 0.0123640559087635
53.5 0.0123640559087635
};
\addplot [semithick, color4]
table {%
55.9 0.0131205496448035
56.5 0.0131205496448035
};
\addplot [semithick, color3]
table {%
-0.3 -0.00024017502795734
0.3 -0.00024017502795734
};
\addplot [semithick, color3]
table {%
2.7 0.0122090835089732
3.3 0.0122090835089732
};
\addplot [semithick, color3]
table {%
5.7 0.00211069576090878
6.3 0.00211069576090878
};
\addplot [semithick, color3]
table {%
8.7 -0.0280894578828796
9.3 -0.0280894578828796
};
\addplot [semithick, color3]
table {%
11.7 -0.0081656251955201
12.3 -0.0081656251955201
};
\addplot [semithick, color3]
table {%
14.7 -0.00284791312493248
15.3 -0.00284791312493248
};
\addplot [semithick, color3]
table {%
17.7 -0.00299764253174255
18.3 -0.00299764253174255
};
\addplot [semithick, color3]
table {%
20.7 -0.00231897489190204
21.3 -0.00231897489190204
};
\addplot [semithick, color3]
table {%
23.7 -0.000388361534623089
24.3 -0.000388361534623089
};
\addplot [semithick, color3]
table {%
26.7 -0.00280383305536334
27.3 -0.00280383305536334
};
\addplot [semithick, color3]
table {%
29.7 0.00165249723260374
30.3 0.00165249723260374
};
\addplot [semithick, color3]
table {%
32.7 0.00687154925693488
33.3 0.00687154925693488
};
\addplot [semithick, color3]
table {%
35.7 0.00429864574707472
36.3 0.00429864574707472
};
\addplot [semithick, color3]
table {%
38.7 0.00555874591989512
39.3 0.00555874591989512
};
\addplot [semithick, color3]
table {%
41.7 0.00192187831466818
42.3 0.00192187831466818
};
\addplot [semithick, color3]
table {%
44.7 -0.00107436349672275
45.3 -0.00107436349672275
};
\addplot [semithick, color3]
table {%
47.7 0.00468179243340577
48.3 0.00468179243340577
};
\addplot [semithick, color3]
table {%
50.7 0.00384157270804658
51.3 0.00384157270804658
};
\addplot [semithick, color3]
table {%
53.7 0.00448430779887661
54.3 0.00448430779887661
};
\addplot [semithick, color3]
table {%
56.7 0.00522361639624535
57.3 0.00522361639624535
};
\addplot [semithick, color6]
table {%
0.5 -0.00421410924876063
1.1 -0.00421410924876063
};
\addplot [semithick, color6]
table {%
3.5 0.00117238715412286
4.1 0.00117238715412286
};
\addplot [semithick, color6]
table {%
6.5 0.00019879712579797
7.1 0.00019879712579797
};
\addplot [semithick, color6]
table {%
9.5 -0.00631437688101056
10.1 -0.00631437688101056
};
\addplot [semithick, color6]
table {%
12.5 -0.00867509925574929
13.1 -0.00867509925574929
};
\addplot [semithick, color6]
table {%
15.5 -0.0048118443293951
16.1 -0.0048118443293951
};
\addplot [semithick, color6]
table {%
18.5 -0.00786744373044118
19.1 -0.00786744373044118
};
\addplot [semithick, color6]
table {%
21.5 -0.00598761965184389
22.1 -0.00598761965184389
};
\addplot [semithick, color6]
table {%
24.5 -0.00448700883541377
25.1 -0.00448700883541377
};
\addplot [semithick, color6]
table {%
27.5 -0.00393023081635323
28.1 -0.00393023081635323
};
\addplot [semithick, color6]
table {%
30.5 -0.0115592925124844
31.1 -0.0115592925124844
};
\addplot [semithick, color6]
table {%
33.5 -4.25028486125134e-05
34.1 -4.25028486125134e-05
};
\addplot [semithick, color6]
table {%
36.5 -0.00177184811666842
37.1 -0.00177184811666842
};
\addplot [semithick, color6]
table {%
39.5 0.00347418090545546
40.1 0.00347418090545546
};
\addplot [semithick, color6]
table {%
42.5 -0.00284473470518297
43.1 -0.00284473470518297
};
\addplot [semithick, color6]
table {%
45.5 -0.00598135310862783
46.1 -0.00598135310862783
};
\addplot [semithick, color6]
table {%
48.5 -0.00247412795086881
49.1 -0.00247412795086881
};
\addplot [semithick, color6]
table {%
51.5 0.00117934367516545
52.1 0.00117934367516545
};
\addplot [semithick, color6]
table {%
54.5 -0.000876836139797066
55.1 -0.000876836139797066
};
\addplot [semithick, color6]
table {%
57.5 0.00302634042974051
58.1 0.00302634042974051
};
\end{axis}

\begin{axis}[
width=\sfwidth,
height=\sfheight,
tick align=outside,
tick pos=right,
xlabel={Video content},
xmin=-3, xmax=60,
xtick={6,21,36,51},
xticklabels={Minecraft,Cities,Tour,Virus popper},
extra x ticks = {13.5, 28.5, 43.5},
extra x tick labels = {},
extra tick style={grid=major, tick align=inside},
ymin=-30, ymax=30,
ymajorticks = false,
]
\end{axis}

\end{tikzpicture}

%% file: img/gen_boxplot_Quantile_6.tex
\begin{tikzpicture}

\begin{axis}[
width=\sfwidth,
height=\sfheight,
legend cell align={left},
legend style={
  fill opacity=0.8,
  draw opacity=1,
  text opacity=1,
  at={(0.03,0.97)},
  anchor=north west,
  draw=white!80!black
},
tick align=outside,
tick pos=left,
ylabel={Relative error},
xlabel={Rate (Mb/s)},
x grid style={white!69.0196078431373!black},
xmin=-3, xmax=60,
xtick={0,3,6,9,12,15,18,21,24,27,30,33,36,39,42,45,48,51,54,57},
xticklabels={
  10,
  20,
  30,
  40,
  50,
  10,
  20,
  30,
  40,
  50,
  10,
  20,
  30,
  40,
  50,
  10,
  20,
  30,
  40,
  50
},
y grid style={white!69.0196078431373!black},
ymin=-1, ymax=0.5,
ytick style={color=black},
ymajorgrids,
ytick={-1,-0.75,-0.5,-0.25,0,0.25,0.5,0.75,1}
]

\addplot [thick, black, forget plot]
table {%
-3 0
-2 0
-1 0
0 0
1 0
2 0
3 0
4 0
5 0
6 0
7 0
8 0
9 0
10 0
11 0
12 0
13 0
14 0
15 0
16 0
17 0
18 0
19 0
20 0
21 0
22 0
23 0
24 0
25 0
26 0
27 0
28 0
29 0
30 0
31 0
32 0
33 0
34 0
35 0
36 0
37 0
38 0
39 0
40 0
41 0
42 0
43 0
44 0
45 0
46 0
47 0
48 0
49 0
50 0
51 0
52 0
53 0
54 0
55 0
56 0
57 0
58 0
59 0
60 0
61 0
62 0
};
\addplot [semithick, color2]
table {%
-1.1 -0.325638114576292
-0.5 -0.325638114576292
-0.5 -0.136797971782813
-1.1 -0.136797971782813
-1.1 -0.325638114576292
};
\addplot [semithick, color2]
table {%
-0.8 -0.325638114576292
-0.8 -0.599576059278866
};
\addplot [semithick, color2]
table {%
-0.8 -0.136797971782813
-0.8 0.145478490103831
};
\addplot [semithick, color2]
table {%
-0.95 -0.599576059278866
-0.65 -0.599576059278866
};
\addplot [semithick, color2]
table {%
-0.95 0.145478490103831
-0.65 0.145478490103831
};
\addplot [semithick, color2]
table {%
1.9 -0.24656909925693
2.5 -0.24656909925693
2.5 -0.127627415142312
1.9 -0.127627415142312
1.9 -0.24656909925693
};
\addplot [semithick, color2]
table {%
2.2 -0.24656909925693
2.2 -0.424245035821436
};
\addplot [semithick, color2]
table {%
2.2 -0.127627415142312
2.2 0.0505771344709839
};
\addplot [semithick, color2]
table {%
2.05 -0.424245035821436
2.35 -0.424245035821436
};
\addplot [semithick, color2]
table {%
2.05 0.0505771344709839
2.35 0.0505771344709839
};
\addplot [semithick, color2]
table {%
4.9 -0.295503949302401
5.5 -0.295503949302401
5.5 -0.162095074681597
4.9 -0.162095074681597
4.9 -0.295503949302401
};
\addplot [semithick, color2]
table {%
5.2 -0.295503949302401
5.2 -0.495197047450341
};
\addplot [semithick, color2]
table {%
5.2 -0.162095074681597
5.2 0.0375377715654083
};
\addplot [semithick, color2]
table {%
5.05 -0.495197047450341
5.35 -0.495197047450341
};
\addplot [semithick, color2]
table {%
5.05 0.0375377715654083
5.35 0.0375377715654083
};
\addplot [semithick, color2]
table {%
7.9 -0.48840556293178
8.5 -0.48840556293178
8.5 -0.15827702699204
7.9 -0.15827702699204
7.9 -0.48840556293178
};
\addplot [semithick, color2]
table {%
8.2 -0.48840556293178
8.2 -0.81263875239479
};
\addplot [semithick, color2]
table {%
8.2 -0.15827702699204
8.2 0.300340385572335
};
\addplot [semithick, color2]
table {%
8.05 -0.81263875239479
8.35 -0.81263875239479
};
\addplot [semithick, color2]
table {%
8.05 0.300340385572335
8.35 0.300340385572335
};
\addplot [semithick, color2]
table {%
10.9 -0.345648915114439
11.5 -0.345648915114439
11.5 -0.155431912727166
10.9 -0.155431912727166
10.9 -0.345648915114439
};
\addplot [semithick, color2]
table {%
11.2 -0.345648915114439
11.2 -0.627863520967304
};
\addplot [semithick, color2]
table {%
11.2 -0.155431912727166
11.2 0.128686325910877
};
\addplot [semithick, color2]
table {%
11.05 -0.627863520967304
11.35 -0.627863520967304
};
\addplot [semithick, color2]
table {%
11.05 0.128686325910877
11.35 0.128686325910877
};
\addplot [semithick, color2]
table {%
13.9 -0.271834792612754
14.5 -0.271834792612754
14.5 -0.100925279558519
13.9 -0.100925279558519
13.9 -0.271834792612754
};
\addplot [semithick, color2]
table {%
14.2 -0.271834792612754
14.2 -0.52660848791714
};
\addplot [semithick, color2]
table {%
14.2 -0.100925279558519
14.2 0.151570831477697
};
\addplot [semithick, color2]
table {%
14.05 -0.52660848791714
14.35 -0.52660848791714
};
\addplot [semithick, color2]
table {%
14.05 0.151570831477697
14.35 0.151570831477697
};
\addplot [semithick, color2]
table {%
16.9 -0.250174604834462
17.5 -0.250174604834462
17.5 -0.123146789106301
16.9 -0.123146789106301
16.9 -0.250174604834462
};
\addplot [semithick, color2]
table {%
17.2 -0.250174604834462
17.2 -0.439699860423024
};
\addplot [semithick, color2]
table {%
17.2 -0.123146789106301
17.2 0.0673267960151388
};
\addplot [semithick, color2]
table {%
17.05 -0.439699860423024
17.35 -0.439699860423024
};
\addplot [semithick, color2]
table {%
17.05 0.0673267960151388
17.35 0.0673267960151388
};
\addplot [semithick, color2]
table {%
19.9 -0.250356769596106
20.5 -0.250356769596106
20.5 -0.116524150835651
19.9 -0.116524150835651
19.9 -0.250356769596106
};
\addplot [semithick, color2]
table {%
20.2 -0.250356769596106
20.2 -0.450390725997575
};
\addplot [semithick, color2]
table {%
20.2 -0.116524150835651
20.2 0.0829909423126952
};
\addplot [semithick, color2]
table {%
20.05 -0.450390725997575
20.35 -0.450390725997575
};
\addplot [semithick, color2]
table {%
20.05 0.0829909423126952
20.35 0.0829909423126952
};
\addplot [semithick, color2]
table {%
22.9 -0.244248640088655
23.5 -0.244248640088655
23.5 -0.124463406162786
22.9 -0.124463406162786
22.9 -0.244248640088655
};
\addplot [semithick, color2]
table {%
23.2 -0.244248640088655
23.2 -0.42224243112511
};
\addplot [semithick, color2]
table {%
23.2 -0.124463406162786
23.2 0.0536777310595231
};
\addplot [semithick, color2]
table {%
23.05 -0.42224243112511
23.35 -0.42224243112511
};
\addplot [semithick, color2]
table {%
23.05 0.0536777310595231
23.35 0.0536777310595231
};
\addplot [semithick, color2]
table {%
25.9 -0.246733541165283
26.5 -0.246733541165283
26.5 -0.124527072393889
25.9 -0.124527072393889
25.9 -0.246733541165283
};
\addplot [semithick, color2]
table {%
26.2 -0.246733541165283
26.2 -0.426019781581095
};
\addplot [semithick, color2]
table {%
26.2 -0.124527072393889
26.2 0.0555082478795662
};
\addplot [semithick, color2]
table {%
26.05 -0.426019781581095
26.35 -0.426019781581095
};
\addplot [semithick, color2]
table {%
26.05 0.0555082478795662
26.35 0.0555082478795662
};
\addplot [semithick, color2]
table {%
28.9 -0.271899540909996
29.5 -0.271899540909996
29.5 -0.100880892904038
28.9 -0.100880892904038
28.9 -0.271899540909996
};
\addplot [semithick, color2]
table {%
29.2 -0.271899540909996
29.2 -0.527663243123342
};
\addplot [semithick, color2]
table {%
29.2 -0.100880892904038
29.2 0.154927824466352
};
\addplot [semithick, color2]
table {%
29.05 -0.527663243123342
29.35 -0.527663243123342
};
\addplot [semithick, color2]
table {%
29.05 0.154927824466352
29.35 0.154927824466352
};
\addplot [semithick, color2]
table {%
31.9 -0.257806093878738
32.5 -0.257806093878738
32.5 -0.11631665303787
31.9 -0.11631665303787
31.9 -0.257806093878738
};
\addplot [semithick, color2]
table {%
32.2 -0.257806093878738
32.2 -0.469633219657832
};
\addplot [semithick, color2]
table {%
32.2 -0.11631665303787
32.2 0.0956977417281393
};
\addplot [semithick, color2]
table {%
32.05 -0.469633219657832
32.35 -0.469633219657832
};
\addplot [semithick, color2]
table {%
32.05 0.0956977417281393
32.35 0.0956977417281393
};
\addplot [semithick, color2]
table {%
34.9 -0.246992453996401
35.5 -0.246992453996401
35.5 -0.125693764021739
34.9 -0.125693764021739
34.9 -0.246992453996401
};
\addplot [semithick, color2]
table {%
35.2 -0.246992453996401
35.2 -0.428627240288078
};
\addplot [semithick, color2]
table {%
35.2 -0.125693764021739
35.2 0.0552204322317401
};
\addplot [semithick, color2]
table {%
35.05 -0.428627240288078
35.35 -0.428627240288078
};
\addplot [semithick, color2]
table {%
35.05 0.0552204322317401
35.35 0.0552204322317401
};
\addplot [semithick, color2]
table {%
37.9 -0.24774729492678
38.5 -0.24774729492678
38.5 -0.132450785313067
37.9 -0.132450785313067
37.9 -0.24774729492678
};
\addplot [semithick, color2]
table {%
38.2 -0.24774729492678
38.2 -0.419760125709509
};
\addplot [semithick, color2]
table {%
38.2 -0.132450785313067
38.2 0.0400869833328899
};
\addplot [semithick, color2]
table {%
38.05 -0.419760125709509
38.35 -0.419760125709509
};
\addplot [semithick, color2]
table {%
38.05 0.0400869833328899
38.35 0.0400869833328899
};
\addplot [semithick, color2]
table {%
40.9 -0.250363941998602
41.5 -0.250363941998602
41.5 -0.126647833663293
40.9 -0.126647833663293
40.9 -0.250363941998602
};
\addplot [semithick, color2]
table {%
41.2 -0.250363941998602
41.2 -0.434325788096223
};
\addplot [semithick, color2]
table {%
41.2 -0.126647833663293
41.2 0.057225947851902
};
\addplot [semithick, color2]
table {%
41.05 -0.434325788096223
41.35 -0.434325788096223
};
\addplot [semithick, color2]
table {%
41.05 0.057225947851902
41.35 0.057225947851902
};
\addplot [semithick, color2]
table {%
43.9 -0.247957883069373
44.5 -0.247957883069373
44.5 -0.125981201196995
43.9 -0.125981201196995
43.9 -0.247957883069373
};
\addplot [semithick, color2]
table {%
44.2 -0.247957883069373
44.2 -0.429766229019938
};
\addplot [semithick, color2]
table {%
44.2 -0.125981201196995
44.2 0.0568539370807829
};
\addplot [semithick, color2]
table {%
44.05 -0.429766229019938
44.35 -0.429766229019938
};
\addplot [semithick, color2]
table {%
44.05 0.0568539370807829
44.35 0.0568539370807829
};
\addplot [semithick, color2]
table {%
46.9 -0.229077395660198
47.5 -0.229077395660198
47.5 -0.134131184145368
46.9 -0.134131184145368
46.9 -0.229077395660198
};
\addplot [semithick, color2]
table {%
47.2 -0.229077395660198
47.2 -0.371289389343623
};
\addplot [semithick, color2]
table {%
47.2 -0.134131184145368
47.2 0.00636188589743975
};
\addplot [semithick, color2]
table {%
47.05 -0.371289389343623
47.35 -0.371289389343623
};
\addplot [semithick, color2]
table {%
47.05 0.00636188589743975
47.35 0.00636188589743975
};
\addplot [semithick, color2]
table {%
49.9 -0.228824193097514
50.5 -0.228824193097514
50.5 -0.137603078196331
49.9 -0.137603078196331
49.9 -0.228824193097514
};
\addplot [semithick, color2]
table {%
50.2 -0.228824193097514
50.2 -0.364482043867782
};
\addplot [semithick, color2]
table {%
50.2 -0.137603078196331
50.2 -0.00166502738748664
};
\addplot [semithick, color2]
table {%
50.05 -0.364482043867782
50.35 -0.364482043867782
};
\addplot [semithick, color2]
table {%
50.05 -0.00166502738748664
50.35 -0.00166502738748664
};
\addplot [semithick, color2]
table {%
52.9 -0.227178443511252
53.5 -0.227178443511252
53.5 -0.137249127521093
52.9 -0.137249127521093
52.9 -0.227178443511252
};
\addplot [semithick, color2]
table {%
53.2 -0.227178443511252
53.2 -0.361751525516296
};
\addplot [semithick, color2]
table {%
53.2 -0.137249127521093
53.2 -0.00313180683204589
};
\addplot [semithick, color2]
table {%
53.05 -0.361751525516296
53.35 -0.361751525516296
};
\addplot [semithick, color2]
table {%
53.05 -0.00313180683204589
53.35 -0.00313180683204589
};
\addplot [semithick, color2]
table {%
55.9 -0.21575527300198
56.5 -0.21575527300198
56.5 -0.145453620653959
55.9 -0.145453620653959
55.9 -0.21575527300198
};
\addplot [semithick, color2]
table {%
56.2 -0.21575527300198
56.2 -0.320440455769139
};
\addplot [semithick, color2]
table {%
56.2 -0.145453620653959
56.2 -0.041371034400087
};
\addplot [semithick, color2]
table {%
56.05 -0.320440455769139
56.35 -0.320440455769139
};
\addplot [semithick, color2]
table {%
56.05 -0.041371034400087
56.35 -0.041371034400087
};
\addplot [semithick, color4]
table {%
-0.3 -0.273707975537553
0.3 -0.273707975537553
0.3 -0.126134721362903
-0.3 -0.126134721362903
-0.3 -0.273707975537553
};
\addplot [semithick, color4]
table {%
0 -0.273707975537553
0 -0.491844693999194
};
\addplot [semithick, color4]
table {%
0 -0.126134721362903
0 0.0945188948189609
};
\addplot [semithick, color4]
table {%
-0.15 -0.491844693999194
0.15 -0.491844693999194
};
\addplot [semithick, color4]
table {%
-0.15 0.0945188948189609
0.15 0.0945188948189609
};
\addplot [semithick, color4]
table {%
2.7 -0.244652597464726
3.3 -0.244652597464726
3.3 -0.126722232404032
2.7 -0.126722232404032
2.7 -0.244652597464726
};
\addplot [semithick, color4]
table {%
3 -0.244652597464726
3 -0.421389449862583
};
\addplot [semithick, color4]
table {%
3 -0.126722232404032
3 0.0494666604877048
};
\addplot [semithick, color4]
table {%
2.85 -0.421389449862583
3.15 -0.421389449862583
};
\addplot [semithick, color4]
table {%
2.85 0.0494666604877048
3.15 0.0494666604877048
};
\addplot [semithick, color4]
table {%
5.7 -0.251422129211133
6.3 -0.251422129211133
6.3 -0.142271567010121
5.7 -0.142271567010121
5.7 -0.251422129211133
};
\addplot [semithick, color4]
table {%
6 -0.251422129211133
6 -0.41332420296923
};
\addplot [semithick, color4]
table {%
6 -0.142271567010121
6 0.0208309156529793
};
\addplot [semithick, color4]
table {%
5.85 -0.41332420296923
6.15 -0.41332420296923
};
\addplot [semithick, color4]
table {%
5.85 0.0208309156529793
6.15 0.0208309156529793
};
\addplot [semithick, color4]
table {%
8.7 -0.285133608229425
9.3 -0.285133608229425
9.3 -0.160225424296625
8.7 -0.160225424296625
8.7 -0.285133608229425
};
\addplot [semithick, color4]
table {%
9 -0.285133608229425
9 -0.469974109872307
};
\addplot [semithick, color4]
table {%
9 -0.160225424296625
9 0.0258150888990639
};
\addplot [semithick, color4]
table {%
8.85 -0.469974109872307
9.15 -0.469974109872307
};
\addplot [semithick, color4]
table {%
8.85 0.0258150888990639
9.15 0.0258150888990639
};
\addplot [semithick, color4]
table {%
11.7 -0.264782477609254
12.3 -0.264782477609254
12.3 -0.148766498316496
11.7 -0.148766498316496
11.7 -0.264782477609254
};
\addplot [semithick, color4]
table {%
12 -0.264782477609254
12 -0.437853721791462
};
\addplot [semithick, color4]
table {%
12 -0.148766498316496
12 0.0251819237294869
};
\addplot [semithick, color4]
table {%
11.85 -0.437853721791462
12.15 -0.437853721791462
};
\addplot [semithick, color4]
table {%
11.85 0.0251819237294869
12.15 0.0251819237294869
};
\addplot [semithick, color4]
table {%
14.7 -0.305170628435277
15.3 -0.305170628435277
15.3 -0.128741327267574
14.7 -0.128741327267574
14.7 -0.305170628435277
};
\addplot [semithick, color4]
table {%
15 -0.305170628435277
15 -0.563079212136931
};
\addplot [semithick, color4]
table {%
15 -0.128741327267574
15 0.135093168269151
};
\addplot [semithick, color4]
table {%
14.85 -0.563079212136931
15.15 -0.563079212136931
};
\addplot [semithick, color4]
table {%
14.85 0.135093168269151
15.15 0.135093168269151
};
\addplot [semithick, color4]
table {%
17.7 -0.291630012378965
18.3 -0.291630012378965
18.3 -0.147749612133192
17.7 -0.147749612133192
17.7 -0.291630012378965
};
\addplot [semithick, color4]
table {%
18 -0.291630012378965
18 -0.50570192562527
};
\addplot [semithick, color4]
table {%
18 -0.147749612133192
18 0.0680635823211928
};
\addplot [semithick, color4]
table {%
17.85 -0.50570192562527
18.15 -0.50570192562527
};
\addplot [semithick, color4]
table {%
17.85 0.0680635823211928
18.15 0.0680635823211928
};
\addplot [semithick, color4]
table {%
20.7 -0.288790336870522
21.3 -0.288790336870522
21.3 -0.141713400051741
20.7 -0.141713400051741
20.7 -0.288790336870522
};
\addplot [semithick, color4]
table {%
21 -0.288790336870522
21 -0.508857653204448
};
\addplot [semithick, color4]
table {%
21 -0.141713400051741
21 0.0777894594619678
};
\addplot [semithick, color4]
table {%
20.85 -0.508857653204448
21.15 -0.508857653204448
};
\addplot [semithick, color4]
table {%
20.85 0.0777894594619678
21.15 0.0777894594619678
};
\addplot [semithick, color4]
table {%
23.7 -0.278499483648295
24.3 -0.278499483648295
24.3 -0.151158282252231
23.7 -0.151158282252231
23.7 -0.278499483648295
};
\addplot [semithick, color4]
table {%
24 -0.278499483648295
24 -0.467283068528188
};
\addplot [semithick, color4]
table {%
24 -0.151158282252231
24 0.0392119956763378
};
\addplot [semithick, color4]
table {%
23.85 -0.467283068528188
24.15 -0.467283068528188
};
\addplot [semithick, color4]
table {%
23.85 0.0392119956763378
24.15 0.0392119956763378
};
\addplot [semithick, color4]
table {%
26.7 -0.280844898303154
27.3 -0.280844898303154
27.3 -0.153235373895677
26.7 -0.153235373895677
26.7 -0.280844898303154
};
\addplot [semithick, color4]
table {%
27 -0.280844898303154
27 -0.469229094572223
};
\addplot [semithick, color4]
table {%
27 -0.153235373895677
27 0.0380151002146871
};
\addplot [semithick, color4]
table {%
26.85 -0.469229094572223
27.15 -0.469229094572223
};
\addplot [semithick, color4]
table {%
26.85 0.0380151002146871
27.15 0.0380151002146871
};
\addplot [semithick, color4]
table {%
29.7 -0.317335922468834
30.3 -0.317335922468834
30.3 -0.139984991675996
29.7 -0.139984991675996
29.7 -0.317335922468834
};
\addplot [semithick, color4]
table {%
30 -0.317335922468834
30 -0.582704098335246
};
\addplot [semithick, color4]
table {%
30 -0.139984991675996
30 0.125754924516495
};
\addplot [semithick, color4]
table {%
29.85 -0.582704098335246
30.15 -0.582704098335246
};
\addplot [semithick, color4]
table {%
29.85 0.125754924516495
30.15 0.125754924516495
};
\addplot [semithick, color4]
table {%
32.7 -0.301396400024753
33.3 -0.301396400024753
33.3 -0.154794615577057
32.7 -0.154794615577057
32.7 -0.301396400024753
};
\addplot [semithick, color4]
table {%
33 -0.301396400024753
33 -0.519554116163402
};
\addplot [semithick, color4]
table {%
33 -0.154794615577057
33 0.064384080840476
};
\addplot [semithick, color4]
table {%
32.85 -0.519554116163402
33.15 -0.519554116163402
};
\addplot [semithick, color4]
table {%
32.85 0.064384080840476
33.15 0.064384080840476
};
\addplot [semithick, color4]
table {%
35.7 -0.291914816251708
36.3 -0.291914816251708
36.3 -0.165925874859837
35.7 -0.165925874859837
35.7 -0.291914816251708
};
\addplot [semithick, color4]
table {%
36 -0.291914816251708
36 -0.478349095686384
};
\addplot [semithick, color4]
table {%
36 -0.165925874859837
36 0.0216660443739269
};
\addplot [semithick, color4]
table {%
35.85 -0.478349095686384
36.15 -0.478349095686384
};
\addplot [semithick, color4]
table {%
35.85 0.0216660443739269
36.15 0.0216660443739269
};
\addplot [semithick, color4]
table {%
38.7 -0.291857409746512
39.3 -0.291857409746512
39.3 -0.174275691054115
38.7 -0.174275691054115
38.7 -0.291857409746512
};
\addplot [semithick, color4]
table {%
39 -0.291857409746512
39 -0.467069430097378
};
\addplot [semithick, color4]
table {%
39 -0.174275691054115
39 0.00171420189231543
};
\addplot [semithick, color4]
table {%
38.85 -0.467069430097378
39.15 -0.467069430097378
};
\addplot [semithick, color4]
table {%
38.85 0.00171420189231543
39.15 0.00171420189231543
};
\addplot [semithick, color4]
table {%
41.7 -0.296407446252739
42.3 -0.296407446252739
42.3 -0.167088757957509
41.7 -0.167088757957509
41.7 -0.296407446252739
};
\addplot [semithick, color4]
table {%
42 -0.296407446252739
42 -0.47453120869532
};
\addplot [semithick, color4]
table {%
42 -0.167088757957509
42 0.0254122135917086
};
\addplot [semithick, color4]
table {%
41.85 -0.47453120869532
42.15 -0.47453120869532
};
\addplot [semithick, color4]
table {%
41.85 0.0254122135917086
42.15 0.0254122135917086
};
\addplot [semithick, color4]
table {%
44.7 -0.214008749515776
45.3 -0.214008749515776
45.3 -0.0905829551185087
44.7 -0.0905829551185087
44.7 -0.214008749515776
};
\addplot [semithick, color4]
table {%
45 -0.214008749515776
45 -0.397600152862452
};
\addplot [semithick, color4]
table {%
45 -0.0905829551185087
45 0.0943705519295617
};
\addplot [semithick, color4]
table {%
44.85 -0.397600152862452
45.15 -0.397600152862452
};
\addplot [semithick, color4]
table {%
44.85 0.0943705519295617
45.15 0.0943705519295617
};
\addplot [semithick, color4]
table {%
47.7 -0.195265814700135
48.3 -0.195265814700135
48.3 -0.0992488663371054
47.7 -0.0992488663371054
47.7 -0.195265814700135
};
\addplot [semithick, color4]
table {%
48 -0.195265814700135
48 -0.337168467376929
};
\addplot [semithick, color4]
table {%
48 -0.0992488663371054
48 0.0446834661087009
};
\addplot [semithick, color4]
table {%
47.85 -0.337168467376929
48.15 -0.337168467376929
};
\addplot [semithick, color4]
table {%
47.85 0.0446834661087009
48.15 0.0446834661087009
};
\addplot [semithick, color4]
table {%
50.7 -0.194441834575241
51.3 -0.194441834575241
51.3 -0.102717890869665
50.7 -0.102717890869665
50.7 -0.194441834575241
};
\addplot [semithick, color4]
table {%
51 -0.194441834575241
51 -0.331053950264618
};
\addplot [semithick, color4]
table {%
51 -0.102717890869665
51 0.0324510093992863
};
\addplot [semithick, color4]
table {%
50.85 -0.331053950264618
51.15 -0.331053950264618
};
\addplot [semithick, color4]
table {%
50.85 0.0324510093992863
51.15 0.0324510093992863
};
\addplot [semithick, color4]
table {%
53.7 -0.192004397595738
54.3 -0.192004397595738
54.3 -0.102980019878553
53.7 -0.102980019878553
53.7 -0.192004397595738
};
\addplot [semithick, color4]
table {%
54 -0.192004397595738
54 -0.325422367840586
};
\addplot [semithick, color4]
table {%
54 -0.102980019878553
54 0.0294297306687139
};
\addplot [semithick, color4]
table {%
53.85 -0.325422367840586
54.15 -0.325422367840586
};
\addplot [semithick, color4]
table {%
53.85 0.0294297306687139
54.15 0.0294297306687139
};
\addplot [semithick, color4]
table {%
56.7 -0.180559876408772
57.3 -0.180559876408772
57.3 -0.112072806173837
56.7 -0.112072806173837
56.7 -0.180559876408772
};
\addplot [semithick, color4]
table {%
57 -0.180559876408772
57 -0.281543521371722
};
\addplot [semithick, color4]
table {%
57 -0.112072806173837
57 -0.0110552690738719
};
\addplot [semithick, color4]
table {%
56.85 -0.281543521371722
57.15 -0.281543521371722
};
\addplot [semithick, color4]
table {%
56.85 -0.0110552690738719
57.15 -0.0110552690738719
};
\addplot [semithick, color6]
table {%
0.5 -0.308575467444446
1.1 -0.308575467444446
1.1 -0.158802611110624
0.5 -0.158802611110624
0.5 -0.308575467444446
};
\addplot [semithick, color6]
table {%
0.8 -0.308575467444446
0.8 -0.532476061907649
};
\addplot [semithick, color6]
table {%
0.8 -0.158802611110624
0.8 0.0657168949910971
};
\addplot [semithick, color6]
table {%
0.65 -0.532476061907649
0.95 -0.532476061907649
};
\addplot [semithick, color6]
table {%
0.65 0.0657168949910971
0.95 0.0657168949910971
};
\addplot [semithick, color6]
table {%
3.5 -0.244501037030674
4.1 -0.244501037030674
4.1 -0.129450943048458
3.5 -0.129450943048458
3.5 -0.244501037030674
};
\addplot [semithick, color6]
table {%
3.8 -0.244501037030674
3.8 -0.416236258428799
};
\addplot [semithick, color6]
table {%
3.8 -0.129450943048458
3.8 0.0421651119807409
};
\addplot [semithick, color6]
table {%
3.65 -0.416236258428799
3.95 -0.416236258428799
};
\addplot [semithick, color6]
table {%
3.65 0.0421651119807409
3.95 0.0421651119807409
};
\addplot [semithick, color6]
table {%
6.5 -0.221086704855507
7.1 -0.221086704855507
7.1 -0.09745210455723
6.5 -0.09745210455723
6.5 -0.221086704855507
};
\addplot [semithick, color6]
table {%
6.8 -0.221086704855507
6.8 -0.405876663575867
};
\addplot [semithick, color6]
table {%
6.8 -0.09745210455723
6.8 0.0876038986995221
};
\addplot [semithick, color6]
table {%
6.65 -0.405876663575867
6.95 -0.405876663575867
};
\addplot [semithick, color6]
table {%
6.65 0.0876038986995221
6.95 0.0876038986995221
};
\addplot [semithick, color6]
table {%
9.5 -0.205002036531796
10.1 -0.205002036531796
10.1 -0.0820622710269489
9.5 -0.0820622710269489
9.5 -0.205002036531796
};
\addplot [semithick, color6]
table {%
9.8 -0.205002036531796
9.8 -0.389382493062814
};
\addplot [semithick, color6]
table {%
9.8 -0.0820622710269489
9.8 0.100292768303741
};
\addplot [semithick, color6]
table {%
9.65 -0.389382493062814
9.95 -0.389382493062814
};
\addplot [semithick, color6]
table {%
9.65 0.100292768303741
9.95 0.100292768303741
};
\addplot [semithick, color6]
table {%
12.5 -0.230693869890522
13.1 -0.230693869890522
13.1 -0.124508101588262
12.5 -0.124508101588262
12.5 -0.230693869890522
};
\addplot [semithick, color6]
table {%
12.8 -0.230693869890522
12.8 -0.389382389577359
};
\addplot [semithick, color6]
table {%
12.8 -0.124508101588262
12.8 0.034372128952311
};
\addplot [semithick, color6]
table {%
12.65 -0.389382389577359
12.95 -0.389382389577359
};
\addplot [semithick, color6]
table {%
12.65 0.034372128952311
12.95 0.034372128952311
};
\addplot [semithick, color6]
table {%
15.5 -0.359774588108077
16.1 -0.359774588108077
16.1 -0.177906030224102
15.5 -0.177906030224102
15.5 -0.359774588108077
};
\addplot [semithick, color6]
table {%
15.8 -0.359774588108077
15.8 -0.631715929108222
};
\addplot [semithick, color6]
table {%
15.8 -0.177906030224102
15.8 0.093763277084238
};
\addplot [semithick, color6]
table {%
15.65 -0.631715929108222
15.95 -0.631715929108222
};
\addplot [semithick, color6]
table {%
15.65 0.093763277084238
15.95 0.093763277084238
};
\addplot [semithick, color6]
table {%
18.5 -0.262699076038408
19.1 -0.262699076038408
19.1 -0.132111297052841
18.5 -0.132111297052841
18.5 -0.262699076038408
};
\addplot [semithick, color6]
table {%
18.8 -0.262699076038408
18.8 -0.457334379369188
};
\addplot [semithick, color6]
table {%
18.8 -0.132111297052841
18.8 0.0635204408354742
};
\addplot [semithick, color6]
table {%
18.65 -0.457334379369188
18.95 -0.457334379369188
};
\addplot [semithick, color6]
table {%
18.65 0.0635204408354742
18.95 0.0635204408354742
};
\addplot [semithick, color6]
table {%
21.5 -0.297829493584512
22.1 -0.297829493584512
22.1 -0.153842989406863
21.5 -0.153842989406863
21.5 -0.297829493584512
};
\addplot [semithick, color6]
table {%
21.8 -0.297829493584512
21.8 -0.511402362455182
};
\addplot [semithick, color6]
table {%
21.8 -0.153842989406863
21.8 0.0617846655783871
};
\addplot [semithick, color6]
table {%
21.65 -0.511402362455182
21.95 -0.511402362455182
};
\addplot [semithick, color6]
table {%
21.65 0.0617846655783871
21.95 0.0617846655783871
};
\addplot [semithick, color6]
table {%
24.5 -0.285412297912196
25.1 -0.285412297912196
25.1 -0.152304352664869
24.5 -0.152304352664869
24.5 -0.285412297912196
};
\addplot [semithick, color6]
table {%
24.8 -0.285412297912196
24.8 -0.484186847418161
};
\addplot [semithick, color6]
table {%
24.8 -0.152304352664869
24.8 0.0461221837275524
};
\addplot [semithick, color6]
table {%
24.65 -0.484186847418161
24.95 -0.484186847418161
};
\addplot [semithick, color6]
table {%
24.65 0.0461221837275524
24.95 0.0461221837275524
};
\addplot [semithick, color6]
table {%
27.5 -0.252597607686222
28.1 -0.252597607686222
28.1 -0.12615830015699
27.5 -0.12615830015699
27.5 -0.252597607686222
};
\addplot [semithick, color6]
table {%
27.8 -0.252597607686222
27.8 -0.436974991050759
};
\addplot [semithick, color6]
table {%
27.8 -0.12615830015699
27.8 0.0631027270500098
};
\addplot [semithick, color6]
table {%
27.65 -0.436974991050759
27.95 -0.436974991050759
};
\addplot [semithick, color6]
table {%
27.65 0.0631027270500098
27.95 0.0631027270500098
};
\addplot [semithick, color6]
table {%
30.5 -0.389681514132676
31.1 -0.389681514132676
31.1 -0.223942874576168
30.5 -0.223942874576168
30.5 -0.389681514132676
};
\addplot [semithick, color6]
table {%
30.8 -0.389681514132676
30.8 -0.637614287898889
};
\addplot [semithick, color6]
table {%
30.8 -0.223942874576168
30.8 0.0245678498441457
};
\addplot [semithick, color6]
table {%
30.65 -0.637614287898889
30.95 -0.637614287898889
};
\addplot [semithick, color6]
table {%
30.65 0.0245678498441457
30.95 0.0245678498441457
};
\addplot [semithick, color6]
table {%
33.5 -0.284010877084583
34.1 -0.284010877084583
34.1 -0.145886597969075
33.5 -0.145886597969075
33.5 -0.284010877084583
};
\addplot [semithick, color6]
table {%
33.8 -0.284010877084583
33.8 -0.489313167748805
};
\addplot [semithick, color6]
table {%
33.8 -0.145886597969075
33.8 0.0609387976386545
};
\addplot [semithick, color6]
table {%
33.65 -0.489313167748805
33.95 -0.489313167748805
};
\addplot [semithick, color6]
table {%
33.65 0.0609387976386545
33.95 0.0609387976386545
};
\addplot [semithick, color6]
table {%
36.5 -0.258860939230148
37.1 -0.258860939230148
37.1 -0.135791573613599
36.5 -0.135791573613599
36.5 -0.258860939230148
};
\addplot [semithick, color6]
table {%
36.8 -0.258860939230148
36.8 -0.442103331389629
};
\addplot [semithick, color6]
table {%
36.8 -0.135791573613599
36.8 0.047236924818619
};
\addplot [semithick, color6]
table {%
36.65 -0.442103331389629
36.95 -0.442103331389629
};
\addplot [semithick, color6]
table {%
36.65 0.047236924818619
36.95 0.047236924818619
};
\addplot [semithick, color6]
table {%
39.5 -0.248209660736207
40.1 -0.248209660736207
40.1 -0.135474599505102
39.5 -0.135474599505102
39.5 -0.248209660736207
};
\addplot [semithick, color6]
table {%
39.8 -0.248209660736207
39.8 -0.41713923524359
};
\addplot [semithick, color6]
table {%
39.8 -0.135474599505102
39.8 0.0328846374650852
};
\addplot [semithick, color6]
table {%
39.65 -0.41713923524359
39.95 -0.41713923524359
};
\addplot [semithick, color6]
table {%
39.65 0.0328846374650852
39.95 0.0328846374650852
};
\addplot [semithick, color6]
table {%
42.5 -0.270948982430572
43.1 -0.270948982430572
43.1 -0.144041190084873
42.5 -0.144041190084873
42.5 -0.270948982430572
};
\addplot [semithick, color6]
table {%
42.8 -0.270948982430572
42.8 -0.459793141698669
};
\addplot [semithick, color6]
table {%
42.8 -0.144041190084873
42.8 0.0436251619335976
};
\addplot [semithick, color6]
table {%
42.65 -0.459793141698669
42.95 -0.459793141698669
};
\addplot [semithick, color6]
table {%
42.65 0.0436251619335976
42.95 0.0436251619335976
};
\addplot [semithick, color6]
table {%
45.5 -0.246645772743949
46.1 -0.246645772743949
46.1 -0.126230059727019
45.5 -0.126230059727019
45.5 -0.246645772743949
};
\addplot [semithick, color6]
table {%
45.8 -0.246645772743949
45.8 -0.425797110742973
};
\addplot [semithick, color6]
table {%
45.8 -0.126230059727019
45.8 0.0543651770789729
};
\addplot [semithick, color6]
table {%
45.65 -0.425797110742973
45.95 -0.425797110742973
};
\addplot [semithick, color6]
table {%
45.65 0.0543651770789729
45.95 0.0543651770789729
};
\addplot [semithick, color6]
table {%
48.5 -0.178361480301108
49.1 -0.178361480301108
49.1 -0.0870387062201
48.5 -0.0870387062201
48.5 -0.178361480301108
};
\addplot [semithick, color6]
table {%
48.8 -0.178361480301108
48.8 -0.313349768728121
};
\addplot [semithick, color6]
table {%
48.8 -0.0870387062201
48.8 0.0494074666094037
};
\addplot [semithick, color6]
table {%
48.65 -0.313349768728121
48.95 -0.313349768728121
};
\addplot [semithick, color6]
table {%
48.65 0.0494074666094037
48.95 0.0494074666094037
};
\addplot [semithick, color6]
table {%
51.5 -0.176279970386959
52.1 -0.176279970386959
52.1 -0.0862233312594425
51.5 -0.0862233312594425
51.5 -0.176279970386959
};
\addplot [semithick, color6]
table {%
51.8 -0.176279970386959
51.8 -0.310480595548356
};
\addplot [semithick, color6]
table {%
51.8 -0.0862233312594425
51.8 0.0484899496181536
};
\addplot [semithick, color6]
table {%
51.65 -0.310480595548356
51.95 -0.310480595548356
};
\addplot [semithick, color6]
table {%
51.65 0.0484899496181536
51.95 0.0484899496181536
};
\addplot [semithick, color6]
table {%
54.5 -0.174757951886913
55.1 -0.174757951886913
55.1 -0.0866724906401639
54.5 -0.0866724906401639
54.5 -0.174757951886913
};
\addplot [semithick, color6]
table {%
54.8 -0.174757951886913
54.8 -0.306812182275436
};
\addplot [semithick, color6]
table {%
54.8 -0.0866724906401639
54.8 0.0445437461576426
};
\addplot [semithick, color6]
table {%
54.65 -0.306812182275436
54.95 -0.306812182275436
};
\addplot [semithick, color6]
table {%
54.65 0.0445437461576426
54.95 0.0445437461576426
};
\addplot [semithick, color6]
table {%
57.5 -0.114486955287448
58.1 -0.114486955287448
58.1 -0.0540045330026008
57.5 -0.0540045330026008
57.5 -0.114486955287448
};
\addplot [semithick, color6]
table {%
57.8 -0.114486955287448
57.8 -0.205103702944805
};
\addplot [semithick, color6]
table {%
57.8 -0.0540045330026008
57.8 0.0362129632658186
};
\addplot [semithick, color6]
table {%
57.65 -0.205103702944805
57.95 -0.205103702944805
};
\addplot [semithick, color6]
table {%
57.65 0.0362129632658186
57.95 0.0362129632658186
};
\addplot [semithick, color2]
table {%
-1.1 -0.220530412035672
-0.5 -0.220530412035672
};
\addplot [semithick, color2]
table {%
1.9 -0.186221263033986
2.5 -0.186221263033986
};
\addplot [semithick, color2]
table {%
4.9 -0.220435976704544
5.5 -0.220435976704544
};
\addplot [semithick, color2]
table {%
7.9 -0.252957385861757
8.5 -0.252957385861757
};
\addplot [semithick, color2]
table {%
10.9 -0.22803222470005
11.5 -0.22803222470005
};
\addplot [semithick, color2]
table {%
13.9 -0.18691495963589
14.5 -0.18691495963589
};
\addplot [semithick, color2]
table {%
16.9 -0.187778449760413
17.5 -0.187778449760413
};
\addplot [semithick, color2]
table {%
19.9 -0.185522867265024
20.5 -0.185522867265024
};
\addplot [semithick, color2]
table {%
22.9 -0.184607690679984
23.5 -0.184607690679984
};
\addplot [semithick, color2]
table {%
25.9 -0.188147949541436
26.5 -0.188147949541436
};
\addplot [semithick, color2]
table {%
28.9 -0.189949382755031
29.5 -0.189949382755031
};
\addplot [semithick, color2]
table {%
31.9 -0.185044493104282
32.5 -0.185044493104282
};
\addplot [semithick, color2]
table {%
34.9 -0.187609909263878
35.5 -0.187609909263878
};
\addplot [semithick, color2]
table {%
37.9 -0.185991224915396
38.5 -0.185991224915396
};
\addplot [semithick, color2]
table {%
40.9 -0.188961542031957
41.5 -0.188961542031957
};
\addplot [semithick, color2]
table {%
43.9 -0.186863406775565
44.5 -0.186863406775565
};
\addplot [semithick, color2]
table {%
46.9 -0.181140267154902
47.5 -0.181140267154902
};
\addplot [semithick, color2]
table {%
49.9 -0.181786848597558
50.5 -0.181786848597558
};
\addplot [semithick, color2]
table {%
52.9 -0.180782652820601
53.5 -0.180782652820601
};
\addplot [semithick, color2]
table {%
55.9 -0.180493798800828
56.5 -0.180493798800828
};
\addplot [semithick, color4]
table {%
-0.3 -0.199480873341492
0.3 -0.199480873341492
};
\addplot [semithick, color4]
table {%
2.7 -0.184284004849825
3.3 -0.184284004849825
};
\addplot [semithick, color4]
table {%
5.7 -0.196881276399013
6.3 -0.196881276399013
};
\addplot [semithick, color4]
table {%
8.7 -0.233289365465452
9.3 -0.233289365465452
};
\addplot [semithick, color4]
table {%
11.7 -0.207812817175495
12.3 -0.207812817175495
};
\addplot [semithick, color4]
table {%
14.7 -0.218733364060527
15.3 -0.218733364060527
};
\addplot [semithick, color4]
table {%
17.7 -0.217501735481327
18.3 -0.217501735481327
};
\addplot [semithick, color4]
table {%
20.7 -0.217414768118258
21.3 -0.217414768118258
};
\addplot [semithick, color4]
table {%
23.7 -0.215174920510324
24.3 -0.215174920510324
};
\addplot [semithick, color4]
table {%
26.7 -0.218851484314537
27.3 -0.218851484314537
};
\addplot [semithick, color4]
table {%
29.7 -0.232306054320168
30.3 -0.232306054320168
};
\addplot [semithick, color4]
table {%
32.7 -0.226670086167027
33.3 -0.226670086167027
};
\addplot [semithick, color4]
table {%
35.7 -0.229601311394526
36.3 -0.229601311394526
};
\addplot [semithick, color4]
table {%
38.7 -0.227928661137397
39.3 -0.227928661137397
};
\addplot [semithick, color4]
table {%
41.7 -0.230856748381878
42.3 -0.230856748381878
};
\addplot [semithick, color4]
table {%
44.7 -0.153367770610735
45.3 -0.153367770610735
};
\addplot [semithick, color4]
table {%
47.7 -0.147808416099081
48.3 -0.147808416099081
};
\addplot [semithick, color4]
table {%
50.7 -0.147654951570057
51.3 -0.147654951570057
};
\addplot [semithick, color4]
table {%
53.7 -0.14668464821617
54.3 -0.14668464821617
};
\addplot [semithick, color4]
table {%
56.7 -0.145753468307405
57.3 -0.145753468307405
};
\addplot [semithick, color6]
table {%
0.5 -0.232120739168276
1.1 -0.232120739168276
};
\addplot [semithick, color6]
table {%
3.5 -0.182883027895124
4.1 -0.182883027895124
};
\addplot [semithick, color6]
table {%
6.5 -0.153295448856146
7.1 -0.153295448856146
};
\addplot [semithick, color6]
table {%
9.5 -0.130647795051584
10.1 -0.130647795051584
};
\addplot [semithick, color6]
table {%
12.5 -0.181070943283206
13.1 -0.181070943283206
};
\addplot [semithick, color6]
table {%
15.5 -0.266017533759162
16.1 -0.266017533759162
};
\addplot [semithick, color6]
table {%
18.5 -0.198197859289411
19.1 -0.198197859289411
};
\addplot [semithick, color6]
table {%
21.5 -0.227338460852841
22.1 -0.227338460852841
};
\addplot [semithick, color6]
table {%
24.5 -0.220438922611346
25.1 -0.220438922611346
};
\addplot [semithick, color6]
table {%
27.5 -0.19089232253753
28.1 -0.19089232253753
};
\addplot [semithick, color6]
table {%
30.5 -0.309395612331283
31.1 -0.309395612331283
};
\addplot [semithick, color6]
table {%
33.5 -0.214628137641132
34.1 -0.214628137641132
};
\addplot [semithick, color6]
table {%
36.5 -0.198593709780133
37.1 -0.198593709780133
};
\addplot [semithick, color6]
table {%
39.5 -0.187302088082268
40.1 -0.187302088082268
};
\addplot [semithick, color6]
table {%
42.5 -0.20843818464618
43.1 -0.20843818464618
};
\addplot [semithick, color6]
table {%
45.5 -0.186865755632164
46.1 -0.186865755632164
};
\addplot [semithick, color6]
table {%
48.5 -0.13349889025452
49.1 -0.13349889025452
};
\addplot [semithick, color6]
table {%
51.5 -0.129641554561446
52.1 -0.129641554561446
};
\addplot [semithick, color6]
table {%
54.5 -0.13038071316196
55.1 -0.13038071316196
};
\addplot [semithick, color6]
table {%
57.5 -0.0833993090963924
58.1 -0.0833993090963924
};
\end{axis}

\begin{axis}[
width=\sfwidth,
height=\sfheight,
tick align=outside,
tick pos=right,
xlabel={Video content},
xmin=-3, xmax=60,
xtick={6,21,36,51},
xticklabels={Minecraft,Cities,Tour,Virus popper},
extra x ticks = {13.5, 28.5, 43.5},
extra x tick labels = {},
extra tick style={grid=major, tick align=inside},
ymin=-30, ymax=30,
ymajorticks = false,
]
\end{axis}

\end{tikzpicture}

%% file: img/legend_sched.tex
\begin{tikzpicture}

\begin{axis}[
    width=0,
    height=0,
    at={(0,0)},
    scale only axis,
    xmin=0,
    xmax=0,
    xtick={},
    ymin=0,
    ymax=0,
    ytick={},
    axis background/.style={fill=white},
    legend style={legend cell align=center, align=center, draw=white!15!black, font=\scriptsize, at={(0, 0)}, anchor=center, /tikz/every even column/.append style={column sep=2em}},
    legend columns=10,
]
\addplot [thick, color2]
table {%
0 1
};
\addlegendentry{\gls{fs}}
\addplot [thick, color6]
table {%
0 1
};
\addlegendentry{\gls{cs}}

\end{axis}

\end{tikzpicture}

%% file: img/schedule_Quantile_latency_granularity.tex
\begin{tikzpicture}

\begin{axis}[
width=\sfwidth,
height=\sfheight,
legend cell align={left},
legend style={at={(0.5,0.02)}, anchor=south, fill opacity=0.8, legend columns=2, draw opacity=1, text opacity=1,
draw=white!80!black, /tikz/every even column/.append style={column sep=1em}},
tick align=outside,
tick pos=left,
x grid style={white!69.0196078431373!black},
xlabel={$S$ (frames)},
xmin=-2, xmax=20,
ymajorgrids,
xtick={0,2,4,6,8,10,12,14,16,18},
xticklabels={1,2,3,4,5,6,7,8,9,10},
y grid style={white!69.0196078431373!black},
ylabel={Latency (ms)},
ymin=5, ymax=20,
ytick style={color=black},
xmajorgrids,
ymajorgrids,
]
\addplot [color2, thick, forget plot]
table {%
-0.7 12.1224511925164
-0.1 12.1224511925164
-0.1 13.9512491922302
-0.7 13.9512491922302
-0.7 12.1224511925164
};
\addplot [color2, thick, forget plot]
table {%
-0.4 12.1224511925164
-0.4 9.38152875650615
};
\addplot [color2, thick, forget plot]
table {%
-0.4 13.9512491922302
-0.4 16.6937271656682
};
\addplot [color2, thick, forget plot]
table {%
-0.55 9.38152875650615
-0.25 9.38152875650615
};
\addplot [color2, thick, forget plot]
table {%
-0.55 16.6937271656682
-0.25 16.6937271656682
};
\addplot [color2, thick, forget plot]
table {%
1.3 12.0269715030949
1.9 12.0269715030949
1.9 13.9505518748054
1.3 13.9505518748054
1.3 12.0269715030949
};
\addplot [color2, thick, forget plot]
table {%
1.6 12.0269715030949
1.6 9.14286623009624
};
\addplot [color2, thick, forget plot]
table {%
1.6 13.9505518748054
1.6 16.8337671026163
};
\addplot [color2, thick, forget plot]
table {%
1.45 9.14286623009624
1.75 9.14286623009624
};
\addplot [color2, thick, forget plot]
table {%
1.45 16.8337671026163
1.75 16.8337671026163
};
\addplot [color2, thick, forget plot]
table {%
3.3 11.9584069802957
3.9 11.9584069802957
3.9 13.9313197499332
3.3 13.9313197499332
3.3 11.9584069802957
};
\addplot [color2, thick, forget plot]
table {%
3.6 11.9584069802957
3.6 9.00530054686766
};
\addplot [color2, thick, forget plot]
table {%
3.6 13.9313197499332
3.6 16.8906061307096
};
\addplot [color2, thick, forget plot]
table {%
3.45 9.00530054686766
3.75 9.00530054686766
};
\addplot [color2, thick, forget plot]
table {%
3.45 16.8906061307096
3.75 16.8906061307096
};
\addplot [color2, thick, forget plot]
table {%
5.3 11.9441786969352
5.9 11.9441786969352
5.9 13.9434878038041
5.3 13.9434878038041
5.3 11.9441786969352
};
\addplot [color2, thick, forget plot]
table {%
5.6 11.9441786969352
5.6 8.94791620620654
};
\addplot [color2, thick, forget plot]
table {%
5.6 13.9434878038041
5.6 16.9406747337071
};
\addplot [color2, thick, forget plot]
table {%
5.45 8.94791620620654
5.75 8.94791620620654
};
\addplot [color2, thick, forget plot]
table {%
5.45 16.9406747337071
5.75 16.9406747337071
};
\addplot [color2, thick, forget plot]
table {%
7.3 11.9163604439775
7.9 11.9163604439775
7.9 13.9298616303462
7.3 13.9298616303462
7.3 11.9163604439775
};
\addplot [color2, thick, forget plot]
table {%
7.6 11.9163604439775
7.6 8.89835881397046
};
\addplot [color2, thick, forget plot]
table {%
7.6 13.9298616303462
7.6 16.9501005674158
};
\addplot [color2, thick, forget plot]
table {%
7.45 8.89835881397046
7.75 8.89835881397046
};
\addplot [color2, thick, forget plot]
table {%
7.45 16.9501005674158
7.75 16.9501005674158
};
\addplot [color2, thick, forget plot]
table {%
9.3 11.9040358725438
9.9 11.9040358725438
9.9 13.9435572840736
9.3 13.9435572840736
9.3 11.9040358725438
};
\addplot [color2, thick, forget plot]
table {%
9.6 11.9040358725438
9.6 8.84540333524359
};
\addplot [color2, thick, forget plot]
table {%
9.6 13.9435572840736
9.6 17.0026117970497
};
\addplot [color2, thick, forget plot]
table {%
9.45 8.84540333524359
9.75 8.84540333524359
};
\addplot [color2, thick, forget plot]
table {%
9.45 17.0026117970497
9.75 17.0026117970497
};
\addplot [color2, thick, forget plot]
table {%
11.3 11.8856460100009
11.9 11.8856460100009
11.9 13.9139319587075
11.3 13.9139319587075
11.3 11.8856460100009
};
\addplot [color2, thick, forget plot]
table {%
11.6 11.8856460100009
11.6 8.84427331505307
};
\addplot [color2, thick, forget plot]
table {%
11.6 13.9139319587075
11.6 16.9561411242237
};
\addplot [color2, thick, forget plot]
table {%
11.45 8.84427331505307
11.75 8.84427331505307
};
\addplot [color2, thick, forget plot]
table {%
11.45 16.9561411242237
11.75 16.9561411242237
};
\addplot [color2, thick, forget plot]
table {%
13.3 11.8743271371187
13.9 11.8743271371187
13.9 13.9268999929087
13.3 13.9268999929087
13.3 11.8743271371187
};
\addplot [color2, thick, forget plot]
table {%
13.6 11.8743271371187
13.6 8.79616964311269
};
\addplot [color2, thick, forget plot]
table {%
13.6 13.9268999929087
13.6 17.0040263334731
};
\addplot [color2, thick, forget plot]
table {%
13.45 8.79616964311269
13.75 8.79616964311269
};
\addplot [color2, thick, forget plot]
table {%
13.45 17.0040263334731
13.75 17.0040263334731
};
\addplot [color2, thick, forget plot]
table {%
15.3 11.8386516054657
15.9 11.8386516054657
15.9 13.9206211577902
15.3 13.9206211577902
15.3 11.8386516054657
};
\addplot [color2, thick, forget plot]
table {%
15.6 11.8386516054657
15.6 8.71670662910332
};
\addplot [color2, thick, forget plot]
table {%
15.6 13.9206211577902
15.6 17.0431293828392
};
\addplot [color2, thick, forget plot]
table {%
15.45 8.71670662910332
15.75 8.71670662910332
};
\addplot [color2, thick, forget plot]
table {%
15.45 17.0431293828392
15.75 17.0431293828392
};
\addplot [color2, thick, forget plot]
table {%
17.3 11.8351194833764
17.9 11.8351194833764
17.9 13.9258530260145
17.3 13.9258530260145
17.3 11.8351194833764
};
\addplot [color2, thick, forget plot]
table {%
17.6 11.8351194833764
17.6 8.69921872523401
};
\addplot [color2, thick, forget plot]
table {%
17.6 13.9258530260145
17.6 17.0577091415097
};
\addplot [color2, thick, forget plot]
table {%
17.45 8.69921872523401
17.75 8.69921872523401
};
\addplot [color2, thick, forget plot]
table {%
17.45 17.0577091415097
17.75 17.0577091415097
};
\addplot [color6, thick, forget plot]
table {%
0.1 12.1224511925164
0.7 12.1224511925164
0.7 13.9512491922302
0.1 13.9512491922302
0.1 12.1224511925164
};
\addplot [color6, thick, forget plot]
table {%
0.4 12.1224511925164
0.4 9.38152875650615
};
\addplot [color6, thick, forget plot]
table {%
0.4 13.9512491922302
0.4 16.6937271656682
};
\addplot [color6, thick, forget plot]
table {%
0.25 9.38152875650615
0.55 9.38152875650615
};
\addplot [color6, thick, forget plot]
table {%
0.25 16.6937271656682
0.55 16.6937271656682
};
\addplot [color6, thick, forget plot]
table {%
2.1 12.5130638551438
2.7 12.5130638551438
2.7 14.5284630577309
2.1 14.5284630577309
2.1 12.5130638551438
};
\addplot [color6, thick, forget plot]
table {%
2.4 12.5130638551438
2.4 9.49168004873199
};
\addplot [color6, thick, forget plot]
table {%
2.4 14.5284630577309
2.4 17.5515072085585
};
\addplot [color6, thick, forget plot]
table {%
2.25 9.49168004873199
2.55 9.49168004873199
};
\addplot [color6, thick, forget plot]
table {%
2.25 17.5515072085585
2.55 17.5515072085585
};
\addplot [color6, thick, forget plot]
table {%
4.1 12.6879748803517
4.7 12.6879748803517
4.7 14.8052944901876
4.1 14.8052944901876
4.1 12.6879748803517
};
\addplot [color6, thick, forget plot]
table {%
4.4 12.6879748803517
4.4 9.5140981522755
};
\addplot [color6, thick, forget plot]
table {%
4.4 14.8052944901876
4.4 17.9750811351843
};
\addplot [color6, thick, forget plot]
table {%
4.25 9.5140981522755
4.55 9.5140981522755
};
\addplot [color6, thick, forget plot]
table {%
4.25 17.9750811351843
4.55 17.9750811351843
};
\addplot [color6, thick, forget plot]
table {%
6.1 12.8491550091019
6.7 12.8491550091019
6.7 15.0426454904013
6.1 15.0426454904013
6.1 12.8491550091019
};
\addplot [color6, thick, forget plot]
table {%
6.4 12.8491550091019
6.4 9.56080572960417
};
\addplot [color6, thick, forget plot]
table {%
6.4 15.0426454904013
6.4 18.3310511962036
};
\addplot [color6, thick, forget plot]
table {%
6.25 9.56080572960417
6.55 9.56080572960417
};
\addplot [color6, thick, forget plot]
table {%
6.25 18.3310511962036
6.55 18.3310511962036
};
\addplot [color6, thick, forget plot]
table {%
8.1 12.9394297828015
8.7 12.9394297828015
8.7 15.1848135993071
8.1 15.1848135993071
8.1 12.9394297828015
};
\addplot [color6, thick, forget plot]
table {%
8.4 12.9394297828015
8.4 9.57483747204532
};
\addplot [color6, thick, forget plot]
table {%
8.4 15.1848135993071
8.4 18.5522294073081
};
\addplot [color6, thick, forget plot]
table {%
8.25 9.57483747204532
8.55 9.57483747204532
};
\addplot [color6, thick, forget plot]
table {%
8.25 18.5522294073081
8.55 18.5522294073081
};
\addplot [color6, thick, forget plot]
table {%
10.1 13.0076623952543
10.7 13.0076623952543
10.7 15.3014097347513
10.1 15.3014097347513
10.1 13.0076623952543
};
\addplot [color6, thick, forget plot]
table {%
10.4 13.0076623952543
10.4 9.57509382409021
};
\addplot [color6, thick, forget plot]
table {%
10.4 15.3014097347513
10.4 18.7391423546245
};
\addplot [color6, thick, forget plot]
table {%
10.25 9.57509382409021
10.55 9.57509382409021
};
\addplot [color6, thick, forget plot]
table {%
10.25 18.7391423546245
10.55 18.7391423546245
};
\addplot [color6, thick, forget plot]
table {%
12.1 13.0479261472529
12.7 13.0479261472529
12.7 15.3640766483598
12.1 15.3640766483598
12.1 13.0479261472529
};
\addplot [color6, thick, forget plot]
table {%
12.4 13.0479261472529
12.4 9.57471853573879
};
\addplot [color6, thick, forget plot]
table {%
12.4 15.3640766483598
12.4 18.8363575089884
};
\addplot [color6, thick, forget plot]
table {%
12.25 9.57471853573879
12.55 9.57471853573879
};
\addplot [color6, thick, forget plot]
table {%
12.25 18.8363575089884
12.55 18.8363575089884
};
\addplot [color6, thick, forget plot]
table {%
14.1 13.0880092944235
14.7 13.0880092944235
14.7 15.444512994295
14.1 15.444512994295
14.1 13.0880092944235
};
\addplot [color6, thick, forget plot]
table {%
14.4 13.0880092944235
14.4 9.55664050189421
};
\addplot [color6, thick, forget plot]
table {%
14.4 15.444512994295
14.4 18.977960968986
};
\addplot [color6, thick, forget plot]
table {%
14.25 9.55664050189421
14.55 9.55664050189421
};
\addplot [color6, thick, forget plot]
table {%
14.25 18.977960968986
14.55 18.977960968986
};
\addplot [color6, thick, forget plot]
table {%
16.1 13.1013984893663
16.7 13.1013984893663
16.7 15.4989005320918
16.1 15.4989005320918
16.1 13.1013984893663
};
\addplot [color6, thick, forget plot]
table {%
16.4 13.1013984893663
16.4 9.50937518257976
};
\addplot [color6, thick, forget plot]
table {%
16.4 15.4989005320918
16.4 19.0948217428632
};
\addplot [color6, thick, forget plot]
table {%
16.25 9.50937518257976
16.55 9.50937518257976
};
\addplot [color6, thick, forget plot]
table {%
16.25 19.0948217428632
16.55 19.0948217428632
};
\addplot [color6, thick, forget plot]
table {%
18.1 13.1193086600292
18.7 13.1193086600292
18.7 15.5679475039044
18.1 15.5679475039044
18.1 13.1193086600292
};
\addplot [color6, thick, forget plot]
table {%
18.4 13.1193086600292
18.4 9.44821753781463
};
\addplot [color6, thick, forget plot]
table {%
18.4 15.5679475039044
18.4 19.2387347771301
};
\addplot [color6, thick, forget plot]
table {%
18.25 9.44821753781463
18.55 9.44821753781463
};
\addplot [color6, thick, forget plot]
table {%
18.25 19.2387347771301
18.55 19.2387347771301
};
\addplot [thick, black, dashed, thick, forget plot]
table {%
-2 16.6666666666667
-1 16.6666666666667
0 16.6666666666667
1 16.6666666666667
2 16.6666666666667
3 16.6666666666667
4 16.6666666666667
5 16.6666666666667
6 16.6666666666667
7 16.6666666666667
8 16.6666666666667
9 16.6666666666667
10 16.6666666666667
11 16.6666666666667
12 16.6666666666667
13 16.6666666666667
14 16.6666666666667
15 16.6666666666667
16 16.6666666666667
17 16.6666666666667
18 16.6666666666667
19 16.6666666666667
20 16.6666666666667
21 16.6666666666667
};
\addplot [color2, thick, forget plot]
table {%
-0.7 13.0404224673954
-0.1 13.0404224673954
};
\addplot [color2, thick, forget plot]
table {%
1.3 12.9843540912789
1.9 12.9843540912789
};
\addplot [color2, thick, forget plot]
table {%
3.3 12.951787692925
3.9 12.951787692925
};
\addplot [color2, thick, forget plot]
table {%
5.3 12.9393858715715
5.9 12.9393858715715
};
\addplot [color2, thick, forget plot]
table {%
7.3 12.9263720236673
7.9 12.9263720236673
};
\addplot [color2, thick, forget plot]
table {%
9.3 12.9163123623126
9.9 12.9163123623126
};
\addplot [color2, thick, forget plot]
table {%
11.3 12.898886368616
11.9 12.898886368616
};
\addplot [color2, thick, forget plot]
table {%
13.3 12.8833484981822
13.9 12.8833484981822
};
\addplot [color2, thick, forget plot]
table {%
15.3 12.8697145971019
15.9 12.8697145971019
};
\addplot [color2, thick, forget plot]
table {%
17.3 12.8672294446065
17.9 12.8672294446065
};
\addplot [color6, thick, forget plot]
table {%
0.1 13.0404224673954
0.7 13.0404224673954
};
\addplot [color6, thick, forget plot]
table {%
2.1 13.5162500377401
2.7 13.5162500377401
};
\addplot [color6, thick, forget plot]
table {%
4.1 13.7286430608016
4.7 13.7286430608016
};
\addplot [color6, thick, forget plot]
table {%
6.1 13.9217232872125
6.7 13.9217232872125
};
\addplot [color6, thick, forget plot]
table {%
8.1 14.0405847078414
8.7 14.0405847078414
};
\addplot [color6, thick, forget plot]
table {%
10.1 14.122391877264
10.7 14.122391877264
};
\addplot [color6, thick, forget plot]
table {%
12.1 14.1752649585641
12.7 14.1752649585641
};
\addplot [color6, thick, forget plot]
table {%
14.1 14.2162754498836
14.7 14.2162754498836
};
\addplot [color6, thick, forget plot]
table {%
16.1 14.2547607432213
16.7 14.2547607432213
};
\addplot [color6, thick, forget plot]
table {%
18.1 14.2818973999326
18.7 14.2818973999326
};

\end{axis}

\begin{axis}[
    width=\sfwidth,
    height=\sfheight,
        tick align=outside,
    tick pos=right,
    xlabel={$\Delta t$ (ms)},
    xmin=0, xmax=183.333,
        ymin=0, ymax=1,
    ymajorticks = false,
    ]
        
    \end{axis}
\end{tikzpicture}

%% file: img/schedule_Quantile_rate_granularity.tex
\begin{tikzpicture}

\begin{axis}[
width=\sfwidth,
height=\sfheight,
legend style={at={(0.5,0.02)}, anchor=south, fill opacity=0.8, legend columns=2, draw opacity=1, text opacity=1,
draw=white!80!black, /tikz/every even column/.append style={column sep=1em}},
tick align=outside,
tick pos=left,
x grid style={white!69.0196078431373!black},
xlabel={$S$ (frames)},
ymajorgrids,
xmin=-2, xmax=20,
xtick={0,2,4,6,8,10,12,14,16,18},
xticklabels={1,2,3,4,5,6,7,8,9,10},
y grid style={white!69.0196078431373!black},
ylabel={Scheduled capacity (Mb/s)},
ymin=25, ymax=45,
ytick style={color=black},
xmajorgrids,
ymajorgrids,
]
\addplot [thick, black, dashed, thick, forget plot]
table {%
-2 30
21 30
};
\addplot [color2, thick, forget plot]
table {%
-0.7 36.6211474796916
-0.1 36.6211474796916
-0.1 39.6266823375145
-0.7 39.6266823375145
-0.7 36.6211474796916
};
\addplot [color2, thick, forget plot]
table {%
-0.4 36.6211474796916
-0.4 32.1133625132256
};
\addplot [color2, thick, forget plot]
table {%
-0.4 39.6266823375145
-0.4 44.1208748890272
};
\addplot [color2, thick, forget plot]
table {%
-0.55 32.1133625132256
-0.25 32.1133625132256
};
\addplot [color2, thick, forget plot]
table {%
-0.55 44.1208748890272
-0.25 44.1208748890272
};
\addplot [color2, thick, forget plot]
table {%
1.3 36.9354908372796
1.9 36.9354908372796
1.9 39.6594987899281
1.3 39.6594987899281
1.3 36.9354908372796
};
\addplot [color2, thick, forget plot]
table {%
1.6 36.9354908372796
1.6 32.8518836881599
};
\addplot [color2, thick, forget plot]
table {%
1.6 39.6594987899281
1.6 43.7378306738241
};
\addplot [color2, thick, forget plot]
table {%
1.45 32.8518836881599
1.75 32.8518836881599
};
\addplot [color2, thick, forget plot]
table {%
1.45 43.7378306738241
1.75 43.7378306738241
};
\addplot [color2, thick, forget plot]
table {%
3.3 37.1840829508745
3.9 37.1840829508745
3.9 39.651062282072
3.3 39.651062282072
3.3 37.1840829508745
};
\addplot [color2, thick, forget plot]
table {%
3.6 37.1840829508745
3.6 33.4844021788634
};
\addplot [color2, thick, forget plot]
table {%
3.6 39.651062282072
3.6 43.3503528805249
};
\addplot [color2, thick, forget plot]
table {%
3.45 33.4844021788634
3.75 33.4844021788634
};
\addplot [color2, thick, forget plot]
table {%
3.45 43.3503528805249
3.75 43.3503528805249
};
\addplot [color2, thick, forget plot]
table {%
5.3 37.3060379994666
5.9 37.3060379994666
5.9 39.6516831526111
5.3 39.6516831526111
5.3 37.3060379994666
};
\addplot [color2, thick, forget plot]
table {%
5.6 37.3060379994666
5.6 33.7903538500622
};
\addplot [color2, thick, forget plot]
table {%
5.6 39.6516831526111
5.6 43.1667707676156
};
\addplot [color2, thick, forget plot]
table {%
5.45 33.7903538500622
5.75 33.7903538500622
};
\addplot [color2, thick, forget plot]
table {%
5.45 43.1667707676156
5.75 43.1667707676156
};
\addplot [color2, thick, forget plot]
table {%
7.3 37.398628611373
7.9 37.398628611373
7.9 39.6503454901451
7.3 39.6503454901451
7.3 37.398628611373
};
\addplot [color2, thick, forget plot]
table {%
7.6 37.398628611373
7.6 34.0221669906972
};
\addplot [color2, thick, forget plot]
table {%
7.6 39.6503454901451
7.6 43.0253280015717
};
\addplot [color2, thick, forget plot]
table {%
7.45 34.0221669906972
7.75 34.0221669906972
};
\addplot [color2, thick, forget plot]
table {%
7.45 43.0253280015717
7.75 43.0253280015717
};
\addplot [color2, thick, forget plot]
table {%
9.3 37.4137316211047
9.9 37.4137316211047
9.9 39.6092731576195
9.3 39.6092731576195
9.3 37.4137316211047
};
\addplot [color2, thick, forget plot]
table {%
9.6 37.4137316211047
9.6 34.1241755729731
};
\addplot [color2, thick, forget plot]
table {%
9.6 39.6092731576195
9.6 42.9012110820478
};
\addplot [color2, thick, forget plot]
table {%
9.45 34.1241755729731
9.75 34.1241755729731
};
\addplot [color2, thick, forget plot]
table {%
9.45 42.9012110820478
9.75 42.9012110820478
};
\addplot [color2, thick, forget plot]
table {%
11.3 37.5441646616044
11.9 37.5441646616044
11.9 39.6659801999716
11.3 39.6659801999716
11.3 37.5441646616044
};
\addplot [color2, thick, forget plot]
table {%
11.6 37.5441646616044
11.6 34.3638452357924
};
\addplot [color2, thick, forget plot]
table {%
11.6 39.6659801999716
11.6 42.8462952440809
};
\addplot [color2, thick, forget plot]
table {%
11.45 34.3638452357924
11.75 34.3638452357924
};
\addplot [color2, thick, forget plot]
table {%
11.45 42.8462952440809
11.75 42.8462952440809
};
\addplot [color2, thick, forget plot]
table {%
13.3 37.6514218522885
13.9 37.6514218522885
13.9 39.6180684248882
13.3 39.6180684248882
13.3 37.6514218522885
};
\addplot [color2, thick, forget plot]
table {%
13.6 37.6514218522885
13.6 34.7048368017501
};
\addplot [color2, thick, forget plot]
table {%
13.6 39.6180684248882
13.6 42.5645651155835
};
\addplot [color2, thick, forget plot]
table {%
13.45 34.7048368017501
13.75 34.7048368017501
};
\addplot [color2, thick, forget plot]
table {%
13.45 42.5645651155835
13.75 42.5645651155835
};
\addplot [color2, thick, forget plot]
table {%
15.3 37.7779230065037
15.9 37.7779230065037
15.9 39.5842058323261
15.3 39.5842058323261
15.3 37.7779230065037
};
\addplot [color2, thick, forget plot]
table {%
15.6 37.7779230065037
15.6 35.0700707914218
};
\addplot [color2, thick, forget plot]
table {%
15.6 39.5842058323261
15.6 42.2920033827437
};
\addplot [color2, thick, forget plot]
table {%
15.45 35.0700707914218
15.75 35.0700707914218
};
\addplot [color2, thick, forget plot]
table {%
15.45 42.2920033827437
15.75 42.2920033827437
};
\addplot [color2, thick, forget plot]
table {%
17.3 37.8725961935293
17.9 37.8725961935293
17.9 39.5653020696279
17.3 39.5653020696279
17.3 37.8725961935293
};
\addplot [color2, thick, forget plot]
table {%
17.6 37.8725961935293
17.6 35.3336330403262
};
\addplot [color2, thick, forget plot]
table {%
17.6 39.5653020696279
17.6 42.1038331511556
};
\addplot [color2, thick, forget plot]
table {%
17.45 35.3336330403262
17.75 35.3336330403262
};
\addplot [color2, thick, forget plot]
table {%
17.45 42.1038331511556
17.75 42.1038331511556
};
\addplot [color6, thick, forget plot]
table {%
0.1 36.6211474796916
0.7 36.6211474796916
0.7 39.6266823375145
0.1 39.6266823375145
0.1 36.6211474796916
};
\addplot [color6, thick, forget plot]
table {%
0.4 36.6211474796916
0.4 32.1133625132256
};
\addplot [color6, thick, forget plot]
table {%
0.4 39.6266823375145
0.4 44.1208748890272
};
\addplot [color6, thick, forget plot]
table {%
0.25 32.1133625132256
0.55 32.1133625132256
};
\addplot [color6, thick, forget plot]
table {%
0.25 44.1208748890272
0.55 44.1208748890272
};
\addplot [color6, thick, forget plot]
table {%
2.1 35.4963974711673
2.7 35.4963974711673
2.7 38.1689059641891
2.1 38.1689059641891
2.1 35.4963974711673
};
\addplot [color6, thick, forget plot]
table {%
2.4 35.4963974711673
2.4 31.4903882078126
};
\addplot [color6, thick, forget plot]
table {%
2.4 38.1689059641891
2.4 42.1737167847751
};
\addplot [color6, thick, forget plot]
table {%
2.25 31.4903882078126
2.55 31.4903882078126
};
\addplot [color6, thick, forget plot]
table {%
2.25 42.1737167847751
2.55 42.1737167847751
};
\addplot [color6, thick, forget plot]
table {%
4.1 35.0568286755957
4.7 35.0568286755957
4.7 37.4751152095093
4.1 37.4751152095093
4.1 35.0568286755957
};
\addplot [color6, thick, forget plot]
table {%
4.4 35.0568286755957
4.4 31.4448731812112
};
\addplot [color6, thick, forget plot]
table {%
4.4 37.4751152095093
4.4 41.0996672805811
};
\addplot [color6, thick, forget plot]
table {%
4.25 31.4448731812112
4.55 31.4448731812112
};
\addplot [color6, thick, forget plot]
table {%
4.25 41.0996672805811
4.55 41.0996672805811
};
\addplot [color6, thick, forget plot]
table {%
6.1 34.6607329781317
6.7 34.6607329781317
6.7 36.9827207239145
6.1 36.9827207239145
6.1 34.6607329781317
};
\addplot [color6, thick, forget plot]
table {%
6.4 34.6607329781317
6.4 31.2073071505745
};
\addplot [color6, thick, forget plot]
table {%
6.4 36.9827207239145
6.4 40.455882454078
};
\addplot [color6, thick, forget plot]
table {%
6.25 31.2073071505745
6.55 31.2073071505745
};
\addplot [color6, thick, forget plot]
table {%
6.25 40.455882454078
6.55 40.455882454078
};
\addplot [color6, thick, forget plot]
table {%
8.1 34.4191795380062
8.7 34.4191795380062
8.7 36.6731781981335
8.1 36.6731781981335
8.1 34.4191795380062
};
\addplot [color6, thick, forget plot]
table {%
8.4 34.4191795380062
8.4 31.041983347203
};
\addplot [color6, thick, forget plot]
table {%
8.4 36.6731781981335
8.4 40.0363239224976
};
\addplot [color6, thick, forget plot]
table {%
8.25 31.041983347203
8.55 31.041983347203
};
\addplot [color6, thick, forget plot]
table {%
8.25 40.0363239224976
8.55 40.0363239224976
};
\addplot [color6, thick, forget plot]
table {%
10.1 34.2326205089623
10.7 34.2326205089623
10.7 36.4044048755283
10.1 36.4044048755283
10.1 34.2326205089623
};
\addplot [color6, thick, forget plot]
table {%
10.4 34.2326205089623
10.4 30.9814647472978
};
\addplot [color6, thick, forget plot]
table {%
10.4 36.4044048755283
10.4 39.6111735728431
};
\addplot [color6, thick, forget plot]
table {%
10.25 30.9814647472978
10.55 30.9814647472978
};
\addplot [color6, thick, forget plot]
table {%
10.25 39.6111735728431
10.55 39.6111735728431
};
\addplot [color6, thick, forget plot]
table {%
12.1 34.2083073305274
12.7 34.2083073305274
12.7 36.2838285022956
12.1 36.2838285022956
12.1 34.2083073305274
};
\addplot [color6, thick, forget plot]
table {%
12.4 34.2083073305274
12.4 31.1023415121712
};
\addplot [color6, thick, forget plot]
table {%
12.4 36.2838285022956
12.4 39.3740080898995
};
\addplot [color6, thick, forget plot]
table {%
12.25 31.1023415121712
12.55 31.1023415121712
};
\addplot [color6, thick, forget plot]
table {%
12.25 39.3740080898995
12.55 39.3740080898995
};
\addplot [color6, thick, forget plot]
table {%
14.1 34.1667730202706
14.7 34.1667730202706
14.7 36.1297832573304
14.1 36.1297832573304
14.1 34.1667730202706
};
\addplot [color6, thick, forget plot]
table {%
14.4 34.1667730202706
14.4 31.2333350217275
};
\addplot [color6, thick, forget plot]
table {%
14.4 36.1297832573304
14.4 39.0722122229047
};
\addplot [color6, thick, forget plot]
table {%
14.25 31.2333350217275
14.55 31.2333350217275
};
\addplot [color6, thick, forget plot]
table {%
14.25 39.0722122229047
14.55 39.0722122229047
};
\addplot [color6, thick, forget plot]
table {%
16.1 34.1732524000775
16.7 34.1732524000775
16.7 35.966743795887
16.1 35.966743795887
16.1 34.1732524000775
};
\addplot [color6, thick, forget plot]
table {%
16.4 34.1732524000775
16.4 31.5314478813985
};
\addplot [color6, thick, forget plot]
table {%
16.4 35.966743795887
16.4 38.6412278912877
};
\addplot [color6, thick, forget plot]
table {%
16.25 31.5314478813985
16.55 31.5314478813985
};
\addplot [color6, thick, forget plot]
table {%
16.25 38.6412278912877
16.55 38.6412278912877
};
\addplot [color6, thick, forget plot]
table {%
18.1 34.1707900134713
18.7 34.1707900134713
18.7 35.8716409377639
18.1 35.8716409377639
18.1 34.1707900134713
};
\addplot [color6, thick, forget plot]
table {%
18.4 34.1707900134713
18.4 31.6201855551225
};
\addplot [color6, thick, forget plot]
table {%
18.4 35.8716409377639
18.4 38.3831157819313
};
\addplot [color6, thick, forget plot]
table {%
18.25 31.6201855551225
18.55 31.6201855551225
};
\addplot [color6, thick, forget plot]
table {%
18.25 38.3831157819313
18.55 38.3831157819313
};
\addplot [color2, thick, forget plot]
table {%
-0.7 38.107560258157
-0.1 38.107560258157
};
\addplot [color2, thick, forget plot]
table {%
1.3 38.2970783785509
1.9 38.2970783785509
};
\addplot [color2, thick, forget plot]
table {%
3.3 38.420312487623
3.9 38.420312487623
};
\addplot [color2, thick, forget plot]
table {%
5.3 38.4926908692817
5.9 38.4926908692817
};
\addplot [color2, thick, forget plot]
table {%
7.3 38.5321986691363
7.9 38.5321986691363
};
\addplot [color2, thick, forget plot]
table {%
9.3 38.5236662776653
9.9 38.5236662776653
};
\addplot [color2, thick, forget plot]
table {%
11.3 38.6232100588778
11.9 38.6232100588778
};
\addplot [color2, thick, forget plot]
table {%
13.3 38.6566305381047
13.9 38.6566305381047
};
\addplot [color2, thick, forget plot]
table {%
15.3 38.7148316688225
15.9 38.7148316688225
};
\addplot [color2, thick, forget plot]
table {%
17.3 38.7801124740236
17.9 38.7801124740236
};
\addplot [color6, thick, forget plot]
table {%
0.1 38.107560258157
0.7 38.107560258157
};
\addplot [color6, thick, forget plot]
table {%
2.1 36.8130800505345
2.7 36.8130800505345
};
\addplot [color6, thick, forget plot]
table {%
4.1 36.2409576131296
4.7 36.2409576131296
};
\addplot [color6, thick, forget plot]
table {%
6.1 35.7921824678359
6.7 35.7921824678359
};
\addplot [color6, thick, forget plot]
table {%
8.1 35.5193109441772
8.7 35.5193109441772
};
\addplot [color6, thick, forget plot]
table {%
10.1 35.3105629514969
10.7 35.3105629514969
};
\addplot [color6, thick, forget plot]
table {%
12.1 35.2317590490666
12.7 35.2317590490666
};
\addplot [color6, thick, forget plot]
table {%
14.1 35.115640240426
14.7 35.115640240426
};
\addplot [color6, thick, forget plot]
table {%
16.1 35.0366495433123
16.7 35.0366495433123
};
\addplot [color6, thick, forget plot]
table {%
18.1 34.9923936817818
18.7 34.9923936817818
};

\end{axis}

\begin{axis}[
    width=\sfwidth,
    height=\sfheight,
        tick align=outside,
    tick pos=right,
    xlabel={$\Delta t$ (ms)},
    xmin=0, xmax=183.333,
        ymin=0, ymax=1,
    ymajorticks = false,
    ]
        
    \end{axis}

\end{tikzpicture}

%% file: img/perc_latency_6.tex
\begin{tikzpicture}

\begin{axis}[
width=\sfwidth,
height=\sfheight,
legend cell align={left},
legend style={at={(0.995,0.99)}, anchor=north east, fill opacity=0.8, legend columns=2, draw opacity=1, text opacity=1, font=\tiny, draw=white!80!black},
tick align=outside,
tick pos=left,
x grid style={white!69.0196078431373!black},
xlabel={$p_s$},
xmin=0.9, xmax=1,
xtick style={color=black},
y grid style={white!69.0196078431373!black},
ylabel={Latency (ms)},
ymin=10, ymax=35,
ytick={0, 5, ..., 50},
ytick style={color=black},
xmajorgrids,
ymajorgrids
]
\addplot[thick, black, dashed, forget plot]
table{
0.9 16.666667
1 16.666667
};

\addplot [thick, color2, mark=o, mark options={solid}]
table {%
0.9 15.7613242779047
0.905 15.6885328898565
0.91 15.6127325478221
0.915 15.5329013738255
0.92 15.4524498789084
0.925 15.3702363554952
0.93 15.2794458494122
0.935 15.1895163839603
0.94 15.0867202381521
0.945 14.9857771694039
0.95 14.8698361420095
0.955 14.7429873490029
0.96 14.6107465941392
0.965 14.4505217191599
0.97 14.2601591540097
0.975 14.0601673130676
0.98 13.8216077689923
0.985 13.5396127684898
0.99 13.1230325004732
0.995 12.4746593906581
};
\addlegendentry{CS, average}

\addplot [thick, dashdotted, color2, mark=x, mark options={solid}]
table {%
0.9 14.5784613009318
0.905 14.500976781304
0.91 14.4226106999314
0.915 14.3377752106377
0.92 14.2527214434427
0.925 14.1622012815075
0.93 14.0669060683639
0.935 13.9662785590525
0.94 13.862931224573
0.945 13.7506720884939
0.95 13.6305833282567
0.955 13.4998736980438
0.96 13.3557378845159
0.965 13.1927759346656
0.97 13.0032370543539
0.975 12.7847047568529
0.98 12.519534724948
0.985 12.1890563569342
0.99 11.7249128543669
0.995 10.905863901774
};
\addlegendentry{FS, average}
%
%

\addplot [thick, color4, mark=o, mark options={solid}]
table {%
0.9 21.5871866073286
0.905 21.4248647499771
0.91 21.2510622782164
0.915 21.0771593140057
0.92 20.8941499980262
0.925 20.7342389372751
0.93 20.5207325497922
0.935 20.3244876992523
0.94 20.0925651396636
0.945 19.8683492949267
0.95 19.6240375718728
0.955 19.360572439117
0.96 19.0578285298147
0.965 18.7091547371938
0.97 18.3289309396206
0.975 17.9419390634813
0.98 17.5418366942289
0.985 17.0195958826213
0.99 16.3831028205683
0.995 15.4779538922047
};
\addlegendentry{CS, 95th perc.}

\addplot [thick, dashdotted, color4, mark=x, mark options={solid}]
table {%
0.9 18.6598406190369
0.905 18.4969813108059
0.91 18.3672659079077
0.915 18.2286576458997
0.92 18.08409875764
0.925 17.9183015760408
0.93 17.769894500201
0.935 17.6080041472099
0.94 17.4396096331188
0.945 17.2647853450739
0.95 17.0817590338882
0.955 16.8864635253391
0.96 16.6787767052782
0.965 16.4464626676535
0.97 16.1784962541958
0.975 15.8912069394523
0.98 15.5526967830979
0.985 15.1317735557681
0.99 14.5680159205636
0.995 13.5776429044096
};
\addlegendentry{FS, 95th perc.}

\addplot [thick, color6, mark=o, mark options={solid}]
table {%
0.9 28.4893929624497
0.905 28.2915066982544
0.91 28.0711939455348
0.915 27.8685624689641
0.92 27.6326095640386
0.925 27.3779838844048
0.93 27.1663244548585
0.935 26.9374975134469
0.94 26.6231209823868
0.945 26.3691931242041
0.95 26.0345222683133
0.955 25.5787910102994
0.96 25.1755361262711
0.965 24.7149405007876
0.97 24.1599862896265
0.975 23.4905822395617
0.98 22.5962975811709
0.985 21.7571671809163
0.99 20.3413720123725
0.995 18.5079497248315
};
\addlegendentry{CS, 99th perc.}

\addplot [thick, dashdotted, color6, mark=x, mark options={solid}]
table {%
0.9 23.2368582146214
0.905 22.9939051476596
0.91 22.7469559677127
0.915 22.5012971515958
0.92 22.2842664835026
0.925 22.0472071771509
0.93 21.8162375183128
0.935 21.5850521459
0.94 21.3128041872271
0.945 21.0397902899049
0.95 20.7254248608243
0.955 20.3576218293724
0.96 20.0009153067236
0.965 19.5845285084415
0.97 19.1009570764521
0.975 18.6905507278605
0.98 18.2002062415277
0.985 17.5710391281853
0.99 16.7610292191154
0.995 15.5160014516879
};
\addlegendentry{FS, 99th perc.}

\end{axis}

\end{tikzpicture}

%% file: img/perc_schedule_6.tex
\begin{tikzpicture}

\begin{axis}[
width=\sfwidth,
height=\sfheight,
legend cell align={left},
legend style={at={(0.005,0.99)}, anchor=north west, fill opacity=0.8, legend columns=2, draw opacity=1, text opacity=1, font=\tiny, draw=white!80!black},
tick align=outside,
tick pos=left,
x grid style={white!69.0196078431373!black},
xlabel={$p_s$},
xmin=0.9, xmax=1,
xtick style={color=black},
y grid style={white!69.0196078431373!black},
ylabel={Scheduled capacity (Mb/s)},
ymin=30, ymax=50,
ytick style={color=black},
xmajorgrids,
ymajorgrids
]
\addplot [thick, color2, mark=o, mark options={solid}]
table {%
0.9 33.2085254749003
0.905 33.2969494350697
0.91 33.3920872371022
0.915 33.4957283340951
0.92 33.6032049311064
0.925 33.7142304915814
0.93 33.8418263594114
0.935 33.9729790057055
0.94 34.1284460437028
0.945 34.2859111834161
0.95 34.4742160479121
0.955 34.6860810111783
0.96 34.915500657475
0.965 35.2049806961068
0.97 35.5692219087571
0.975 35.9735999948712
0.98 36.4894073189377
0.985 37.1384477260227
0.99 38.1894143121075
0.995 40.0242881529187
};
\addlegendentry{CS, average}

\addplot [thick, dashdotted, color2, mark=x, mark options={solid}]
table {%
0.9 34.7659298176137
0.905 34.9099166184
0.91 35.0600092164221
0.915 35.2258483781938
0.92 35.3962713568873
0.925 35.5833393311604
0.93 35.7843483967268
0.935 36.0028747787191
0.94 36.2333486257974
0.945 36.4908691951357
0.95 36.7750829869019
0.955 37.0935707506687
0.96 37.4579059862318
0.965 37.8845377125683
0.97 38.3997833058044
0.975 39.0204686709216
0.98 39.8125439149144
0.985 40.8575060275876
0.99 42.4457566058064
0.995 45.6134004693183
};
\addlegendentry{FS, average}
%
%

\addplot [thick, color4, mark=o, mark options={solid}]
table {%
0.9 36.8704951418864
0.905 36.8990749685905
0.91 36.9209012061906
0.915 36.9605891821469
0.92 36.9846565084916
0.925 37.0391862113173
0.93 37.1100832125508
0.935 37.2328976492923
0.94 37.3206921890681
0.945 37.4428279592155
0.95 37.5625191878589
0.955 37.7000000112502
0.96 37.8773856219687
0.965 38.0931752726566
0.97 38.3580299516182
0.975 38.6936476176794
0.98 39.1601499888461
0.985 39.6921596174781
0.99 40.5884536773951
0.995 42.1758767210809
};
\addlegendentry{CS, 95th perc.}

\addplot [thick, dashdotted, color4, mark=x, mark options={solid}]
table {%
0.9 37.7126835655917
0.905 37.8184279702371
0.91 37.9486026375049
0.915 38.0901068576218
0.92 38.2269101176611
0.925 38.3880255146387
0.93 38.5480457280778
0.935 38.7251571180675
0.94 38.9134917063097
0.945 39.1336806261039
0.95 39.3923963375488
0.955 39.6676767170225
0.96 39.9873754906848
0.965 40.3759664990905
0.97 40.8309837409471
0.975 41.3786220796682
0.98 42.104978711927
0.985 43.0243504555176
0.99 44.4707426511026
0.995 47.4203959216762
};
\addlegendentry{FS, 95th perc.}

\addplot [thick, color6, mark=o, mark options={solid}]
table {%
0.9 41.9595451306577
0.905 41.957687227686
0.91 41.9696897280286
0.915 41.9877813518752
0.92 41.9775411204027
0.925 41.9601247935856
0.93 41.9555358450294
0.935 41.9181616513728
0.94 41.8696891464254
0.945 41.8226494691525
0.95 41.7866030508514
0.955 41.7267847143181
0.96 41.7051202434034
0.965 41.6692042090519
0.97 41.8182773103021
0.975 41.8734070503595
0.98 41.9656206286542
0.985 42.1663347422155
0.99 42.5150567774223
0.995 43.7882258058093
};
\addlegendentry{CS, 99th perc.}

\addplot [thick, dashdotted, color6, mark=x, mark options={solid}]
table {%
0.9 41.9862463643567
0.905 41.8820666118853
0.91 41.784555557544
0.915 41.7206424203568
0.92 41.7123344067742
0.925 41.7800671800957
0.93 41.8172187471249
0.935 41.8219458263538
0.94 41.8517589880621
0.945 41.9323783503822
0.95 42.0252542690691
0.955 42.1114916385313
0.96 42.3426808627325
0.965 42.6284744428477
0.97 42.9659381246199
0.975 43.3876520702499
0.98 43.9925204901377
0.985 44.7572811987104
0.99 45.9777718736776
0.995 48.5834062361514
};
\addlegendentry{FS, 99th perc.}
\end{axis}

\end{tikzpicture}